\documentclass[twocolumn]{aastex63}
\usepackage{graphicx}
\usepackage{wrapfig}
\usepackage{natbib}
\usepackage{color}
\graphicspath{{./figures/}} 
\usepackage{mathtools}
\usepackage{epstopdf}
\usepackage{hyperref} 
\def    \apjl  		{\rm {ApJL}}

\def    \apjl  		{\rm {ApJL}}

\def	\cm		{\,{\rm {cm}}}
\def	\K		{\,{\rm K}}
\def	\g		{\,{\rm {g}}}
\def	\mum	{\,{\mu \rm{m}}}
\def	\erg		{\,{\rm {erg}}}
\def	\s		{\,{\rm {s}}}

\def \bea {\begin{eqnarray}}
\def \ena {\end{eqnarray}}

\begin{document}
\shorttitle{Grain Disruption Effects on SNe Ia}
\shortauthors{Giang, Hoang, and Tram}
\title{Time-Varying Extinction, Polarization, and Colors of Type Ia Supernovae due to Rotational Disruption of Dust Grains}

\author{Nguyen Chau Giang}
\affiliation{University of Science and Technology of Hanoi, VAST, 18 Hoang Quoc Viet, Vietnam}

\author{Thiem Hoang}
\affiliation{Korea Astronomy and Space Science Institute, Daejeon 34055, South Korea}
\affiliation{Korea University of Science and Technology, 217 Gajeong-ro, Yuseong-gu, Daejeon, 34113, South Korea}

\author{Le Ngoc Tram}
\affiliation{SOFIA-USRA, NASA Ames Research Center, MS 232-11, Moffett Field, 94035 CA, USA}
\affiliation{University of Science and Technology of Hanoi, VAST, 18 Hoang Quoc Viet, Vietnam}

\correspondingauthor{Thiem Hoang}
\email{thiemhoang@kasi.re.kr}

\begin{abstract}
Photometric and polarimetric observations toward type Ia supernovae (SNe Ia) frequently report an unusually low total-to-selective extinction ratio ($R_{\rm V} < 2$) and small peak wavelength of polarization ($\lambda_{\rm max}< 0.4 \mum$). Recently, Hoang et al. proposed that the increase in the abundance of small grains relative to large grains near SNe Ia due to RAdiative Torque Disruption (RATD) can explain this puzzle. To test this scenario, we will perform detailed modeling of dust extinction and polarization of SNe Ia accounting for grain disruption by RATD and grain alignment by RAdiative Torques (RATs). For dust clouds at distance $d< 4$ pc from the source, we find that $R_{\rm V}$ decreases rapidly from the standard value of $3.1$ to $\sim 1.5$ after a disruption time $t_{\rm disr}< 40$ days. We then calculate the observed SNe Ia light curve and find that the colors of SNe Ia would change with time due to time-varying extinction for dust clouds at distance $d<4$ pc. We also calculate the wavelength-dependence polarization produced by grains aligned with the magnetic fields by RATs. We find that $ \lambda_{\rm max}$ decreases rapidly from $\sim 0.55 \mum$ to $\sim 0.15 \mum$ over an alignment time of $t_{\rm align}< 10$ days due to the enhanced alignment of small grains. By fitting the theoretical polarization curve with the Serkowski law, we find that the parameter $K$ from the Serkowski law increases when large grains are disrupted by RATD which can explain the $K$ vs. $\lambda_{\rm max}$ data observed for SNe Ia. Finally, we discover an anti-correlation between $K$ and $R_{\rm V}$ which might already be supported by SNe Ia observational data. Our results demonstrate the important effect of rotational disruption of dust grains by radiative torques on the time-dependent extinction, polarization, and colors of SNe Ia.
\end{abstract}
\keywords{supernovae: general, kilonova, dust, extinction, massive stars; stars: neutron}

\section{Introduction}\label{sec:intro}
Type Ia supernovae (hereafter SNe Ia) is an explosion of the white dwarf star in the binary system when the accretion of material from the evolving companion star onto the white dwarf increases its mass to $\sim 1.4M_{\odot}$, i.e., the Chandrasekhar limit (\citealt{Hill00}). Beyond this mass limit, the electron degeneracy pressure is insufficient to support against the gravity, and the white dwarf collapses, releasing a large amount of material and energy into the surrounding environment. SNe Ia have similar intrinsic luminosity because they explode at similar mass, for which SNe Ia are considered "standard candles" to measure cosmological distances (\citealt{1998AJ....116.1009R}). SNe Ia are also useful to study the physical and chemical properties of the interstellar medium of external galaxies (\citealt{Nobi08}; \citealt{Brown15}; \citealt{Foley14}).

Dust along the line of sight toward SNe Ia absorbs, scatters, and polarizes the SNe radiation in the UV-NIR wavelength range (see, e.g., \citealt{Hoang17}). Photometric observations of SNe Ia show that the total-to-selective extinction ratio $R_{\rm V}=A_{\rm V}/E(B-V)$, with $A_{\rm V}$ the optical extinction and $E(B-V)=A_{\rm B}-A_{\rm V}$ the color excess, is much lower than the typical value of interstellar dust in the Milky Way of $R_{\rm V}\sim 3.1$. For example, 80 SNe Ia studied by \cite{Nobi08} have an average value of $R_{\rm V} \sim 1.75$. Using Hubble space telescope and Swift satellites, \cite{Ama15} and \cite{Wang12} estimated a range of $R_{\rm V} \sim 1.4-3$. These examples can be explained by the peculiar properties of dust in the SNe Ia's host galaxy. Other suggestions include the scattering of circumstellar (CS) material in the vicinity of SNe Ia (\citealt{Wang08}; \citealt{Goobar08}) and the enhancement of small grains (smaller than $ 0.05 \mum$) relative to large grains (size above $0.1\mum$) in the local interstellar environment of SNe Ia (\citealt{Phillips13}; \citealt{Hoang17}).

Moreover, polarimetric observations also reveal unusually low values of the wavelength at the maximum polarization curve of SNe Ia, i.e., $ \lambda_{\rm max}< 0.4 \mum$. For instance, four SNe SN 1986G, 2006X, 2008fp and 2014J shows $\lambda_{\rm max}$ from $0.05\mum$ to $0.43\mum$ (\citealt{Kawa14}; \citealt{Patat15}; \citealt{Hoang17}), and nine SNe Ia listed in \cite{Zelaya17} have $\lambda_{\rm max}< 0.4 \mum$, much lower than the standard value of the ISM of $\lambda_{\rm max}=0.55\mum$ (\citealt{Whit92}). Based on simultaneous fitting to extinction and polarization of SNe Ia, \cite{Hoang17} found that the unusual values of $\lambda_{\rm max}$ can be reproduced if small grains  can be aligned efficiently as large grains. 
 
The enhancement in the abundance of small grains relative to large grains around SNe Ia is the popular explanation for the anomalous values of $R_{\rm V}$ and $\lambda_{\rm max}$ \citep{Hoang17}. Several mechanisms were proposed to explain the disruption of large grain such as grain-grain collision, thermal sublimation. However, the thermal sublimation mechanism only destroys grains within radius of $R_{\rm sub}\simeq 0.015(L_{\rm UV}/10^{9}L_{\rm \odot})^{1/2}(T_{\rm sub}/1800\K)^{-2.8}$ pc where $L_{\rm UV}$ is the total luminosity of SNe Ia in the NUV-optical band and $T_{\rm sub}$ is the grain sublimation temperature (\citealt{Wax20}; \citealt{Hoang19}). The grain-grain collision mechanism destroys dust by the strong interaction between dust and gas in the interstellar medium (ISM). This process takes near 100 years for gas release from SNe Ia reach the dust in the ISM at 1 pc if we assume gas moves at 0.028 c. However, the values of $R_{\rm V}$ and $\lambda_{\rm max}$ are measured during the early-time observations (less than a few weeks after the explosion). Thus, the question of why small grains are predominant around SNe Ia remains mysterious.
 
The mystery in the predominance of small grains around SNe Ia might be resolved when \cite{Hoang19} introduced a new mechanism of grain disruption, so-called Radiative Torque Disruption (RATD), which is based on centrifugal force within rapidly spinning grains spun-up by radiative torques. This RATD mechanism can disrupt large grains into small fragments on a characteristic timescale of tens of days, which dramatically changes the grain size distribution and observable properties of dust near SNe Ia. Therefore, the first goal of this paper is to model the time-dependent extinction and polarization curves due to dust grains which are being disrupted by RATD. We will demonstrate that RATD can indeed reproduce the low values of $R_{\rm V}$ and $\lambda_{\rm max}$ as observed toward SNe Ia.

Intrinsic colors and light curves of SNe Ia are critically crucial for achieving an accurate measurement of cosmological constant and better constraint of dark energy, which are the next frontiers of SNe Ia cosmology. Yet, the major systematic uncertainty arises from uncertainty in dust extinction toward SNe Ia (see, e.g., \citealt{Scolnic:2019ug}). Moreover, intrinsic lightcurve is essential for understanding the trigger mechanism of SNe Ia. The second goal of this paper is to quantify the time-variation of the SNe Ia colors due to time-varying extinction caused by RATD. 
 
The structure of this paper is as follows. In Section \ref{sec:disrupt}, we review the RATD mechanism and calculate the disruption size of dust grains as functions of time and cloud distance toward SNe Ia. In Sections \ref{sec:ext} and \ref{sec:pol}, we will model dust extinction and polarization curves induced by grains aligned with the magnetic fields using the grain size distribution determined by RATD and grain alignment function by RAT alignment theory. In Section \ref{sec:lightcurve}, we will predict the observed light curve of SNe Ia, which vary over time due to RATD. Discussion and summary of our main findings are given in Section \ref{sec:discuss} and \ref{sec:summary}, respectively.

\section{Rotational Disruption of dust grains by SNe Ia}\label{sec:disrupt}
\subsection{The RATD mechanism}

When dust grains of irregular shapes are located in an intense radiation field, such as around SNe Ia, radiative torques induced by the interaction of anisotropic radiation with the irregular grain can spin the grain up to extremely fast rotation (\citealt{Draine96}; \citealt{Laza07}; \citealt{Hoang19}). Such suprathermal rotation of grains produces a strong centrifugal force, which can exceed the tensile strength of grain material, so that grains are disrupted into small fragments. Note that the grain shape is assumed to be irregular, but we can approximate it as an equivalent spherical grain of the effective size $a$ which has the same volume as the irregular one (see, e.g., \citealt{Draine96}; \citealt{Laza07}). This RATD mechanism was described in detail in \cite{Hoang19}. Therefore, here we briefly describe the RATD mechanism for reference.

The luminosity of SNe Ia varies over time, which can be approximately given by an analytical formula \citep{Zheng17}:
\begin{equation}\label{eq:1}
L_{\rm SNIa}(t)= L_{0} \left(\frac{t-t_{0}}{t_{p}}\right)^{\alpha_{r}} \left[1+\left(\frac{t-t_{\rm 0}}{t_{p}}\right)^{s\alpha_{d}}\right]^{-2/s},
\end{equation}
where $L_{\rm 0}$ is the scaling parameter, and we take $ L_{\rm 0}=2 \times 10^{10} L_{\odot}$. The first term describes the rising luminosity of SNe Ia (\citealt{Riess99}) while the second term is the decrease luminosity function after the peak around 20 days. We take $\alpha_{\rm r}=$ 2, $ \alpha_{\rm d}=$ 2.5, $s=$ 1 , $t_{0}=$ 0 days and $ t_{\rm p}=$ 23 days (\citealt{Hoang19}).

Since the luminosity of SNe Ia is time-dependent, the grain angular velocity can only be obtained by numerically solving the equation of motion (see \citealt{Hoang19}): 
\begin{equation} \label{eq:2}
\frac{I d\omega}{dt} = \Gamma_{\rm RAT}-\frac{I\omega}{\tau_{\rm damp}},
\end{equation}
where $I$ is the grain inertia moment with $I=8 \pi \rho a^{5}/15$, $\Gamma_{\rm RAT}$ is radiative torques, and $\tau_{\rm damp}$ is the characteristic timescale of grain rotational damping. 

Usually, $\Gamma_{\rm RAT}$ is the radiative torque averaged over the radiation spectrum as given by: 
\bea \label{eq:3}
\Gamma_{\rm RAT} = \pi a^{2} \gamma u_{\rm rad} \left(\frac{\bar{\lambda}}{2\pi}\right) \overline{Q}_{\Gamma},
\ena
where $\pi a^{2}$ is geometric cross-section, $\gamma$ is the anisotropy degree of the radiation field ($ 0\leq \gamma \leq 1$), $\bar{\lambda}$ is the mean wavelength of the radiation spectrum, $\overline{Q}_{\Gamma}$ is the averaged RAT efficiency. Here $u_{\rm rad}$ is the radiation energy density of the radiation field which is equal to:
\begin{align} \label{eq:4}
u_{\rm rad} =  \frac{L_{\rm bol}e^{\rm -\tau}}{4 \pi c d^{\rm 2}},
\end{align}
where $\tau$ the effective optical depth defined by $e^{-\tau}=\int u_{\rm \lambda} \times e^{-\tau_{\rm \lambda}} d\lambda/u_{\rm rad}$. Let $U=u_{\rm rad}/u_{\rm ISRF}$ where $u_{\rm ISRF}=8.64\times 10^{-13}\erg\cm^{-3}$ is the energy density of the average radiation field in the solar neighborhood (see \citealt{Hoang19}).

In general, $Q_{\Gamma}$ depends on the grain size and shape, the wavelength and the angle between the direction of the grain and incoming radiation (\citealt{Laza07}; \citealt{Herranen:2019kj}). Spectacularly, when the anisotropic radiation direction is parallel to the axis of maximum moment of the inertia of grain, \cite{Laza07}, \cite{Hoang08} and \cite{Hoang14} showed that $Q_{\Gamma}$ could be approximated by a power law:
\bea \label{eq:5}
\overline{Q}_{\Gamma}=\left\{
\begin{array}{l l}	
     \approx 2\left(\frac{\bar{\lambda}}{a} \right)^{-2.7} & {\rm for~} a \lesssim a_{\rm trans}\\
     \sim 0.4 & {\rm for~} a > a_{\rm trans},
\end{array}\right\}
\ena
where $a_{\rm trans}=\frac{\bar{\lambda}}{1.8}$.

The damping of grain rotation can arise from collisions with gas atoms followed by evaporation and re-emission of infrared radiation. The total damping rate can be written as (\citealt{Hoang19}):
\bea \label{eq:6}
\tau_{\rm damp} = \frac{\tau_{\rm gas}}{1+F_{\rm IR}},
\ena
where $\tau_{\rm gas}$ is the rotational damping timescale due to collisions given by
\bea \label{eq:7}
\tau_{\rm gas} \simeq 8.74\times 10^{4}a_{-5} \hat{\rho} \left(\frac{30 \cm^{-3}}{n_{\rm H}}\right) \left(\frac{100\K}{T_{\rm gas}}\right)^{1/2}\rm yr
\ena
with $a_{-5} = a/(10^{-5}\cm)$ and $\hat{\rho}=\rho/3\g\cm^{-3}$ with $\rho$ the dust mass density, $n_{\rm H}$ the gas density and the gas temperature $T_{\rm gas}$ (see \citealt{Draine96}; \citealt{Hoan09}): 

Assuming that grains are in thermal equilibrium between heating by stellar radiation and cooling by IR re-emission, the IR damping coefficient can be estimated as (see \citealt{Draine98}):
\bea \label{eq:8}
F_{\rm IR} \simeq \frac{0.4U^{2/3}}{a_{-5}} \frac{30\cm^{-3}}{n_{\rm H}} \left(\frac{100\K}{T_{\rm gas}}\right)^{1/2}.
\ena

A spinning grain of angular velocity $\omega$ is disrupted when the centrifugal stress $S=\rho a^{2} \omega^{2}/4$ exceeds the maximum tensile strength of grain material $S_{\rm max}$. The critical velocity at which the disruption occurs is determined by $S\equiv S_{\rm max}$, which yields
\bea \label{eq:9}
\omega_{\rm disr} = \frac{2}{a} \left(\frac{S_{\rm max}}{\rho}\right)^{1/2},
\ena
where $S_{\rm max}$ depends on the grain material, internal structure, and the grain size (see \citealt{Hoang19}). 

The disruption time is roughly estimated by
\bea
t_{\rm disr}&=&\frac{I\omega_{\rm disr}}{dJ/dt}=\frac{I\omega_{\rm disr}}{\Gamma_{\rm RAT}}\nonumber\\
&\simeq& 38(\gamma U_{6})^{-1}\bar{\lambda}_{0.5}^{1.7}\hat{\rho}^{1/2}S_{\max,7}^{1/2}a_{-5}^{-0.7}{~\rm days}\label{eq:tdisr}
\ena
for $a_{\rm disr}<a \lesssim a_{\rm trans}$, and
\bea
t_{\rm disr}\simeq& 2.5(\gamma U_{6})^{-1}\bar{\lambda}_{0.5}^{-1}\hat{\rho}^{1/2}S_{\max,7}^{1/2}a_{-5}^{2}{~\rm days}
\ena
for $a>a_{\rm trans}$ where $U_{6}=U/10^{6}$.

Assuming the mass density $\rho=3.5\g\cm^{-3}$ and $2.2\g\cm^{-3}$ for silicate and graphite grains, one can see that the disruption time of graphite grains is shorter than that of silicate grains due to its lower mass density. 

\subsection{Grain disruption sizes}
The grain disruption size by the RATD mechanism depends on the tensile strength of grain material $S_{\rm max}$, the strength of the radiation field which is determined by the cloud distance to SNe Ia, the spectrum of the radiation field, and the irradiation time. As in \cite{Hoang19}, SNe Ia are considered a black-body radiation source and have a constant effective temperature of SNe Ia of $15000 \K$. This yields the mean wavelength of $\bar{\lambda}\approx 0.35\mum$ . We assume that the radiation field is unidirectional, i.e., $\gamma =1$.  

We solve Equation (\ref{eq:2}) numerically to obtain $\omega_{\rm RAT}(t)$ for a grid of grain sizes $a$ and compare it with $\omega_{\rm disr}(a)$ (Eq. \ref{eq:9}). The grain disruption size $a_{\rm disr}$ and disruption time $t_{\rm disr}$ are then determined (see \citealt{Hoang19} for details). As discussed in \cite{Hoang19}, small grains acquire lower radiative torques than large grains such that they can be spun-up to lower rotation rates. In addition, Equation (\ref{eq:9}) reveals that small grains have the higher critical angular velocity to be disrupted. As a result, large grains can be disrupted into smaller ones, but small grains may survive RATD.

\begin{figure}[h]
        \includegraphics[width=0.5\textwidth]{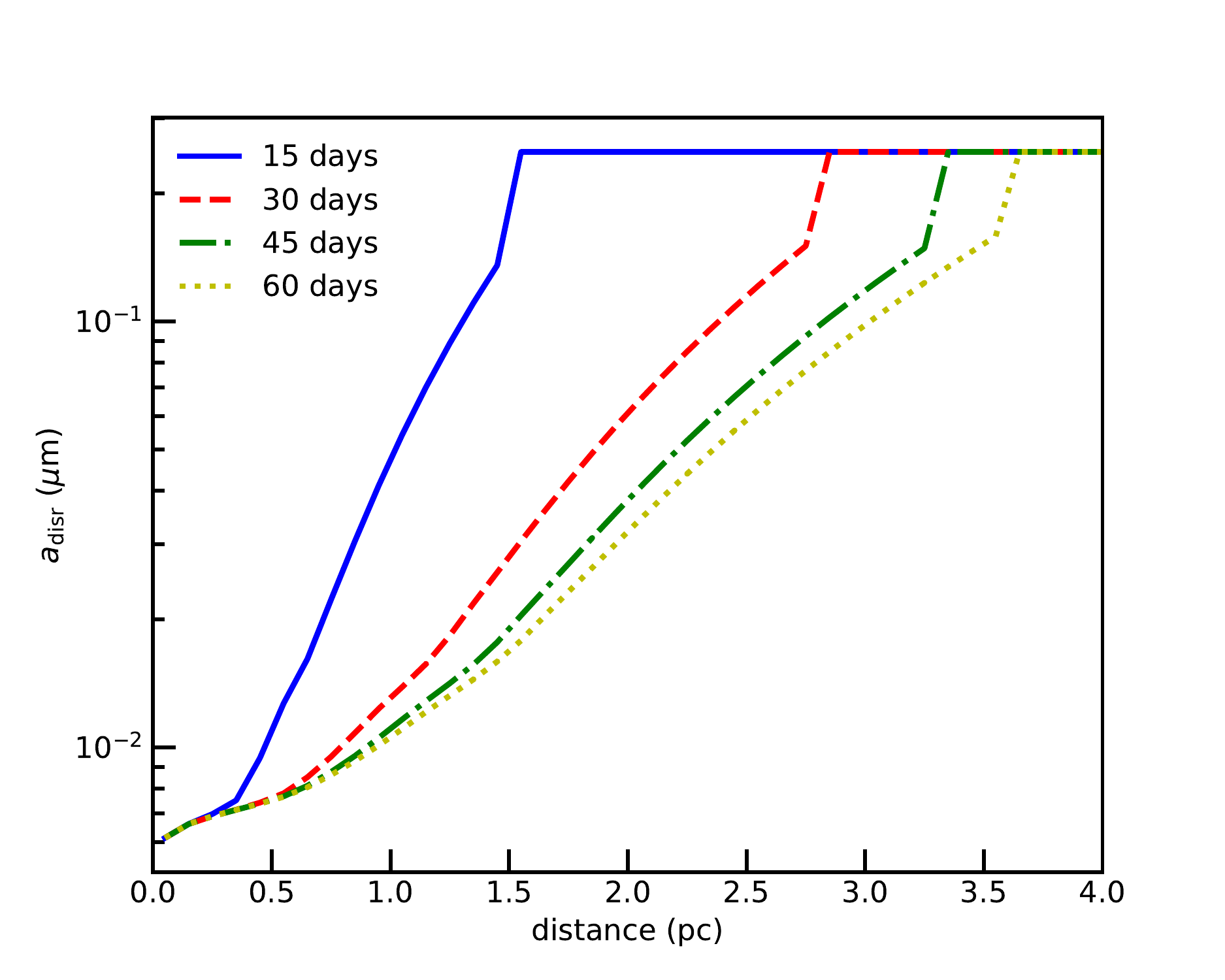}
        \includegraphics[width=0.5\textwidth]{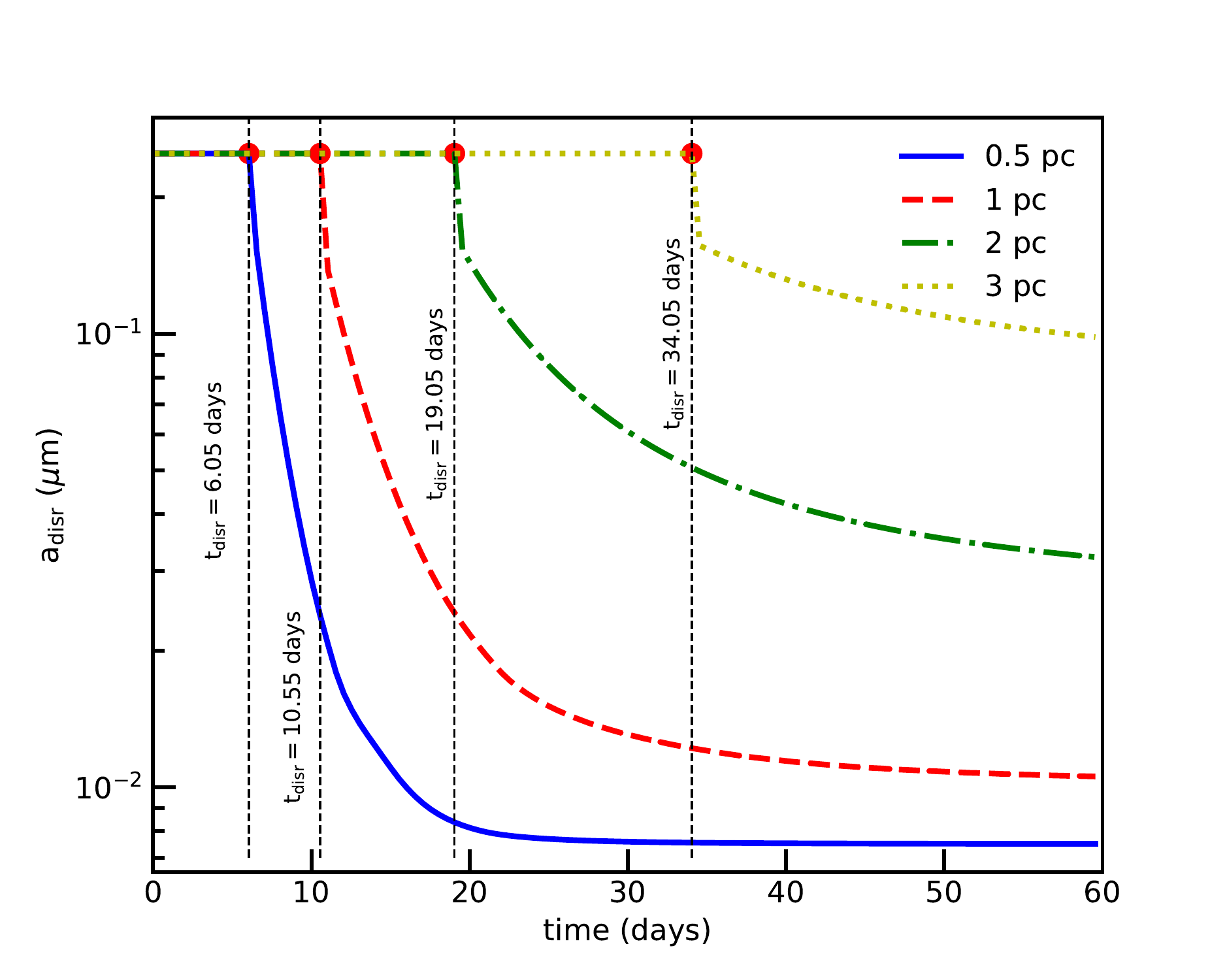}
        \caption{Upper panel: disruption size of silicate grains as a function of cloud distance evaluated at 15 days, 30 days, 45 days and 60 days, assuming the tensile strength $S_{\rm max}=10^{7}\erg\cm^{-3}$. Lower panel: disruption size of silicate vs. time for the dust cloud located between $0.5-3$ pc. The disruption size starts to decrease from an original value of $0.25\mum$ after some disruption time $t_{\rm disr}$ marked by the vertical dotted line.}
        \label{fig:adisr_d}
\end{figure}

Figure \ref{fig:adisr_d} (upper panel) shows the variation of grain disruption size with cloud distance at $t =$ 15, 30, 45 and 60 days, assuming $S_{\rm max}=10^{7}\erg\cm^{-3}$ and the typical gas density $ n_{\rm H}=30\cm^{-3}$. For distant clouds ($d>3.5$ pc), grain disruption cannot occur because of insufficient radiation intensity, and we set $a_{\rm disr}$ to the popular upper cutoff of the grain size distribution of $a_{\rm max}=0.25\mum$ (\citealt{Mathis77}). As the cloud distance decreases (i.e., radiation energy density increases as $d^{-2}$), the grain disruption size increases rapidly. Moreover, within 15 days, only grains within $d\sim 1.5$ pc can be disrupted, and after 60 days, even grains located at $d\sim $ 3 pc can be disrupted.

Figure \ref{fig:adisr_d} (lower panel) shows the variation of the grain disruption size with time for the different cloud distances where the initial value is equal to $0.25\mum$. For a given distance, the disruption size starts to decrease rapidly with time from the original value after some disruption time $t_{\rm disr}$, and it achieves saturated values when the disruption ends. Dust grains closer to SNe Ia receive higher radiation flux and have shorter $t_{\rm disr}$. As a result, one sees the decrease of $a_{\rm disr}$ begins earlier, but its saturation also ends earlier than grains at farther distances. For example, from Figure \ref{fig:adisr_d} (lower panel), one can obtain $t_{\rm disr}\sim 6, 10, 19$ and 34 days for silicate grains at $d=0.5, 1, 2$ and 3 pc, respectively (see more details in Figure 3 in \cite{Hoang19}. The decrease of $a_{\rm disr}$ starts from $t\sim t_{\rm disr}$ and achieves its saturation of $a_{\rm disr}\sim 0.01-0.03\mum$. We note that for all considered cloud distances in Figure \ref{fig:adisr_d}, the grain disruption ceases after about about 60 days because the SNe luminosity already fade away substantially after its peak.

\begin{figure}[h]
        \includegraphics[width=0.5\textwidth]{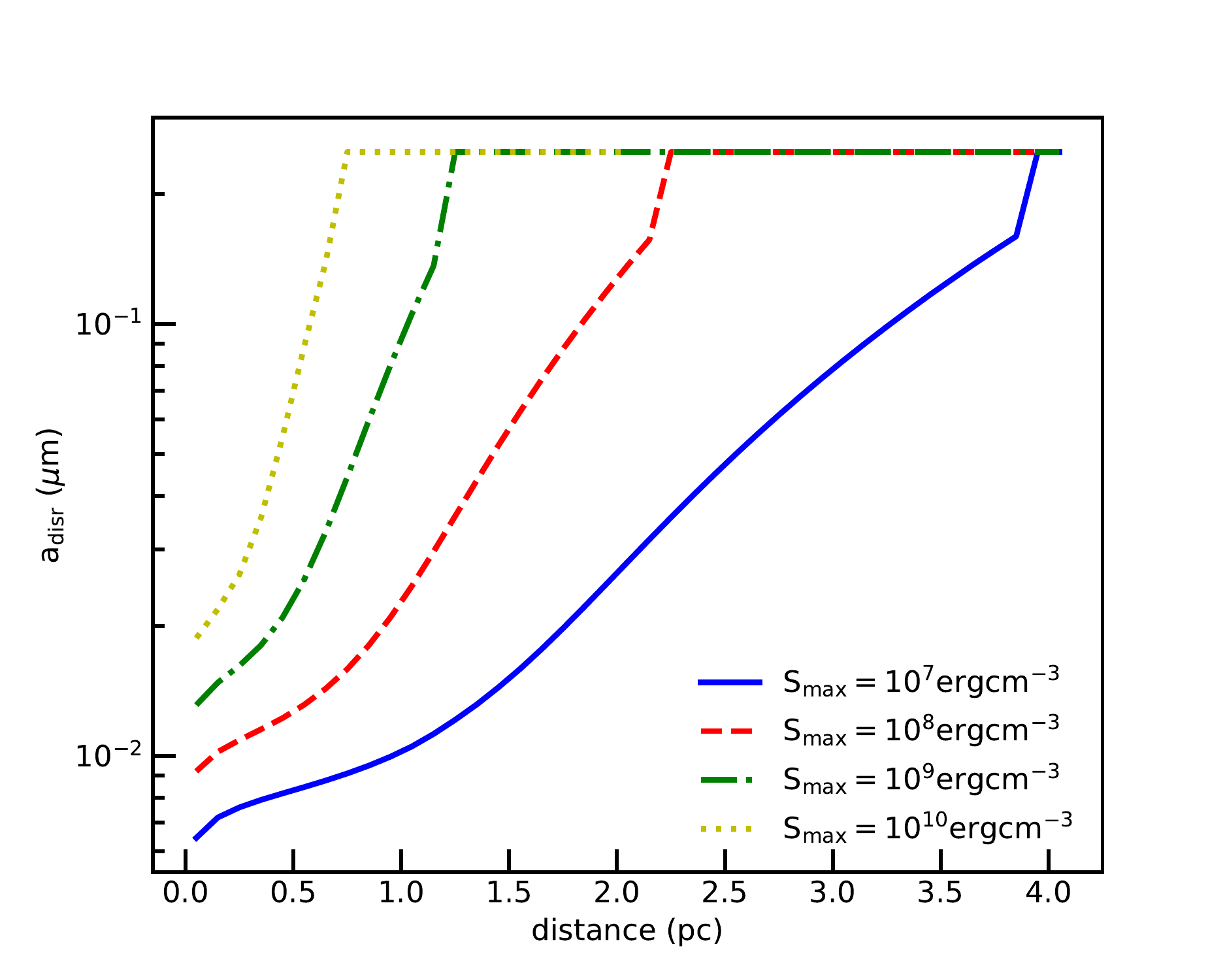}
        \includegraphics[width=0.5\textwidth]{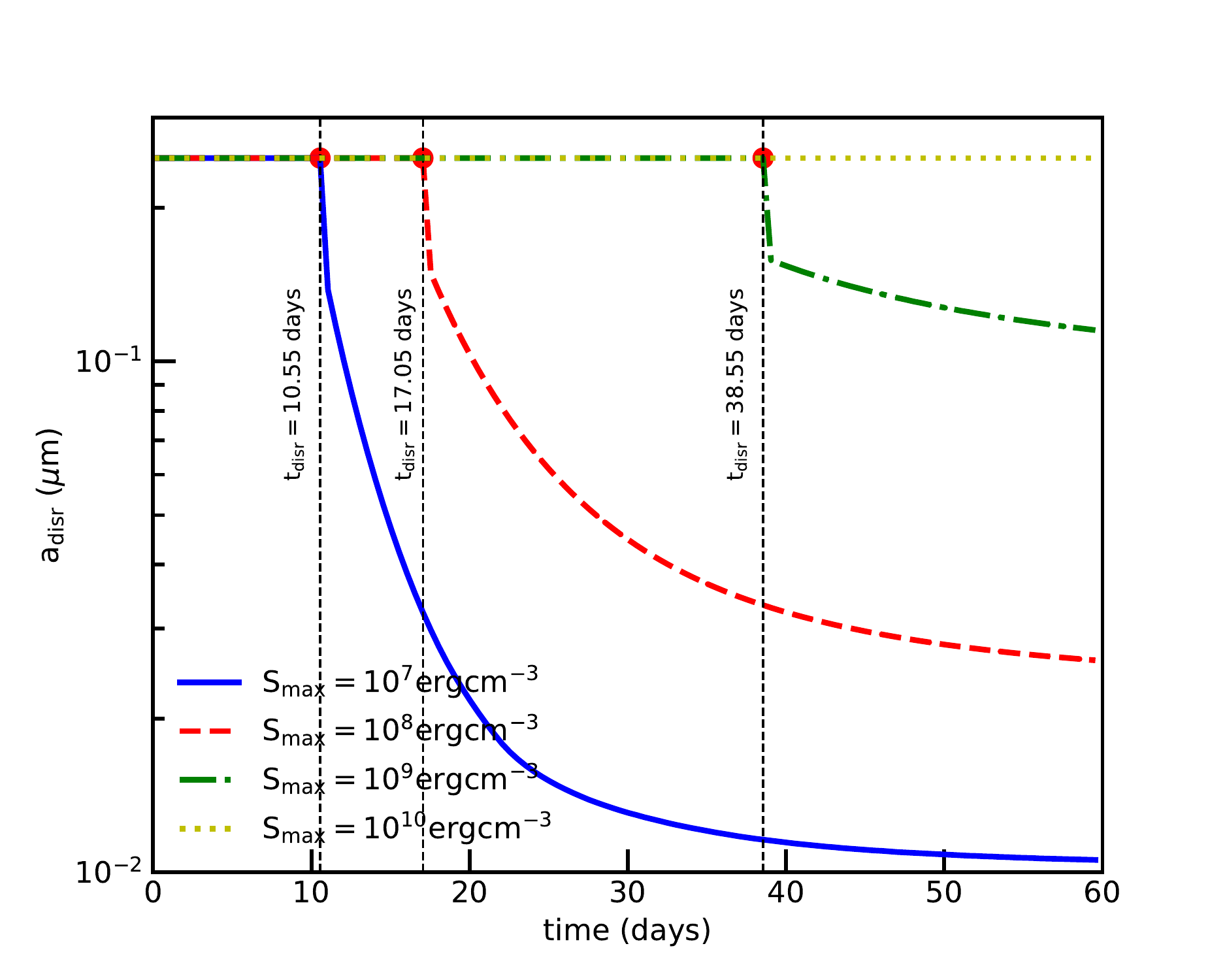}
        \caption{Disruption size of silicate vs. cloud distance evaluated at 200 days (upper panel) and disruption size vs. time for clouds at 1 pc for different values of $S_{\rm max}$. Vertical lines mark the disruption time by RATD.}
           \label{fig:adisr_Smax}
\end{figure}

Figure \ref{fig:adisr_Smax} (upper panel) presents the disruption size of silicate grains as a function of the cloud distance for different maximum tensile strengths. Grains with lower tensile strength can be disrupted out to a larger distance than those with higher $S_{\rm max}$. This can be seen from Equation (\ref{eq:9}) that grains with higher $S_{\rm max}$ have a higher critical angular velocity and require a higher radiation energy (i.e., smaller distance) to be disrupted. For example, at a given cloud distance of 1 pc, the disruption size $a_{\rm disr}\sim 0.009\mum$ for $S_{\rm max}=10^{7}\erg\cm^{-3}$, it increases to $a_{\rm disr}\sim 0.07\mum$ for $S_{\rm max}= 10^{9}\erg\cm^{-3}$, and there is no disruption for very strong material of $\rm S_{\rm max}=10^{10}\erg\cm^{-3}$. Moreover, the active region $D_{\rm disr}$ of RATD, i.e., disruption region, is smaller when grains have higher maximum tensile strength. For example, the disruption distance decreases from $D_{\rm disr}=4$ pc for $ S_{\rm max}=10^{7}\erg\cm^{-3}$ to $D_{\rm disr}=0.5$ pc for $S_{\rm max}=10^{10}\erg\cm^{-3}$.

Figure \ref{fig:adisr_Smax} (lower panel) shows the disruption size of silicate as a function of time for a dust cloud at $1$ pc for different $S_{\rm max}$. The disruption time of grains with higher $S_{\rm max}$ is longer and the grain disruption stops earlier at a higher saturated grain size. For example, the disruption time is $t_{\rm disr}\sim 10, 17$ and $38$ days for $S_{\rm max}=10^{7}, 10^{8}$ and $10^{9}\erg\cm^{-3}$, respectively. The final disruption sizes are $a_{\rm disr}\sim 0.01, 0.03$ and $0.1\mum$ for these tensile strengths.

We also calculated the grain disruption size for the different gas density $n_{\rm H}\sim 10-10^{4} \cm^{-3}$ and found that the disruption size is almost the same for clouds within 4 pc from the source. This arises from the fact that the strong radiation field of SNe Ia induces $F_{\rm IR} \textgreater \textgreater 1$, IR damping dominates over gas damping, so that $n_{\rm H}$ from $\tau_{\rm gas}$ and $F_{\rm IR}$ are canceled out in Equation \ref{eq:6}. As a result, the rotational damping rate $\tau_{\rm damp}$ only depends on the radiation strength.

\section{Extinction of SNe Light in the presence of RATD effect}\label{sec:ext}
\subsection{Grain size distribution}
For simplicity, we assume that dust grains follow a power law size distribution:
\begin{align} \label{eq:12}
\frac{dn^{j}}{da} = C_{j} n_{\rm H} a^{\alpha},
\end{align} 
where $j$ denotes the grain composition of silicate and graphite, $C_{j}$ is the constant, and $\alpha$ is the power slope. 

For the standard ISM in our galaxy, \cite{Mathis77}) derives the slope $\alpha=-3.5$, and $C_{\rm sil}=10^{-25.11}\cm^{2.5}$ for silicate grains and $C_{\rm gra}=10^{-25.14}\cm^{2.5}$ for graphite grains. The size distribution has a lower cutoff $a_{\rm min}=3.5$~\AA~determined by thermal sublimation due to temperature fluctuations of very small grains (see e.g., \citealt{Draine07}), and an upper cutoff of $a_{\rm max}=0.25\mum$ (\citealt{Mathis77}). 

In the presence of RATD, largest grains are disrupted into smaller ones, resulting in the decrease of $a_{\rm max}$, and we can set $a_{\rm max}=a_{\rm disr}$ for the diffuse ISM. Since the dust mass is conserved in the RATD picture, obviously, the grain size distribution ${dn/da}$ must be modified. 

Due to the lack of experimental study on the size distribution of resulting fragments from rotational disruption, we will assume that the new grain size distribution also follows a power law. One can imagine there are three possible ways to model the size distribution of the new size distribution. First, the power slope should vary, but the normalization constant $C$ is assumed to be fixed (model 1). Second, the slope is assumed to be fixed, but the normalization constant $C$ varies (model 2). Third, both $C$ and the slope should vary (model 3). Model 1 corresponds to the case in which the original large grain is made up of small constituents that have various sizes. The model 2 corresponds to the case where an original large grain is disrupted into smaller fragments which follows the same size distribution as the original one. Model 3 would require a detailed understanding of the internal structure of dust grains which is uncertain and thus beyond the scope of this paper. So we consider here model 1 and will discuss its difference with model 2 in Appendix \ref{apdx:sizedist}. 

For a given grain disruption size $a_{\rm disr}$, the new value of the size distribution slope is determined by the conservation of the dust mass:
\bea \label{eq:13}
\int\limits_{a_{\rm min}}^{a_{\rm max}} a^{3} C a^{-3.5} da =\int\limits_{a_{\rm min}}^{a_{\rm disr}} a^{3} C_{\rm new} a^{\alpha} da,
\ena
which yields
\begin{align} \label{eq:19}
\frac{a_{\rm disr}^{4+\alpha} - a_{\rm min}^{4+\alpha}}{4+\alpha}=\frac{a_{\rm max}^{0.5} - a_{\rm min}^{0.5}}{0.5}.
\end{align}

The new slope $\alpha$ is obtained by solving numerically Equation (\ref{eq:19}). Note that the unit of the constant $C_{\rm new}$ must be changed accordingly to conform with the new slope $\alpha$.

\subsection{Extinction curve}

The radiation intensity of SNe Ia is reduced due to the absorption and scattering (i.e., extinction) of dust and gas along the line of sight. The extinction efficiency of light by a dust grain is defined by:

\begin{align} \label{eq:10}
Q_{\rm ext} = \frac{C_{\rm ext}}{S} = \frac{C_{\rm ext}}{\pi a^{2}},
\end{align}

where $C_{\rm ext}$ is the extinction cross-section. We use a mixed-dust model comprising silicate and graphite materials (\citealt{Wein01}; \citealt{Draine07}) and take $C_{\rm ext}$ calculated for oblate spheroidal grains of axial ratio $a/b=2$ from \cite{Hoang13}.  

The extinction of supernova light at wavelength $\lambda$ in a unit of magnitude per atom is given by (see e.g., \citealt{Hoang13}):

\begin{align} \label{eq:11}
\frac{A(\lambda)}{N_{\rm H}} = \sum_{j=sil,gra} 1.086 \displaystyle\int\limits_{a_{\rm min}}^{a_{\rm max}}  C_{\rm ext}^{j}(a)\left(\frac{1}{n_{\rm H}}\frac{dn^{j}}{da}\right)da,
\end{align}

where $N_{\rm H}=\int n_{\rm H}dz=n_{\rm H}L$ with $L$ the path length is the column density, $dn^{j}/da$ is the grain size distribution of dust component $j$, and $a_{\rm max}=a_{\rm disr}$.

For a given $a_{\rm disr}$, the new grain size distribution function is obtained, and we can calculate the wavelength-dependence extinction by dust being modified by RATD using Equation (\ref{eq:11}).

\begin{figure}[!htb]
        \includegraphics[width=0.5\textwidth]{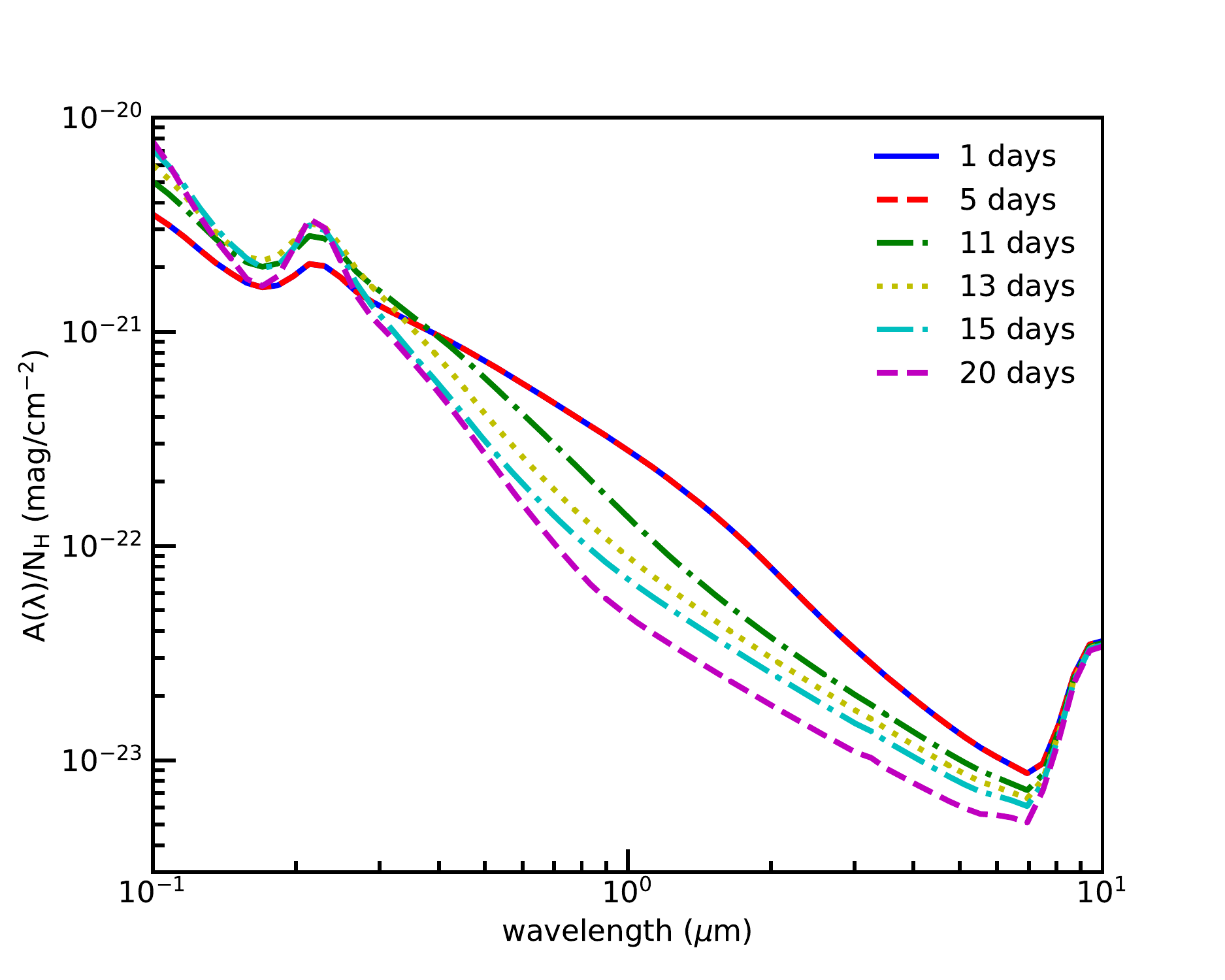}
        \includegraphics[width=0.5\textwidth]{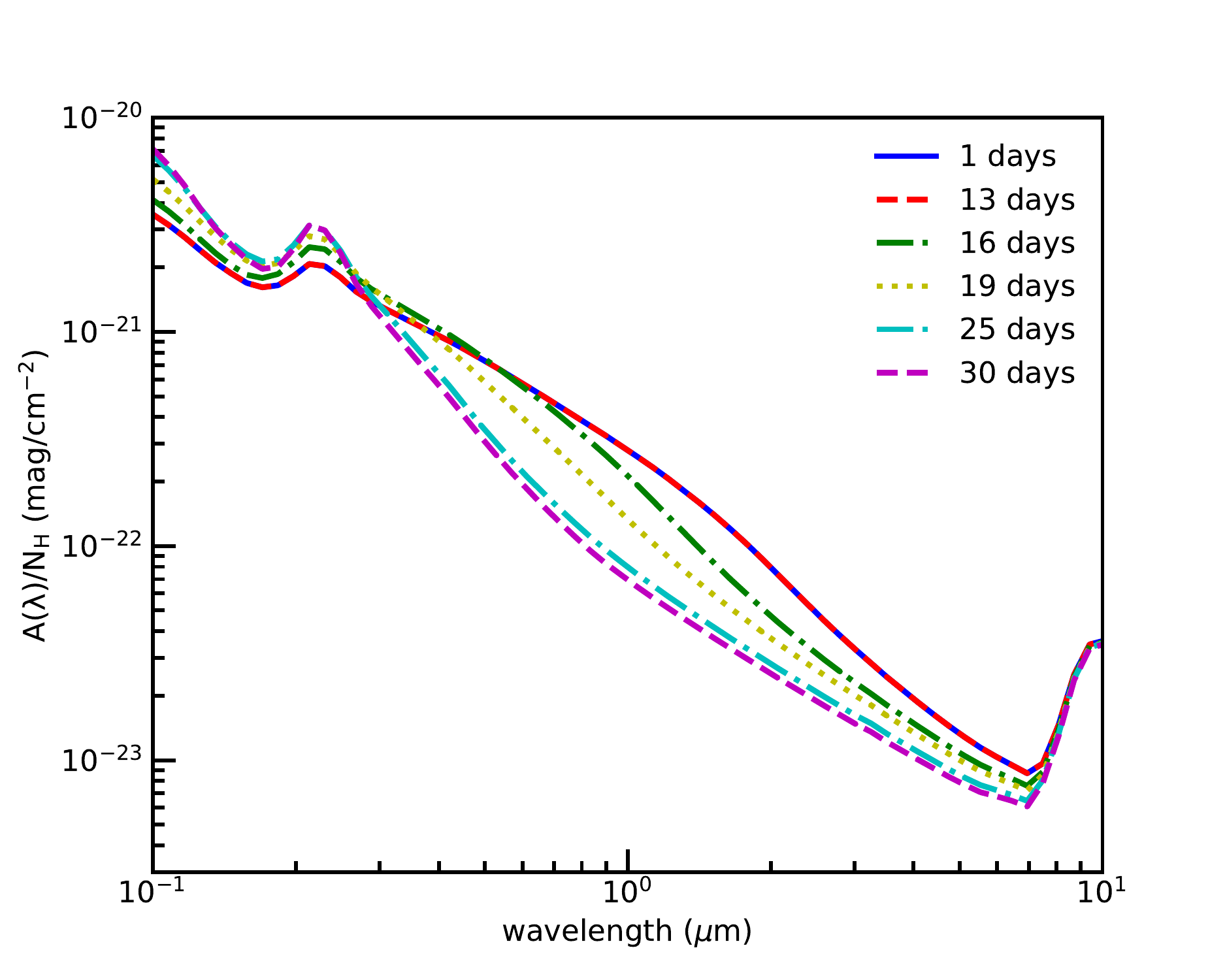}
        \caption{Extinction curves evaluated at different times for $S_{\rm max}=10^{7}\erg\cm^{-3}$ (upper panel) and $S_{\rm max}=10^{8}\erg\cm^{-3}$ (lower panel), assuming the dust cloud at 1 pc from SNe Ia. Extinction at $ \lambda>0.4\mum$ decreases while extinction at $ \lambda<0.4\mum$ increases over time.}
           \label{fig:Alamda_t}
\end{figure}

Figure \ref{fig:Alamda_t} shows the extinction curves evaluated at different times for dust grains located at distance $d= 1$ pc from the source and the maximum tensile strength $S_{\rm max}=10^{7}\erg\cm^{-3}$ (upper panel) and $S_{\rm max}=10^{8}\erg\cm^{-3}$ (lower panel). In the upper panel, the extinction curve at $\rm t = 5$ days (red dashed line) is the same as the extinction at $t=1$ days because $t<t_{\rm disr}$ (see Eq. \ref{eq:tdisr}). For $t> t_{\rm disr}\sim 10$ days (see Figure \ref{fig:adisr_d}, lower panel), the optical-NIR extinction decreases rapidly with time due to the removal of large grains by RATD. On the other hand, the UV extinction is increased due to the enhancement in the abundance of small grains by RATD. The extinction at $\lambda > 7\mum$ is essentially unchanged because of $\lambda\gg (2\pi a)$.

The lower panel of Figure \ref{fig:Alamda_t} shows the extinction curve for grains with $S_{\rm max}=10^{8}\erg\cm^{-3}$. The variation in the extinction curve with time is similar to the upper panel, but it starts later because of larger disruption time with $t_{\rm disr}\sim 15$ days (see Figure \ref{fig:adisr_Smax} (lower panel)). 

\begin{figure}[h]
        \includegraphics[width=0.5\textwidth]{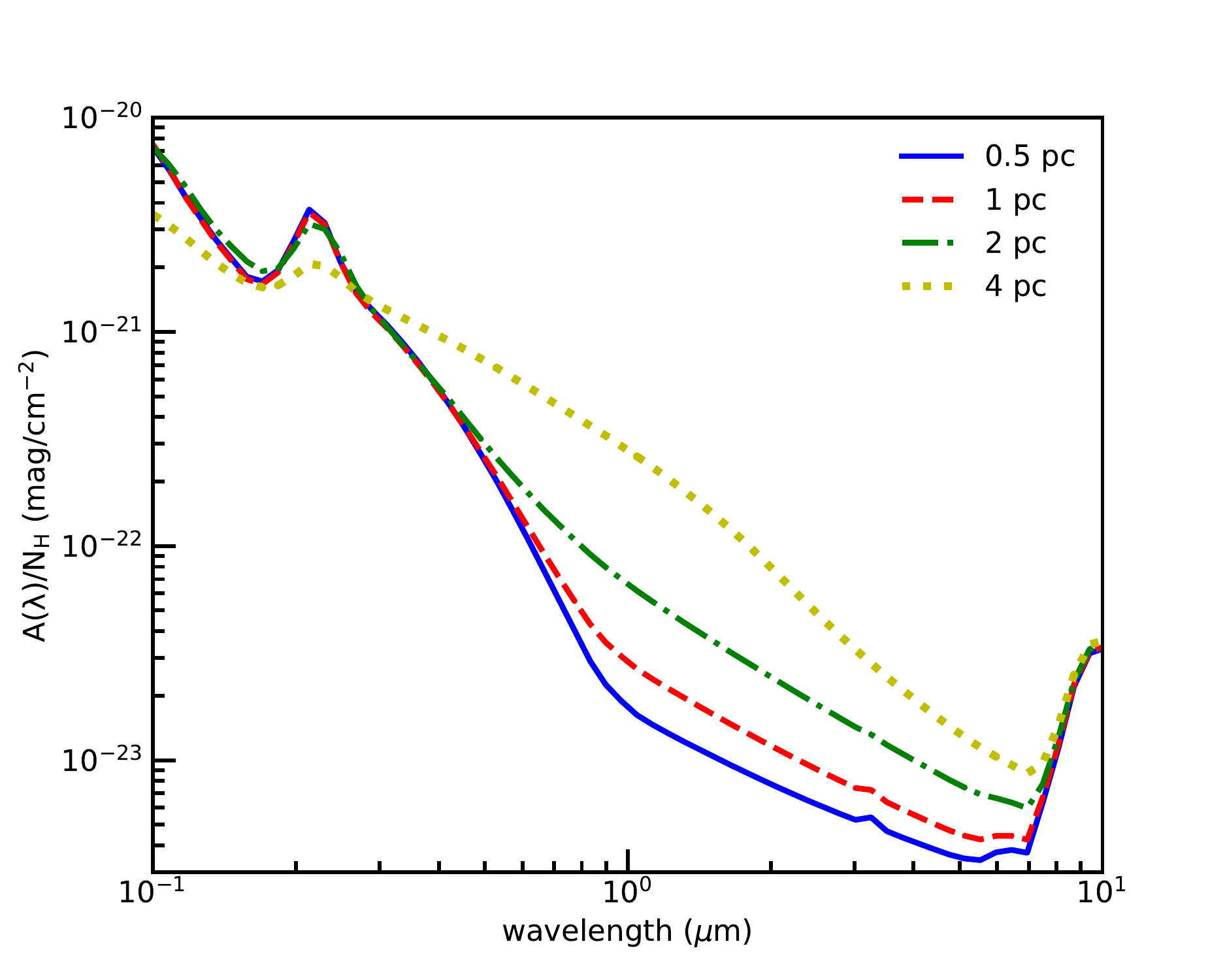}
        \includegraphics[width=0.5\textwidth]{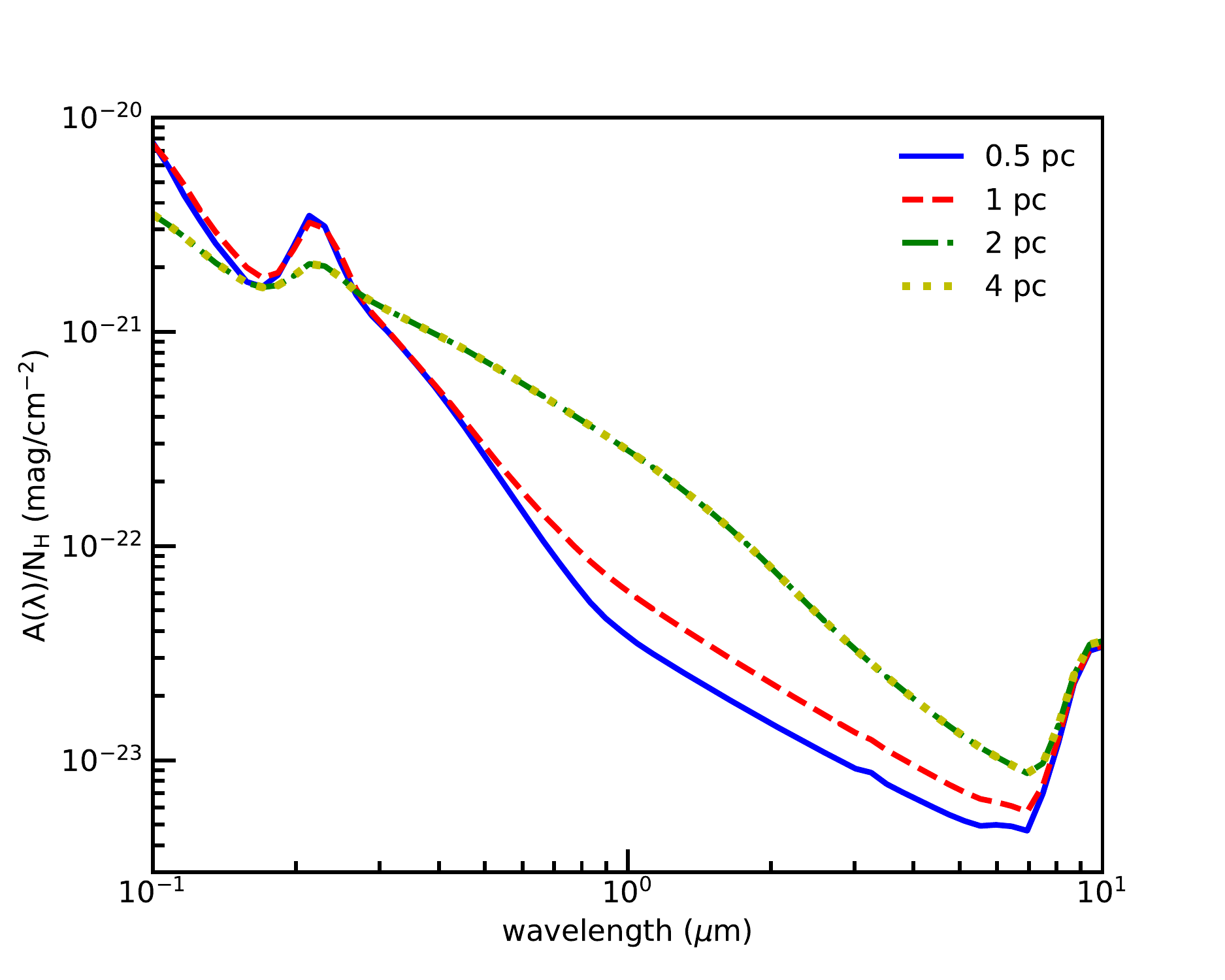}
        \caption{Extinction curve evaluated at $t=45$ days for the different cloud distance $d= 0.5, 1, 2$ and 4 pc for $S_{\rm max}=10^{7}\erg\cm^{-3}$ (upper panel) and $S_{\rm max}=10^{8}\erg\cm^{-3}$ (lower panel). Extinction curve at $d=4$ pc is closely similar to that of the original dust without RATD.}
           \label{fig:Alamda_d}
\end{figure}

Figure \ref{fig:Alamda_d} shows the extinction curve for the different cloud distances at $t=45$ days, assuming $S_{\rm max}=10^{7}\erg\cm^{-3}$ (upper panel) and $S_{\rm max}=10^{8}\erg\cm^{-3}$ (lower panel). The optical-NIR extinction decreases while the NUV extinction increases with decreasing the cloud distance. This is a direct consequence of the dependence of the grain disruption size on the cloud distance as shown in Figure \ref{fig:adisr_d}.

\begin{figure*}
    \includegraphics[width=0.5\textwidth]{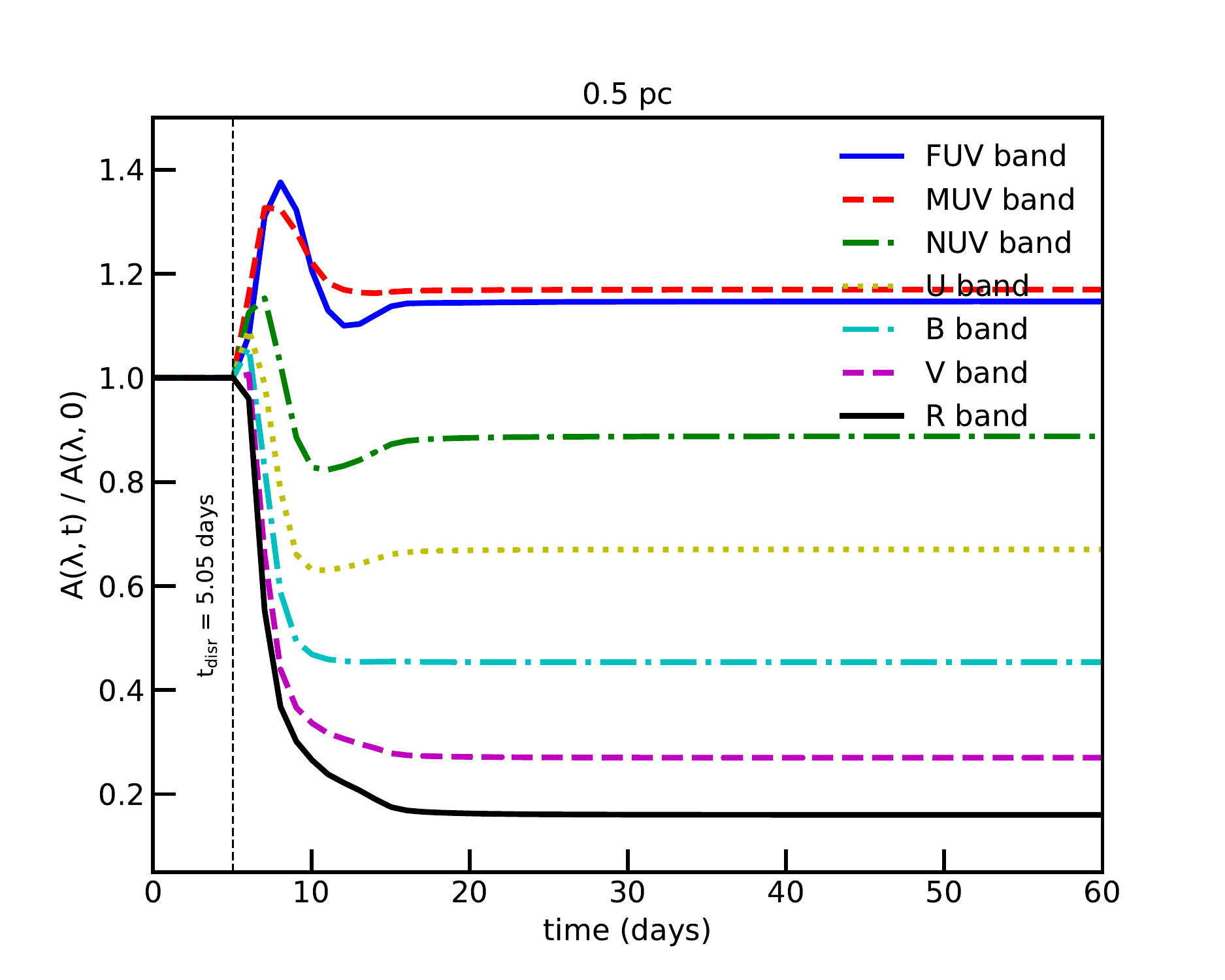}
    \includegraphics[width=0.5\textwidth]{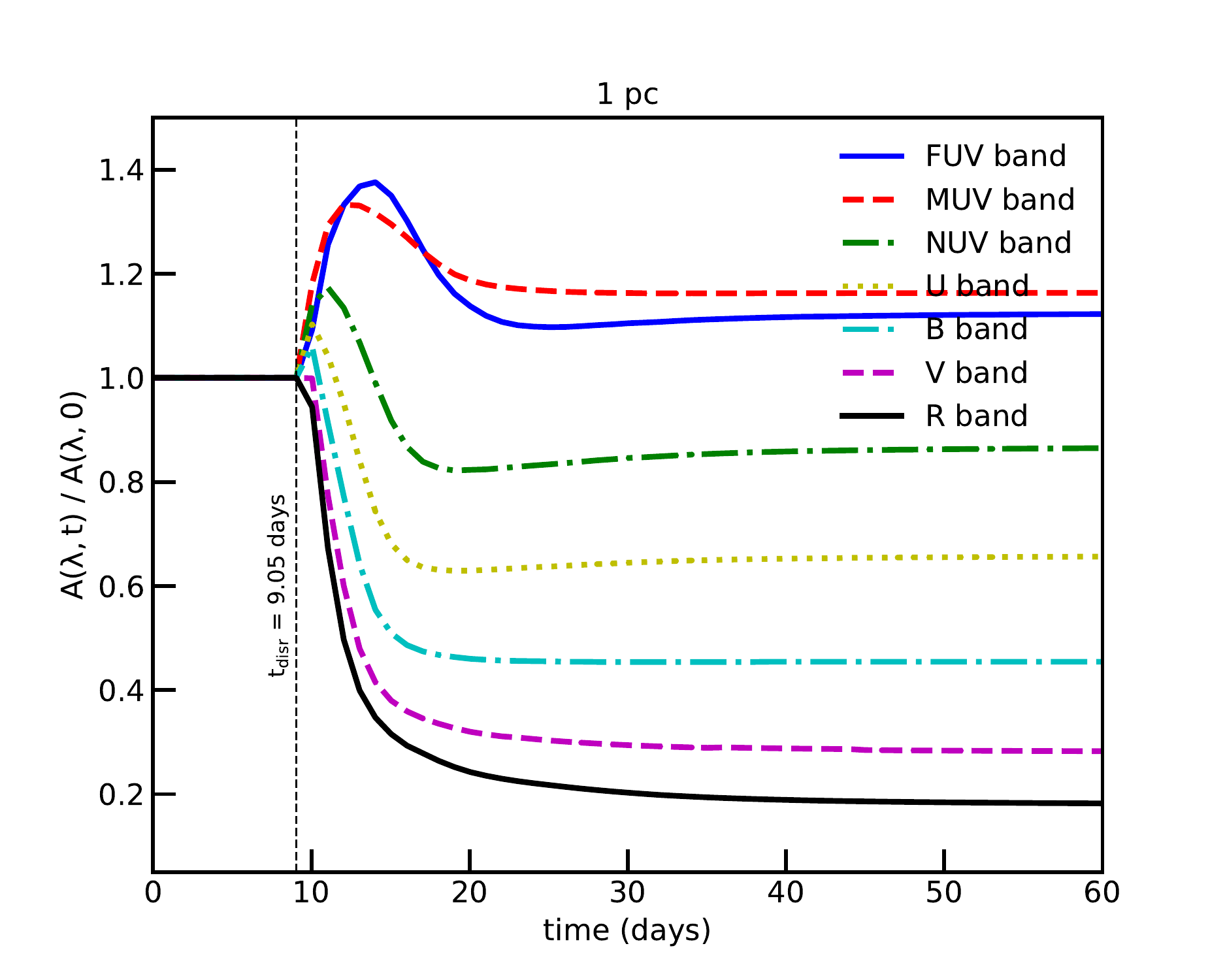}
    \includegraphics[width=0.5\textwidth]{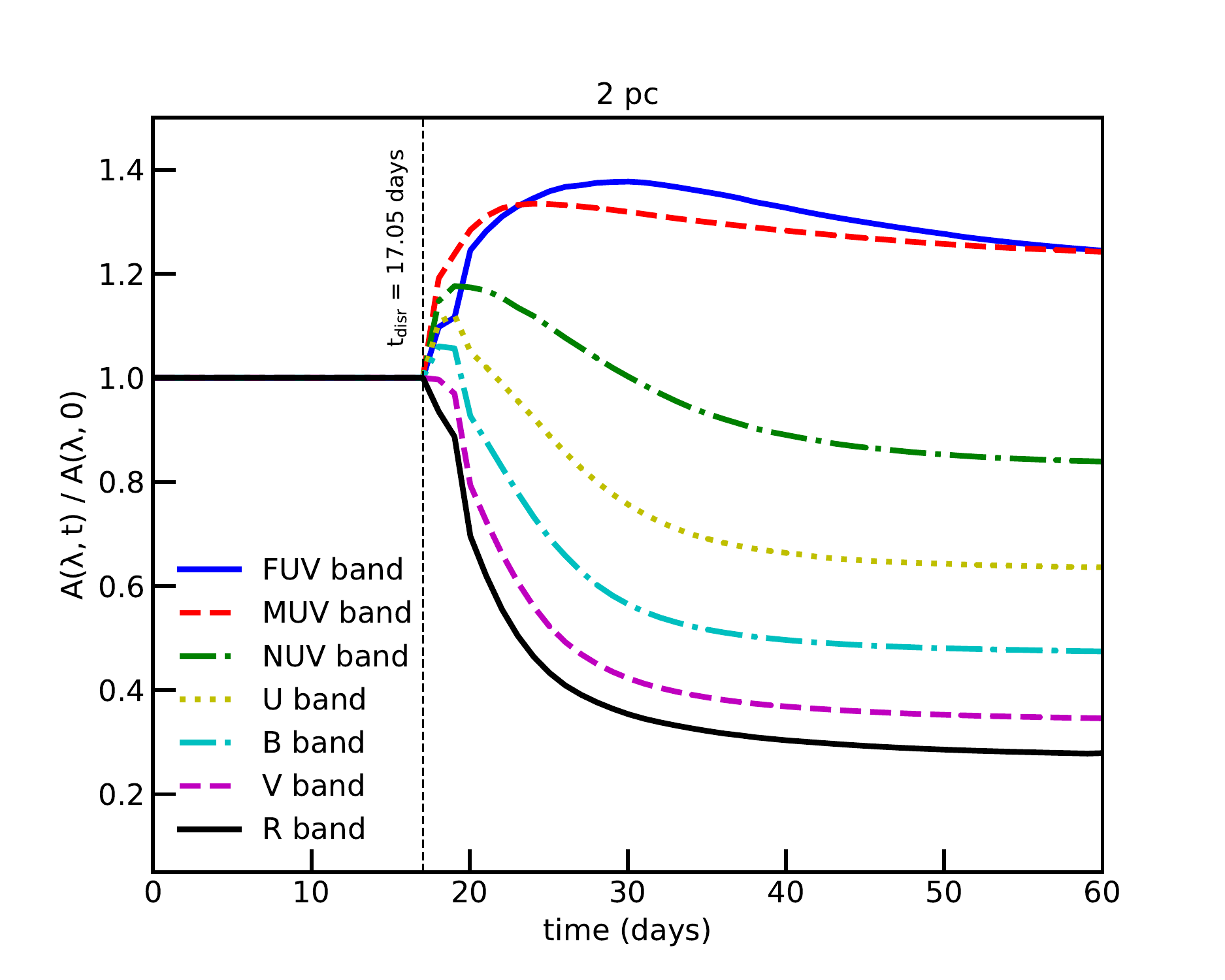}
    \includegraphics[width=0.5\textwidth]{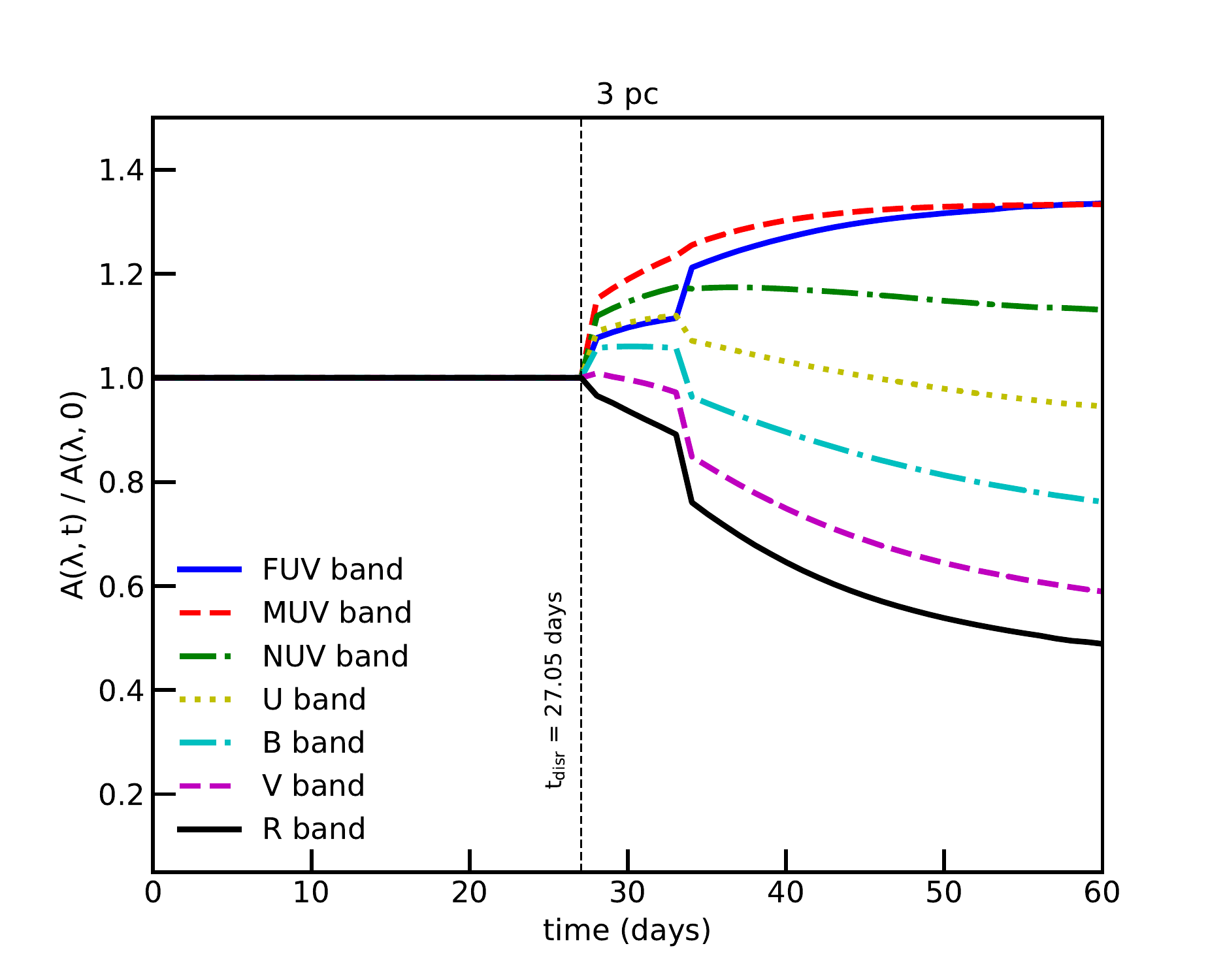}
    \caption{Variation of the ratio $A(\lambda,t)/A(\lambda,0)$ from UV to R bands with time for the different cloud distances. Vertical lines mark the disruption time of graphite which occurs earlier than silicates. The ratio is constant initially and starts to vary with time when RATD begins at $t_{\rm disr}$. After the disruption ceases, the ratio is constant again.}
       \label{fig:Alamda}
\end{figure*}

Figure \ref{fig:Alamda} shows the time-dependence of the ratio $ A(\lambda,t)/A(\lambda,0)$ for the different photometric bands (FUV to R bands) and different cloud distances. Here we choose $ \lambda=0.15\mum$ for the far-UV band (FUV), $\lambda=0.25\mum$ for the mid-UV band (MUV) and $ \lambda=0.3\mum$ for the near-UV band (NUV). As shown, the ratio $A(\lambda,t)/A(\lambda,0)$ is constant during the initial stage of $t<t_{\rm disr}$ before grain disruption, and it starts to rapidly change when RATD begins at $t_{\rm disr}$. For example, optical-NIR extinction (e.g., $R$-band, black line) decreases rapidly with time, but the UV extinction (FUV-U bands) increases first and then decreases with time. The reason is that the optical extinction is produced by large grains which have short disruption time by RATD (see \citealt{Hoang19}), and the production of small grains by RATD results in higher UV extinction. We also see that the amplitude of the extinction variation is smaller for more distant clouds.

\subsection{Time variation of color excess $E(B-V)$ and $R_{\rm V}$}\label{sec:RV}
From $A(\lambda,t)$ we can calculate the color excess $E(\rm B-V)=A_{\rm B}-A_{\rm V}$ and the total-to-selective extinction ratio $R_{\rm V}=A_{\rm V}/E(B-V)$ to study how these quantities vary with time due to RATD.

\begin{figure}
    \includegraphics[width=0.5\textwidth]{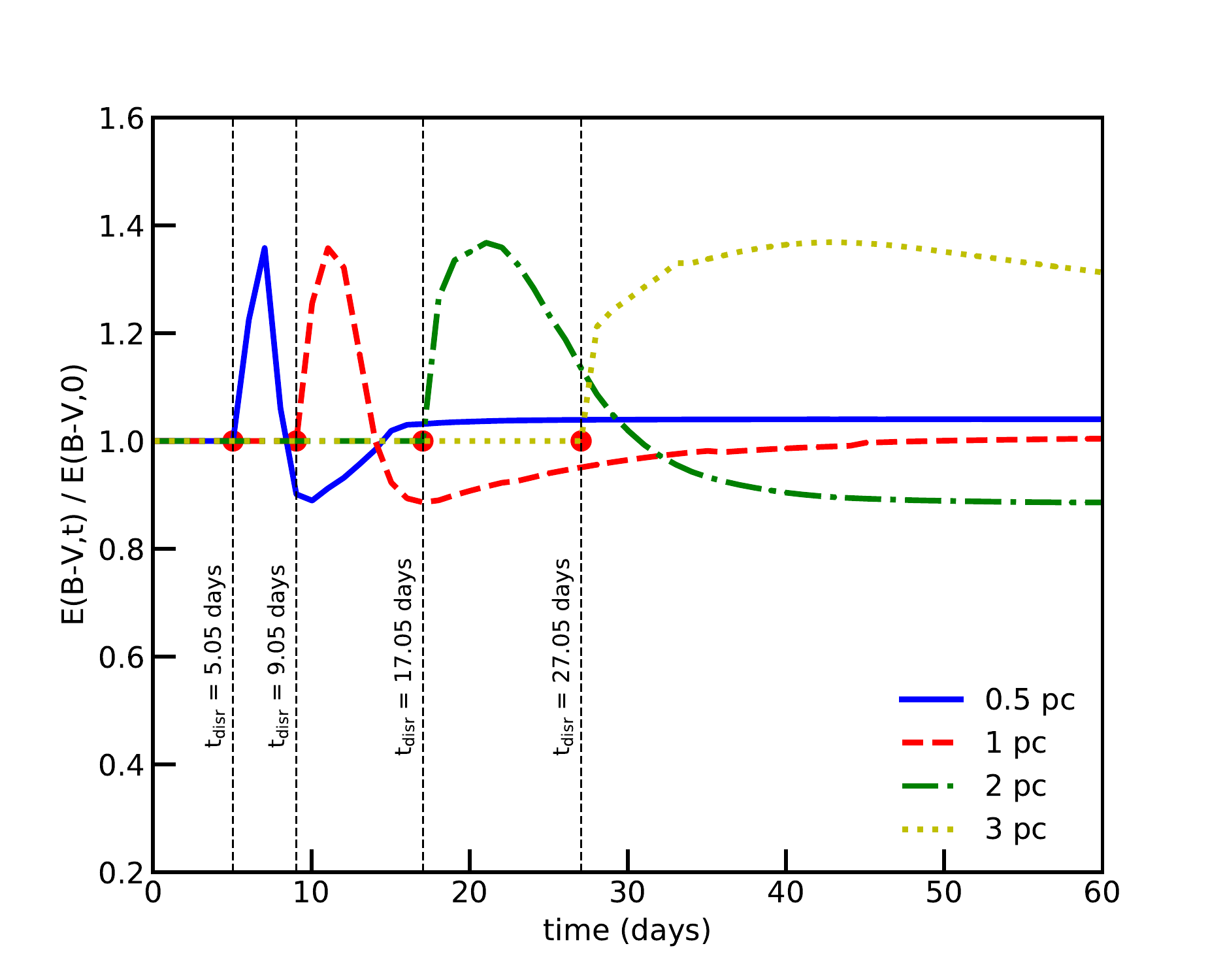}
    \includegraphics[width=0.5\textwidth]{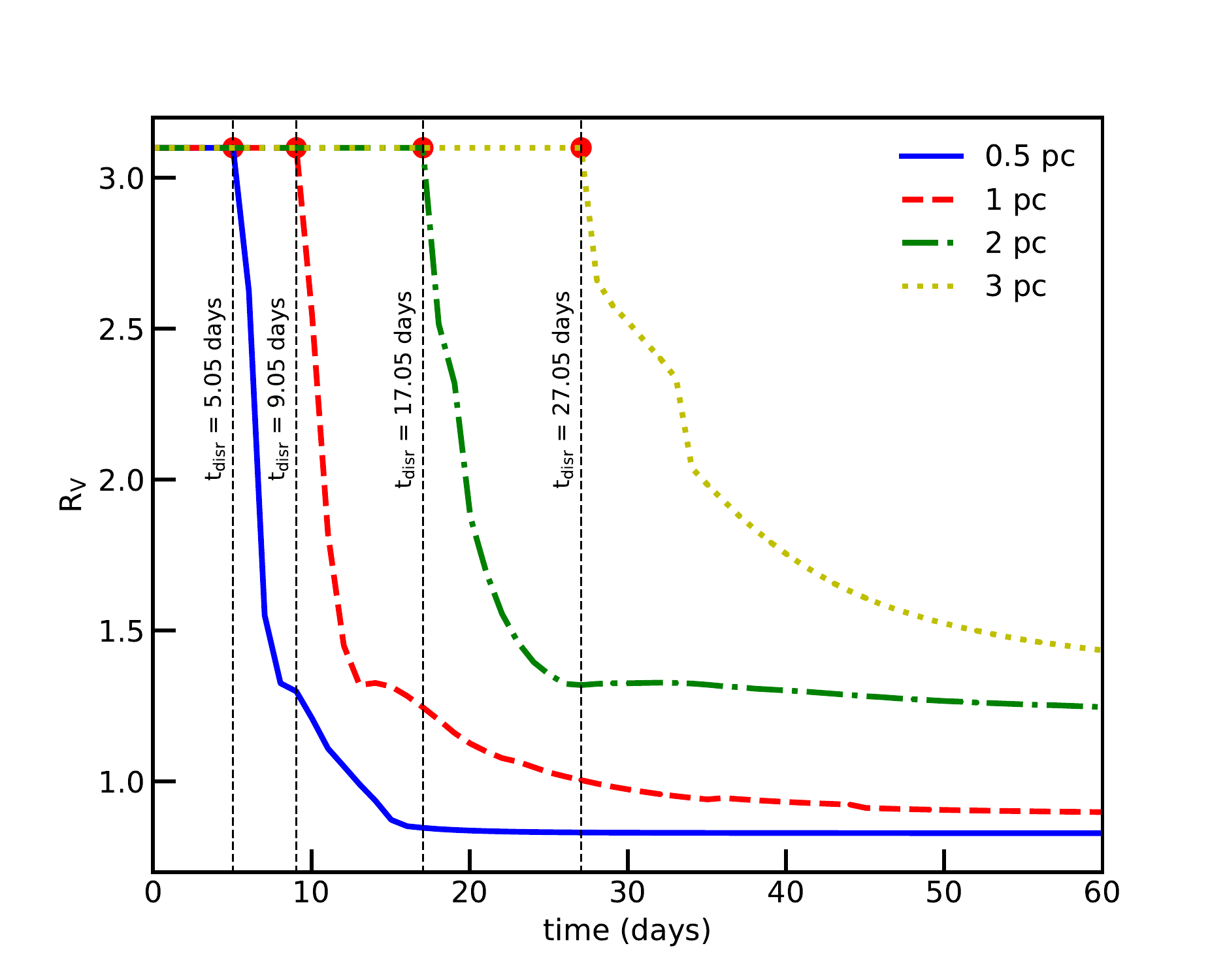}
    \caption{Time variation of $E(B-V)$ (upper panel) and $R_{\rm V}$ (lower panel) for different cloud distances and $S_{\rm max}=10^{7} \erg \cm^{-3}$. Both $E(B-V)$ and $R_{V}$ begin to change when grain disruption begin at $t\sim t_{\rm disr}$ (marked by vertical dotted lines). $R_{\rm V}$ decreases rapidly from their original values from $t=t_{\rm disr}$ to 40 days and then almost saturates when RATD ceases.} 
    \label{fig:Rv}
\end{figure}

Figure \ref{fig:Rv} (upper panel) shows the variation of $E(B-V,t)/E(B-V,0)$ with time for the different cloud distances. For a given cloud distance, the color excess remains constant until grain disruption begins at $t\sim t_{\rm disr}$. Subsequently, the ratio increases rapidly and then decreases to a saturated level when RATD ceases. For example, at distance $d=2$ pc, the color excess starts to rise at $t\sim 20$ days and declines again to the saturated value at $t\sim 40$ days.

Figure \ref{fig:Rv} (lower panel) shows the variation of $R_{\rm V}$ with time. The value $R_{\rm V}$ starts to rapidly decrease from the initial value of $R_{\rm V}=3.1$, corresponding with the average $R_{\rm V}$ in the diffuse medium of the Milky Way, to $R_{V}\sim 1-1.5$ after less than $40$ days. The moment where $R_{V}$ starts to decline coincides with that of $E(B-V)$, which is determined by the grain disruption time $t_{\rm disr}$ (see, e.g., Figure \ref{fig:adisr_d}). We note that although $E(B-V)$ has some fluctuations over time, $R_{\rm V}$ tends to decrease smoothly because $A_{V}$ always decreases with time due to the disruption of large grains by RATD (see Figure \ref{fig:adisr_d}). The decrease of $R_{\rm V}$ starts earlier for grains closer to the source due to shorter $t_{\rm disr}$ and also achieve a smaller terminal value. For example, a dust cloud located at 0.5 pc will have $R_{\rm V}$ to decreases after 5 days and gives the minimum value of 0.5 after 15 days, while at 2 pc, it takes 20 days to drop $R_{\rm V}$ down and the final value $\sim$ 1.4.
 
\begin{figure}[!htb]
        \includegraphics[width=0.5\textwidth]{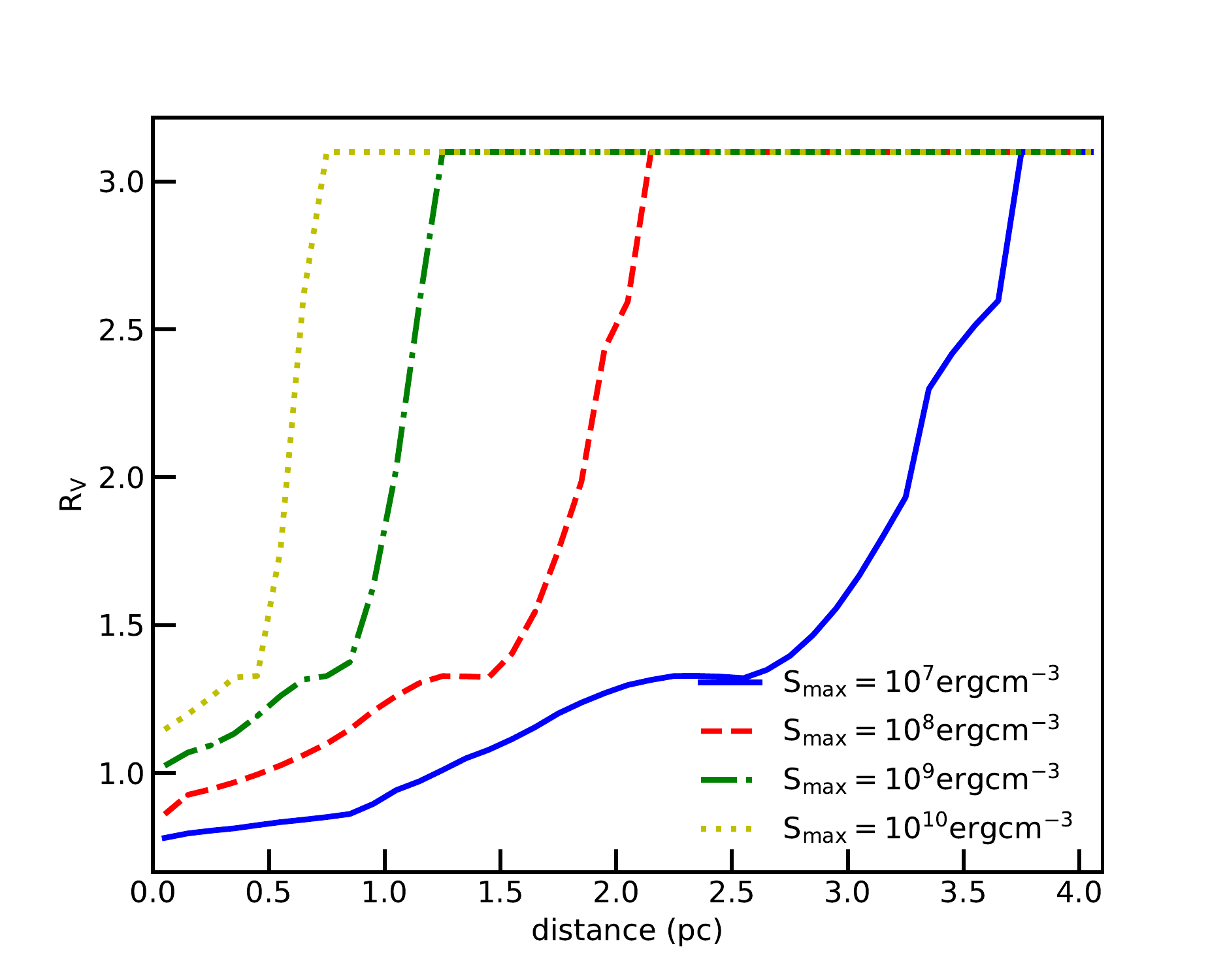}
        \includegraphics[width=0.5\textwidth]{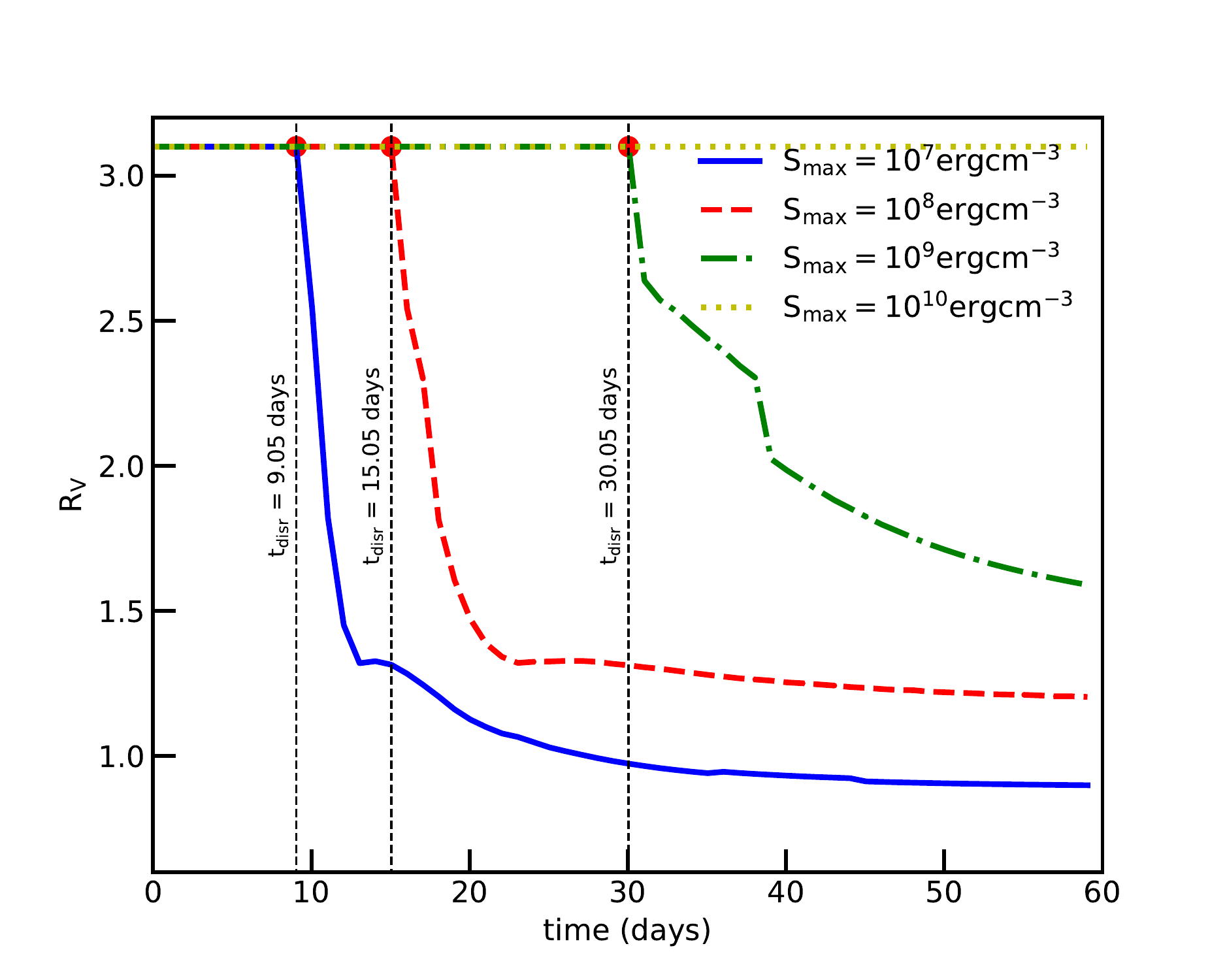}    
        \caption{$R_{\rm V}$ vs. cloud distance at $t= 45$ days (upper panel), and $R_{\rm V}$ vs. time for the cloud distance of 1 pc (lower panel), assuming the different values of $S_{\rm max}$.}
           \label{fig:Rv_Smax}
\end{figure}

Figure \ref{fig:Rv_Smax} (upper panel) shows the variation of $R_{\rm V}$ evaluated at $t=45$ days with cloud distance for the different tensile strength $S_{\rm max}$. Dust grains with a lower $S_{\rm max}$ can be disrupted out to a larger distance (see Figure \ref{fig:adisr_Smax}), so that $R_{\rm V}$ starts to decrease from a larger distance. On the other hand, grains with very high $S_{\rm max}\ge 10^{9} \erg\cm^{-3}$ can be disrupted within a small distance of $d\sim 1$ pc, thus one only can observe the smaller value of $R_{\rm V}$ if the cloud is close to the source.

Figure \ref{fig:Rv_Smax} (lower panel) shows the variation of $R_{\rm V}$ with time for a dust cloud at distance $d=1$ pc. Weaker grains have $R_{\rm V}$ decreasing sooner and achieve a lower terminal value than stronger ones due to the dependence of RATD on the material strength (see Figure \ref{fig:adisr_Smax}). For example, dust in the cloud at 1.5 pc after 45 days will give $R_{\rm V}=1.1$ for $S_{\rm max}=10^{7}\erg\cm^{-3}$, $\rm 1.4$ for $S_{\rm max}=10^{8}\erg\cm^{-3}$ and $R_{\rm V}=3.1$ with all grains of $S_{\rm max} \geq 10^{9}\erg\cm^{-3}$. 

\section{Polarization of SNe light in the presence of RAT alignment and RATD}\label{sec:pol}
\subsection{Grain alignment by RATs}

An anisotropic radiation field can align dust grains via the RAT mechanism (see \citealt{Ander15} and \citealt{Laza15} for reviews). In the unified theory of RAT alignment, grains are first spun-up to suprathermal rotation and then driven to be aligned with the ambient magnetic fields by superparamagnetic relaxation within grains having iron inclusions \citep{Hoang16}. Therefore, grains are only efficiently aligned when they can rotate suprathermally. One can adopt the suprathermal condition as follows (\citealt{Hoang08}):
\begin{align} \label{eq:14}
\omega_{\rm RAT}(t) \gtrsim 3 \omega_{T},
\end{align}
where $\omega_{\rm T}$ is the thermal angular velocity of dust grains at gas temperature $T_{\rm gas}$ as given by

\begin{align}
\omega_{\rm T} = \sqrt{\frac{2 k T_{\rm gas}}{I}}\simeq 2.3\times 10^{5}\hat{\rho}^{-1/2}T_{2}^{1/2}a_{-5}^{-5/2}\rm rad \rm s^{-1},\label{eq:omegaT}
\end{align}
where $T_{2}=T_{\rm gas}/100\K$. 

\begin{figure}[h]
        \includegraphics[width=0.5\textwidth]{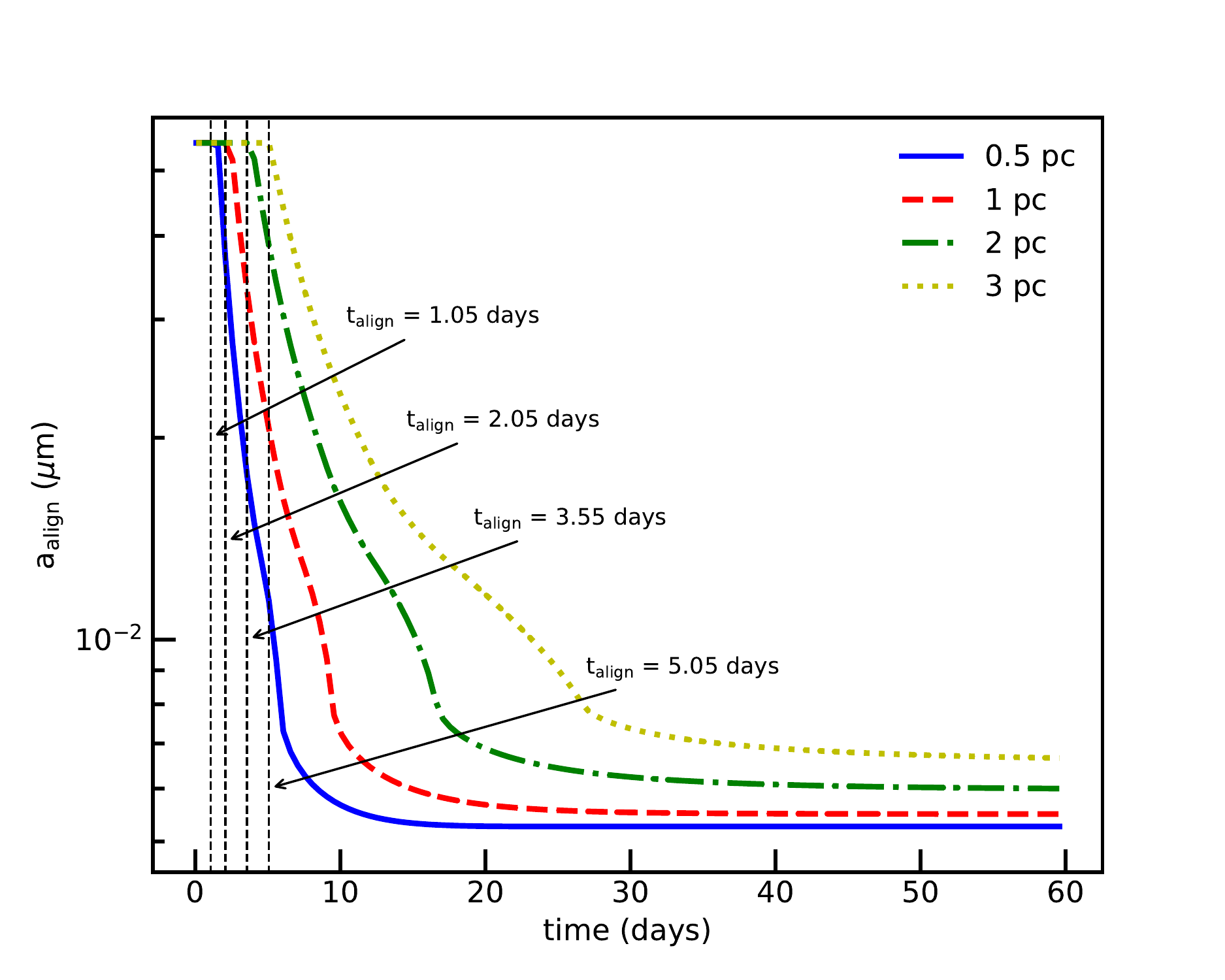}
        \caption{Critical size of grain alignment by RATs vs. time for the different cloud distances The alignment size starts to decrease rapidly from the original value when enhanced alignment by SNe radiation begins marked by vertical dotted lines. The decrease continues to their terminal values.}
           \label{fig:align} 
\end{figure}

Let $a_{\rm align}$ be the critical size of aligned grains. Then, $a_{\rm align}$ can be determined by using the condition $\omega_{\rm RAT}(t)=3\omega_{T}$.

The grain alignment timescale is given the timescale required for RATs to spin-up grains to the suprathermal rotation:
\bea
\tau_{\rm align}\equiv \frac{3I\omega_{\rm T}}{\Gamma_{\rm RAT}}
\simeq 0.6 \left(\frac{d_{\rm pc}^{2}}{L_{8}e^{-\tau}} \right)\bar{\lambda}_{0.5}^{1.7}a_{-5}^{-2.2}T_{2}^{1/2} {~\rm days},~~~\label{eq:tali}
\ena
where $\Gamma_{\rm RAT}$ is given by Equation (\ref{eq:3}) with $a\lesssim \bar{\lambda}/1.8$ (see also \citealt{Hoang17}).

In Figure \ref{fig:align}, we show the variation of alignment size with time for the different cloud distances. During the first stage, the alignment size is constant, which is equal to the typical value $a_{\rm align}\sim 0.055 \mum$ induced by the average ISRF. Due to intense SNe radiation, the alignment size starts to decrease after some alignment time ($t_{\rm align}$) from the original value to smaller values. One can see that the alignment time by SNe Ia is $t_{\rm align}\sim 3-7$ days for $d=0.5-3$ pc. Grains in a more distant cloud have larger alignment sizes, such as small grains at 0.5 pc can be aligned up to 0.005 $\mum$ after 10 days but it is still $0.05 \mum$ with the cloud father than 1 pc (see also Eq. \ref{eq:tali}).
 
\subsection{Modeling SNe Polarization}

We assume that only silicate grains can be aligned with the magnetic field, whereas graphite grains are not efficiently aligned (\citealt{Chiar06}; see \citealt{Hoang16} for explanation).\footnote{Although carbonaceous grains are expected to be aligned via k-RAT mechanism (see \citealt{Laza18}), their degree of alignment is not yet quantified, in contrast for silicate grains which have alignment degree quantified in \cite{Hoang16} using numerical simulations.} For the magnetic field in the plane of the sky, the degree of starlight polarization per H atom due to aligned grains in the unit of $\%$ is given by (\citealt{Hoang17}):
\begin{align} \label{eq:16}
\frac{P(\lambda)}{N_{\rm H}} = 100 \displaystyle\int\limits_{a_{\rm align}}^{a_{\rm max}} \frac{1}{2} C_{\rm pol}(a) f(a) \left(\frac{1}{n_{\rm H}}\frac{dn}{da}\right)da,
\end{align}
where $C_{\rm pol}$ = $Q_{\rm pol} \pi a^{2}$ is the polarization cross-section with $Q_{\rm pol}$ the polarization efficiency, and $f(a)$ is the alignment function describing the grain-size dependence of the grain alignment degree, and $a_{\rm max}\equiv a_{\rm disr}$. For our modeling, we consider the oblate grain shape and take data of $Q_{\rm pol}$ computed by \cite{Hoang13}. 

The alignment function can be modeled by the following function:
\begin{align} \label{eq:17}
f(a) = 1-\exp\left[-\left(\frac{0.5 a}{a_{\rm align}}\right)^{3}\right],
\end{align}
which yields the perfect alignment $f(a)=1$ for large grains of $a\gg a_{\rm align}$ and adequately approximates the numerical results from \cite{Hoang16} as well as results from inverse modeling of starlight polarization (\citealt{Hoan14}; \citealt{Hoang17}).

\subsection{Polarization curves} 
\begin{figure}[h]
        \includegraphics[width=0.5\textwidth]{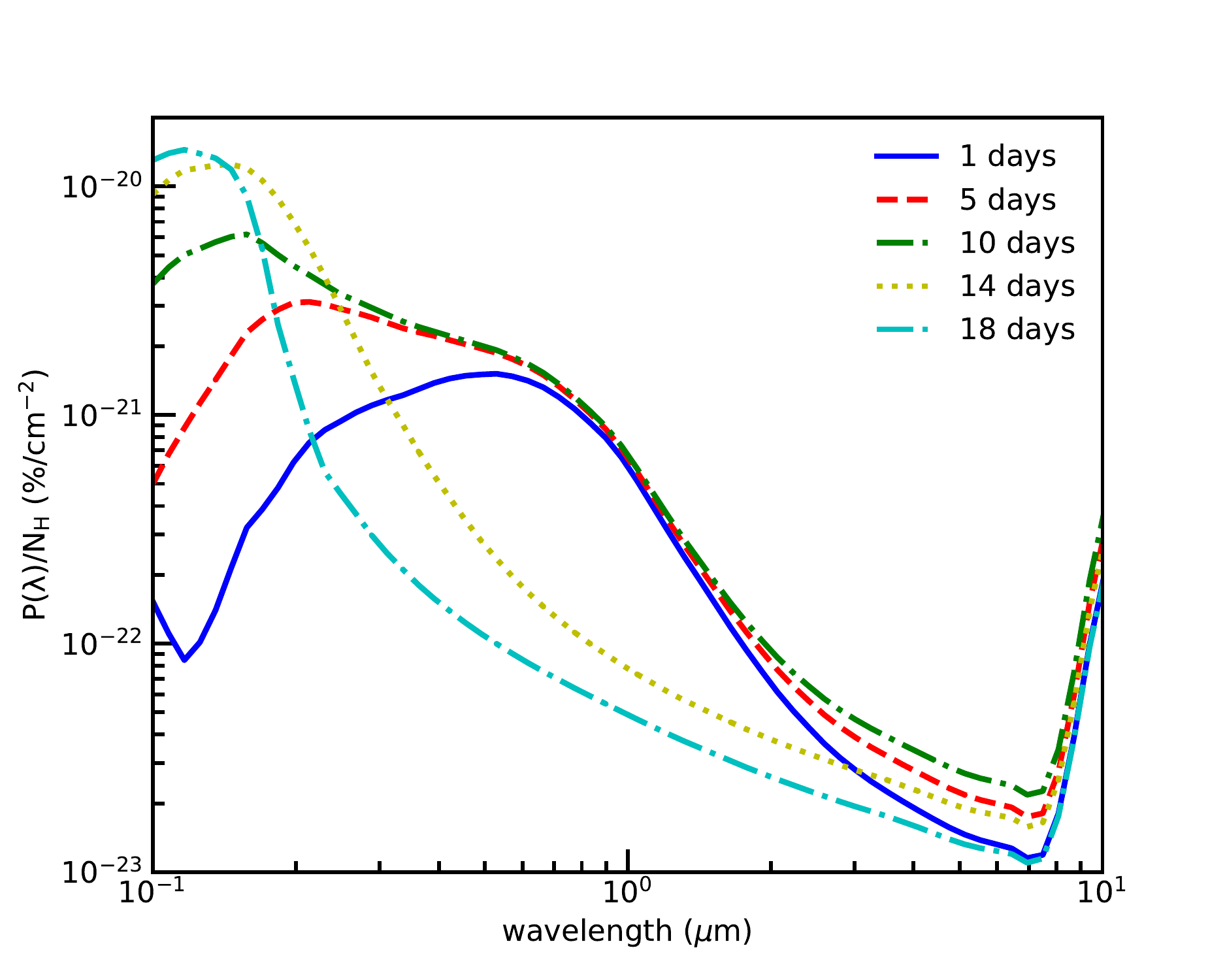}
        \includegraphics[width=0.5\textwidth]{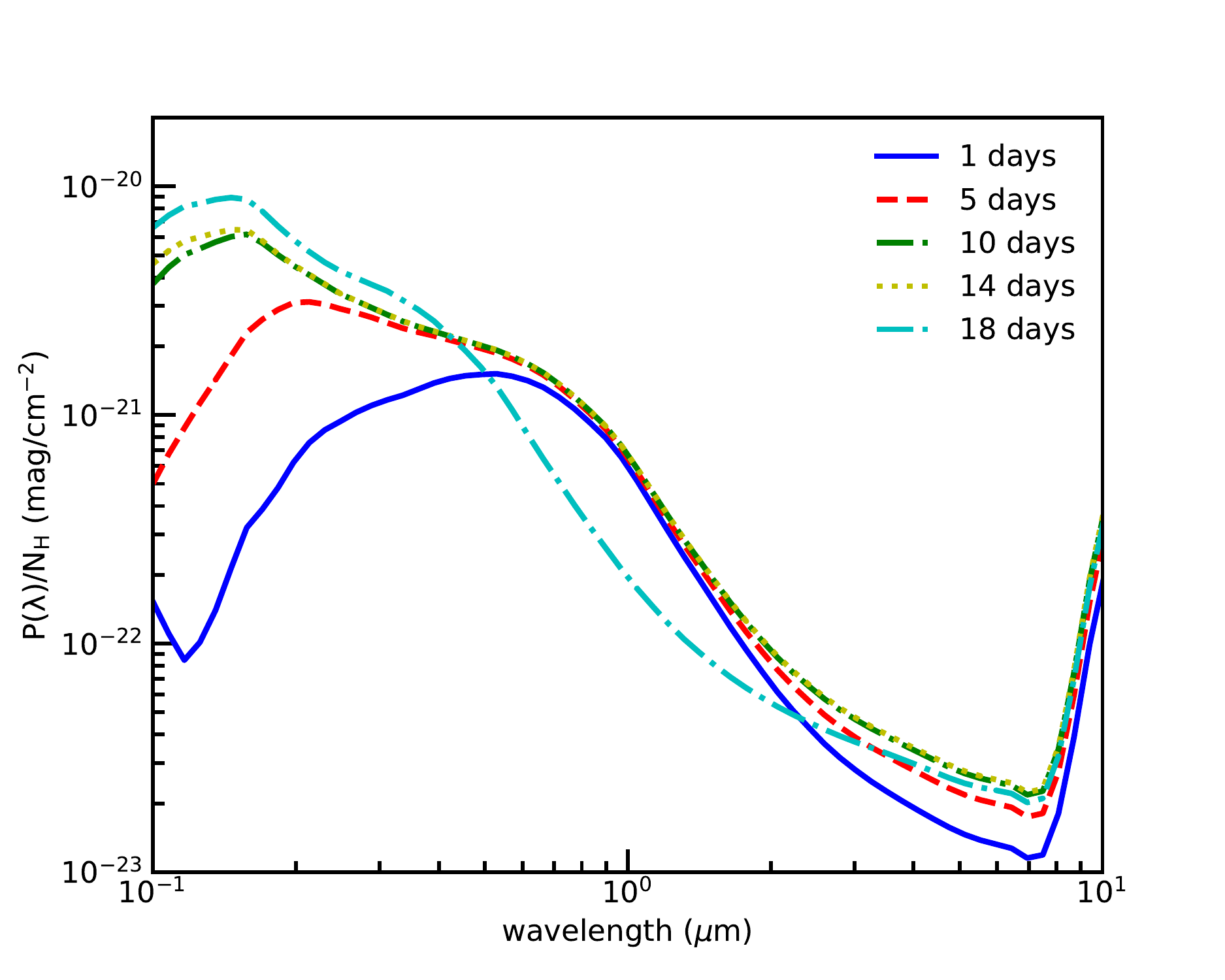}
        \caption{Polarization curves evaluated at different times for dust cloud at 1pc, assuming $S_{\rm max} = 10^{7} \erg\cm^{-3}$ (upper panel) and $S_{\rm max}=10^{8}\erg\cm^{-3}$ (lower panel). Enhanced alignment of small grains induces blueshift of the peak wavelength. The RATD effect reduces polarization at $\lambda>0.3\mum$, and the efficiency is weaker for higher $S_{\rm max}$. }
           \label{fig:Plamda_t} 
\end{figure}

Figure \ref{fig:Plamda_t} (upper panel) shows the polarization curve given by the cloud at 1 pc with time for  $S_{\rm max}=10^{7}\erg\cm^{-3}$ (upper panel) and $S_{\rm max}=10^{8}\erg\cm^{-3}$ (lower panel) as a result of RAT alignment and RATD. At $t\lesssim1$ days, dust grains are aligned by the diffuse interstellar radiation, so the maximum polarization at $\lambda_{\rm max}$ is around $0.55\mum$. After that, SNe radiation dominates and makes smaller grains to be aligned. As a result, the UV polarization is increased rapidly, and the peak wavelength is shifted toward the blue. The degree of polarization at long wavelengths ($\lambda > 0.5 \mum$) is slightly increased. After $t\sim 10$ days, grain disruption by RATD begins (see Figure \ref{fig:adisr_d}), reducing the abundance of large grains. Therefore, the degree of optical-NIR polarization decreases substantially, which results in a narrower the polarization profile compared to the original polarization curve. 

Figure \ref{fig:Plamda_t} (lower panel) shows the similar results, but with $S_{\rm max}=10^{8}\erg\cm^{\rm -3}$. The same trend as the upper panel is observed, but the significant decrease of optical-NIR polarization is only seen at a later time of $t \gtrsim 14$ days because grains of stronger material require more time to disrupt (see Figure \ref{fig:adisr_Smax}). One note that the polarization curves at 10 days and 14 days are very similar, revealing that grain alignment is saturated. 
 
\begin{figure}[h]
        \includegraphics[width=0.5\textwidth]{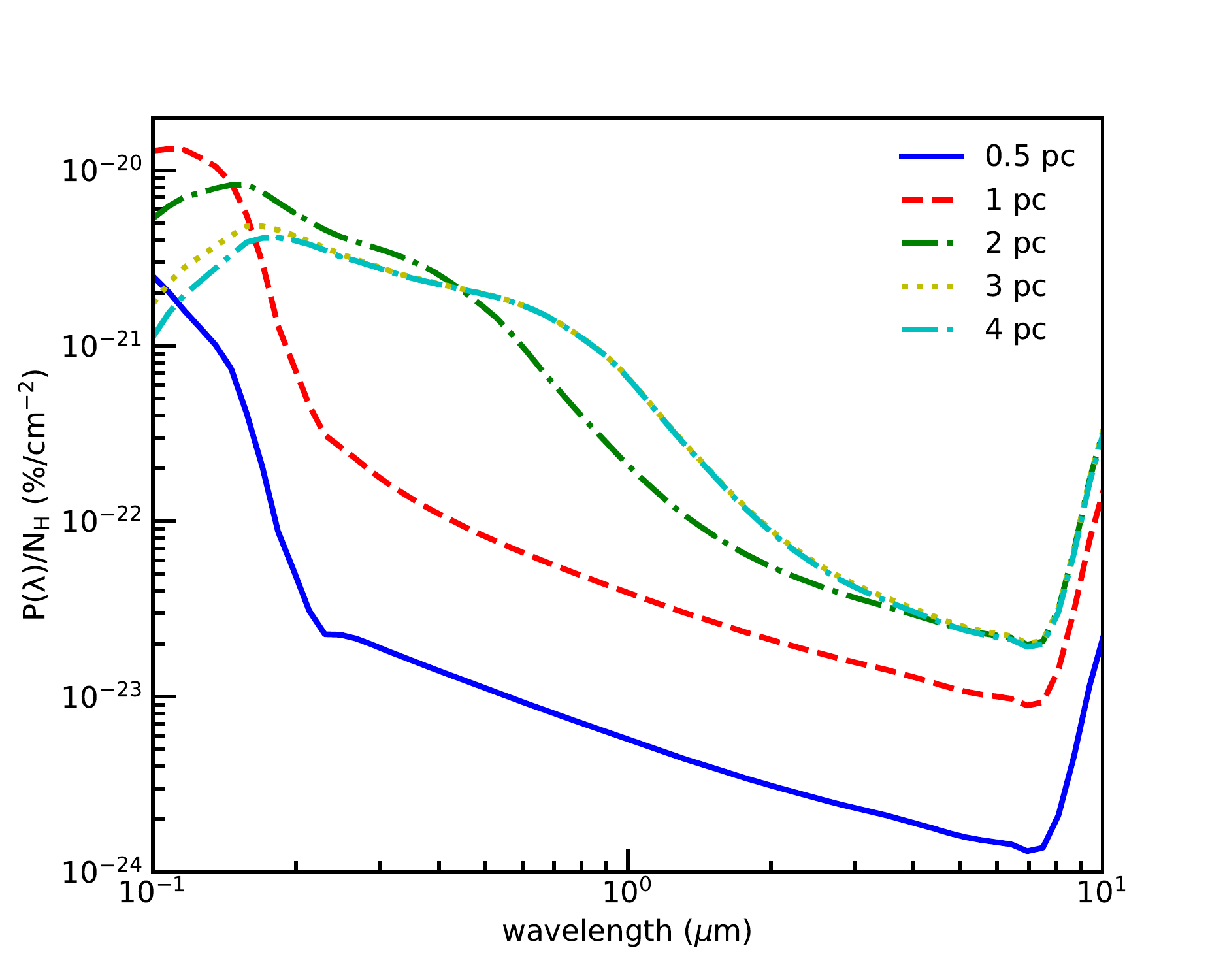}
        \includegraphics[width=0.5\textwidth]{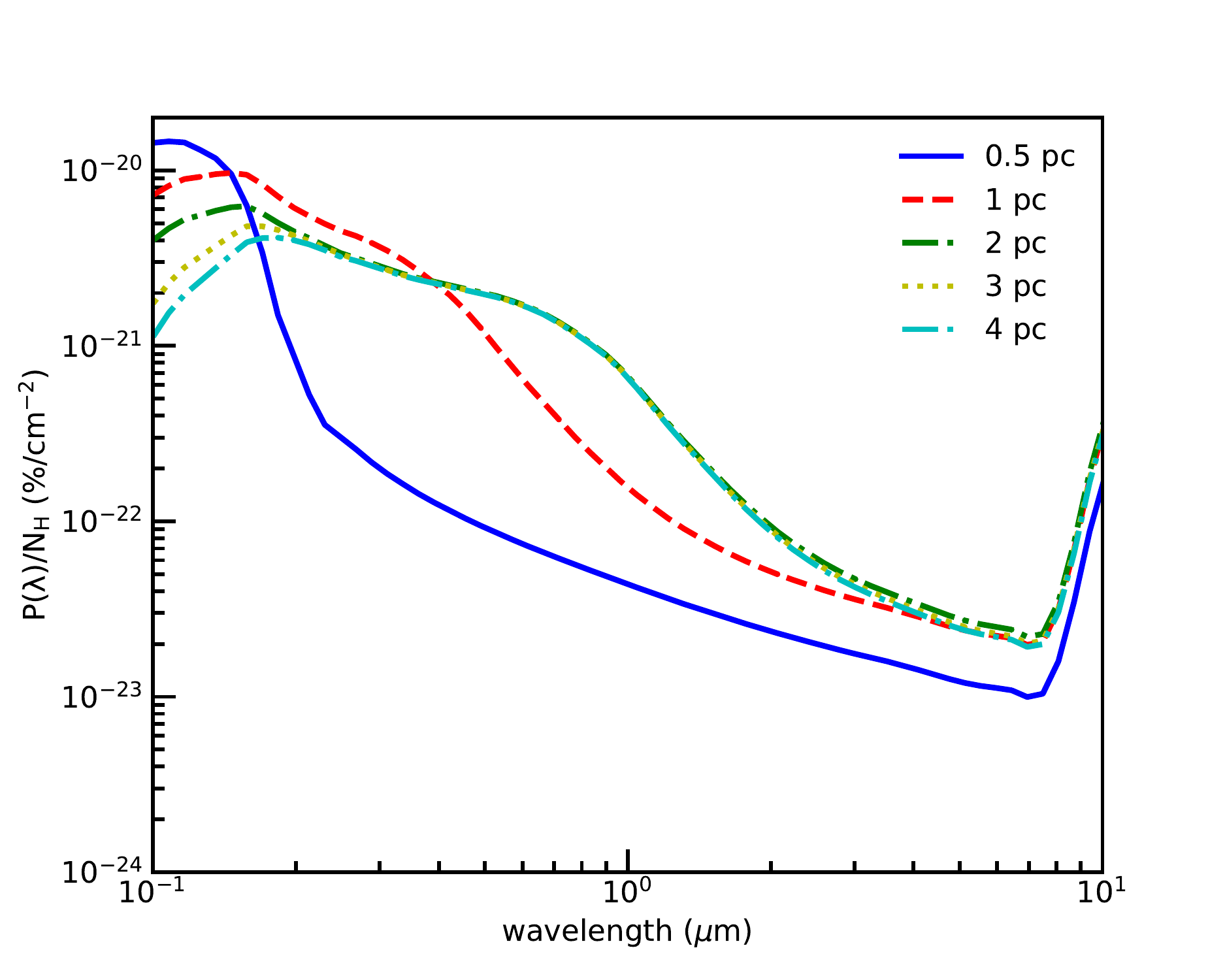}
        \caption{Polarization curve evaluated at $t=20$ days for the different cloud distance, assuming $S_{\rm max} = 10^{7} \erg\cm^{-3}$ (upper panel) and $S_{\rm max}=10^{8}\erg\cm^{-3}$ (lower panel). }
           \label{fig:Plamda_d}
\end{figure}

Figure \ref{fig:Plamda_d} shows the polarization curve calculated at $t=20$ days for the different cloud distances, assuming $S_{\rm max}=10^{7}\erg\cm^{-3}$ (upper panel) and $S_{\rm max}=10^{8}\erg\cm^{-3}$ (lower panel). The optical-NIR polarization decreases significantly with decreasing the cloud distance. This arises from the fact that grains closer to the source experience stronger disruption due to the dependence of the radiation flux as $1/d^{2}$ (see Figure \ref{fig:adisr_Smax}). At the same time, the peak wavelength $\lambda_{\rm max}$ is smaller for grains at smaller distances due to more efficient alignment of grains by RATs. 

Above we have assumed that only silicate grains are aligned with the magnetic field.  Nevertheless, the peak wavelength of polarization would be not much different when carbonaceous grains are assumed to be aligned because it mostly depends on the alignment function.

\subsection{Time-variation of dust polarization and peak wavelength}

\begin{figure*}
    \includegraphics[width=0.5\textwidth]{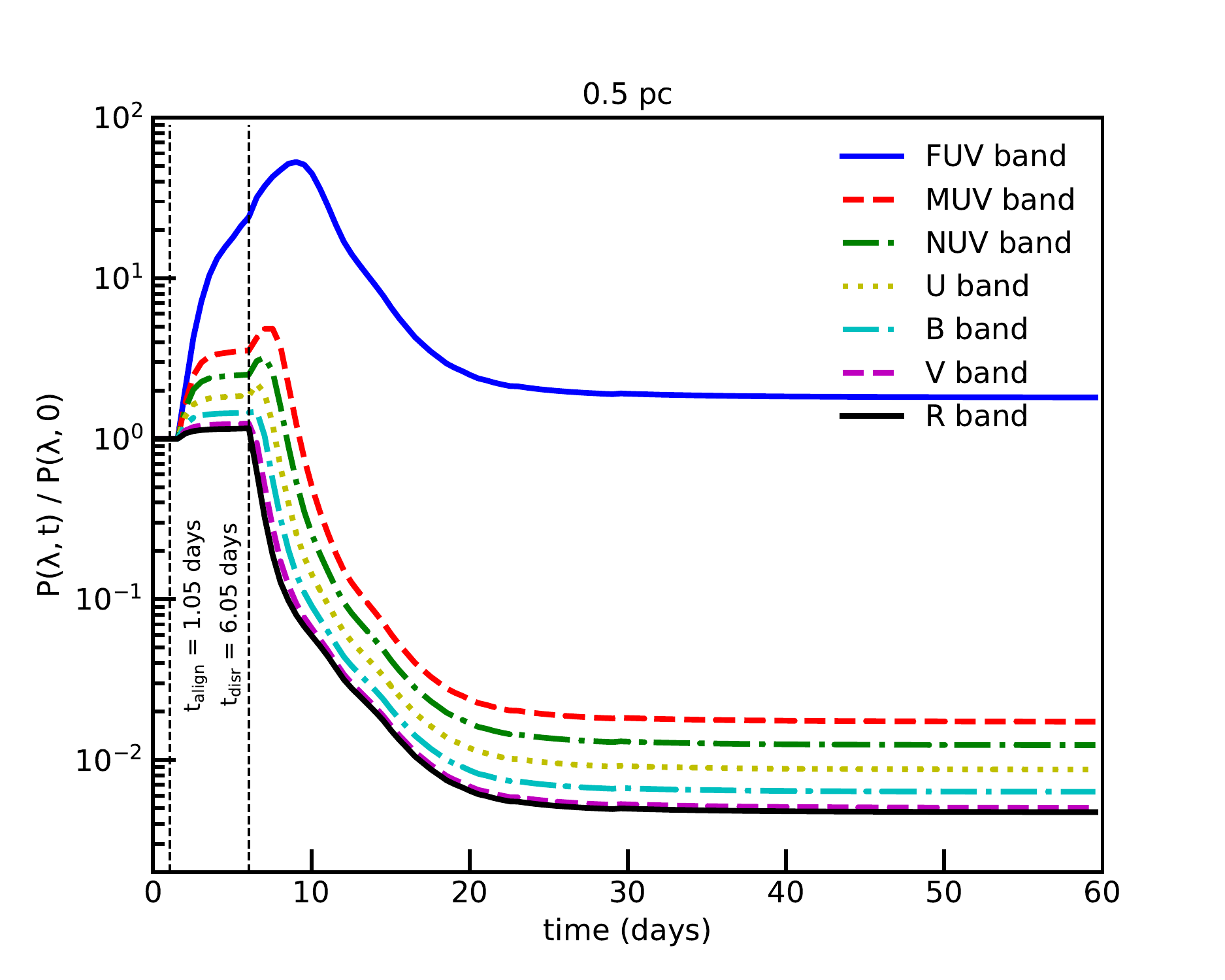}
    \includegraphics[width=0.5\textwidth]{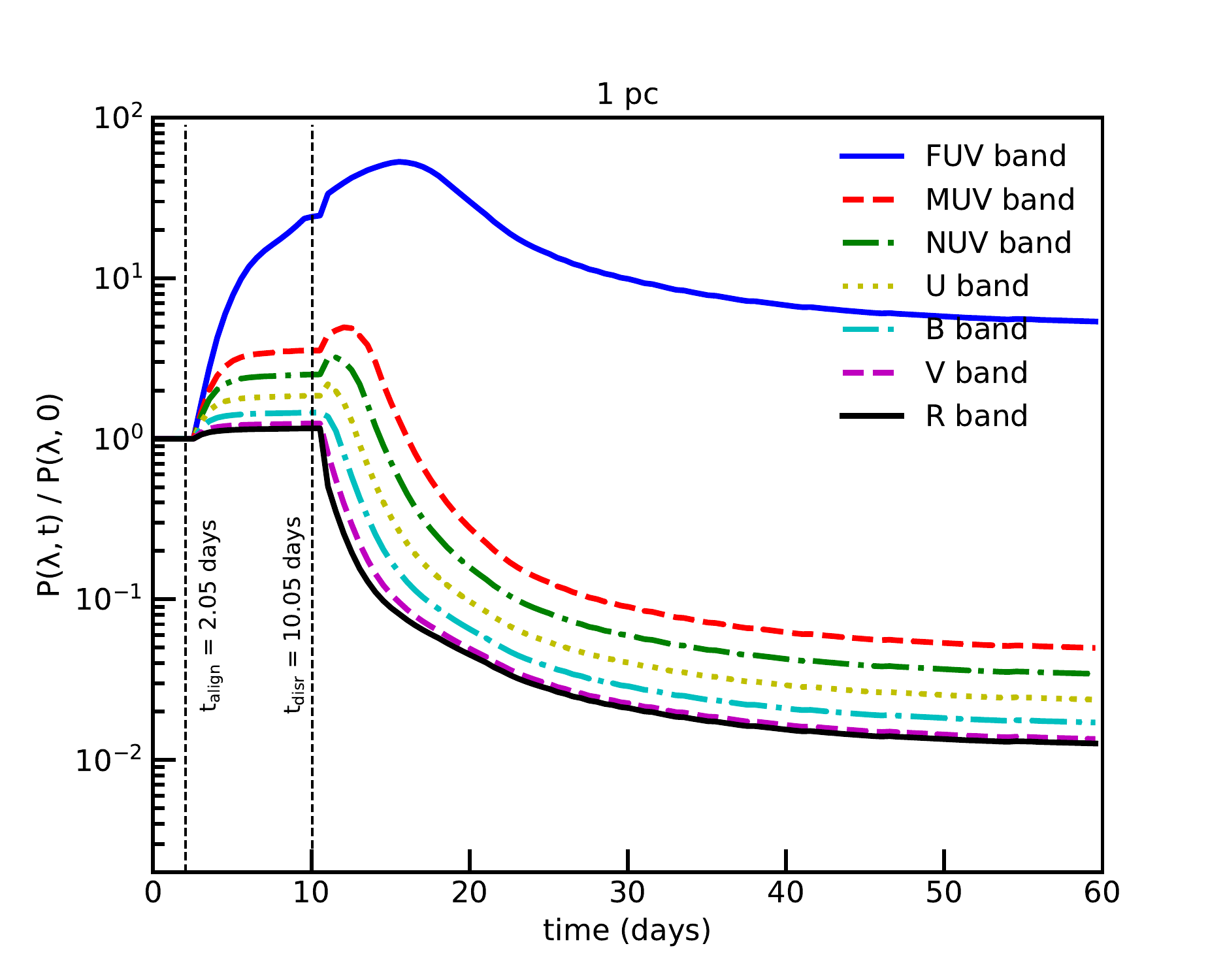}
    \includegraphics[width=0.5\textwidth]{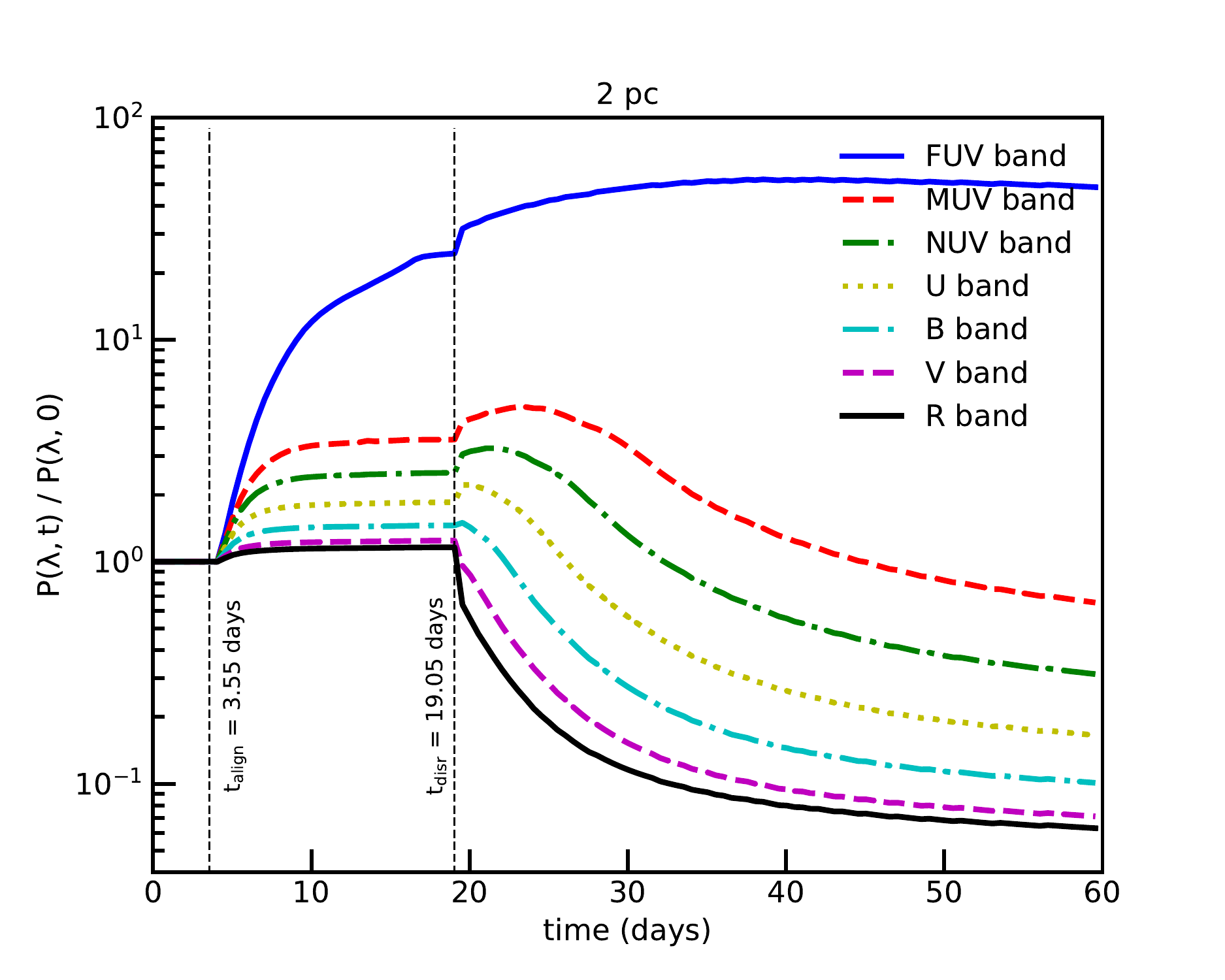}
    \includegraphics[width=0.5\textwidth]{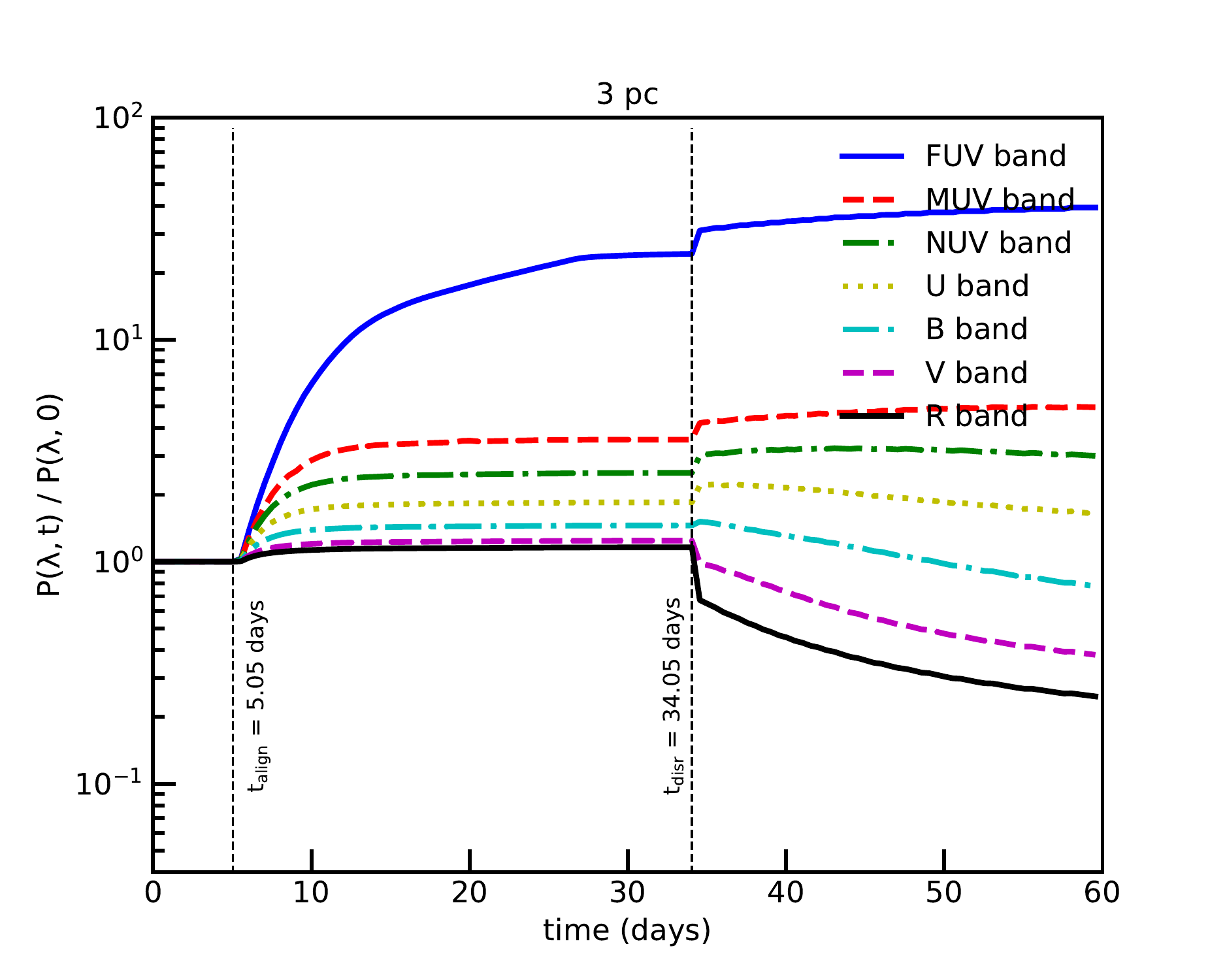}
    \caption{Ratio $P(\lambda,t)/P(\lambda,0)$ vs. time from FUV to V band for different cloud distances assuming $S_{\rm max}=10^{7}\erg\cm^{-3}$. Optical/NIR polarization degree first increases due to enhanced alignment by RATs and then declines when grain disruption by RATD starts. Dotted vertical lines mark alignment time ($t_{\rm align}$) and disruption time ($t_{\rm disr}$) of silicate grains.}
       \label{fig:Plamda}
\end{figure*}

Figure \ref{fig:Plamda} shows the temporal variation of $P(\lambda,t)/P(\lambda,0)$ from FUV to R bands for the different cloud distances. During the initial stage, the ratio $P(\lambda,t)/P(\lambda,0)$ is constant, however, this stage is rather short, between $1-5$ days corresponding to the alignment timescale $t_{\rm align}$ (see Figure \ref{fig:align}; also \citealt{Hoang17}). After that, the polarization degree increases gradually and this rising period continues until $t\sim 5-30$ days until the grain disruption begins (i.e., at $t=t_{\rm disr}$) for $d=0.5-3$ pc. The polarization degree then declines rapidly when grain disruption by RATD begins to occur. The polarization degree achieves a saturated level when RATD ceases, which occurs after $t\sim 20$ days for $d=0.5$ pc, $40$ days for $d=1$ pc, respectively. In summary, due to RAT alignment and RATD, the polarization degree increases from $t_{\rm align}$ to $t_{\rm disr}$, and it decreases rapidly at $t> t_{\rm disr}$.

\begin{figure}[!htb]
        \includegraphics[width=0.5\textwidth]{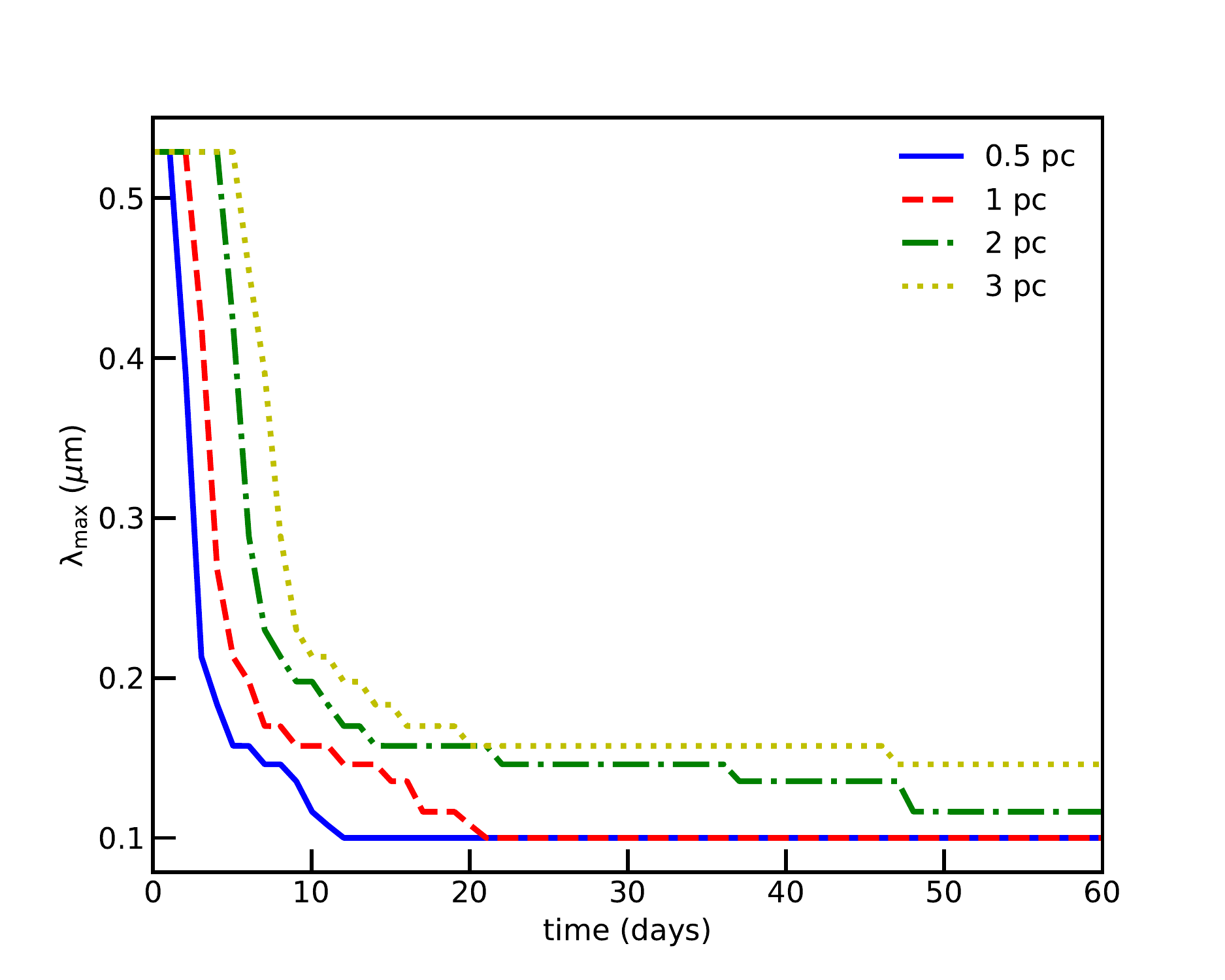}
        \includegraphics[width=0.5\textwidth]{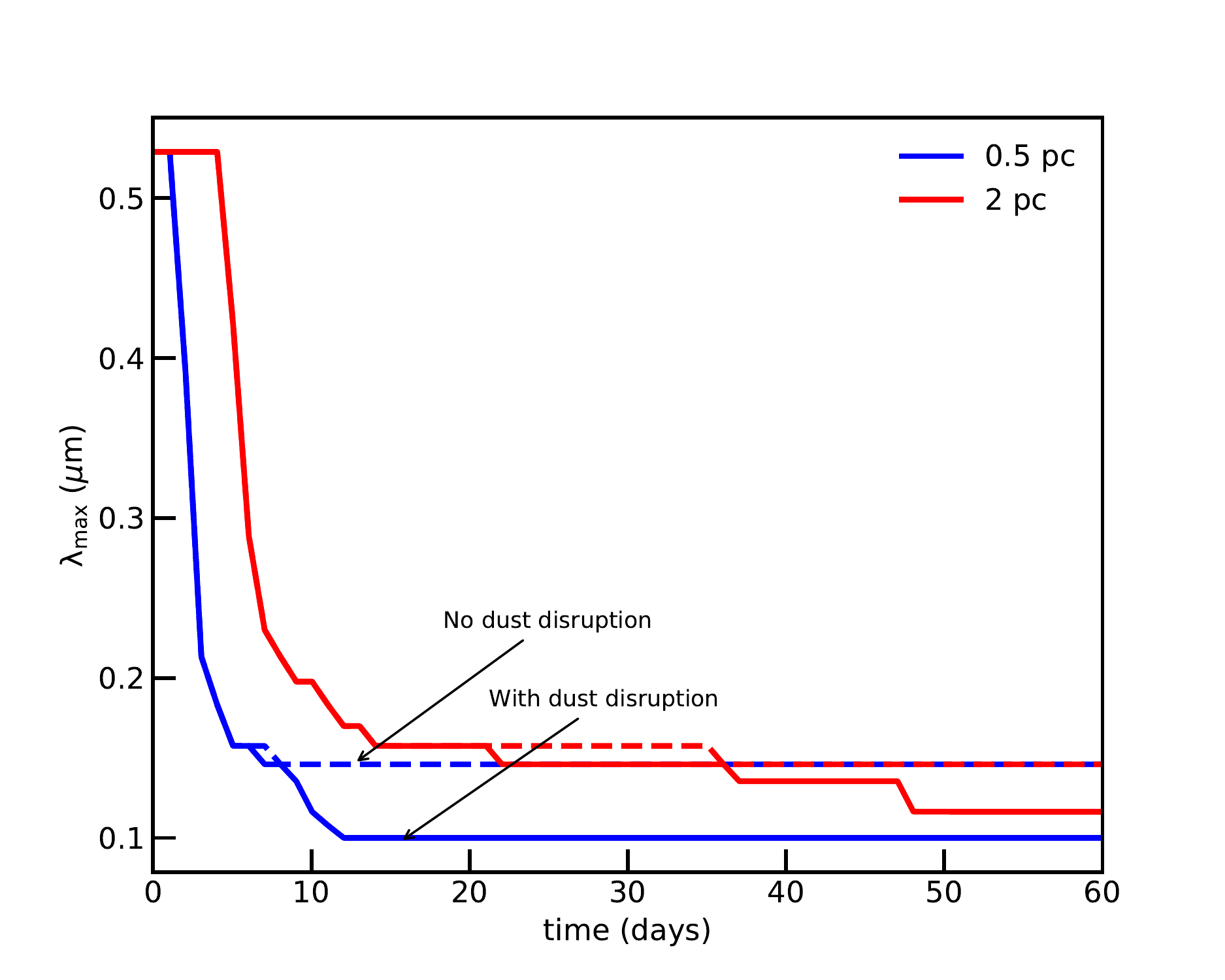}
        \caption{Upper panel: decrease of the peak wavelength $\lambda_{\rm max}$ with time during the first 20 days due to RAT alignment, assuming dust cloud between $0.5- 3$ pc and $S_{\rm max}=10^{7}\erg\cm^{-3}$. Lower panel: Comparison of $\lambda_{\rm max}$ with time for cloud distances of 0.5 pc and 2 pc with RATD (solid line) and without RATD (dashed line). RATD has a small effect on the peak wavelength.}
           \label{fig:lamdamax}
\end{figure}

Figure \ref{fig:lamdamax} (upper panel) presents the variation of the peak wavelength with time for dust clouds at different locations and $S_{\rm max}=10^{7}\erg\cm^{-3}$. Initially, $\lambda_{\rm max}\sim 0.55\mum$ is produced by grains aligned by the interstellar radiation field (\citealt{Hoang14}). In the presence of strong SNe radiation, smaller grains can be aligned, resulting in a rapid decrease of $\lambda_{\rm max}$. Dust grains at farther distances receive lower radiation energy density such that $\lambda_{\rm max}$ start to decrease later and stops at higher $\lambda_{\rm max}$. For example, dust at 0.5 pc give $\lambda_{\rm max} \leq 0.5 \mum$ after $\sim$ 1 day and stop at $0.1\mum$, while dust at 3 pc needs $\sim$ 8 days to begin and give the minimum $\lambda_{\rm max} \sim 0.15\mum$.

Figure \ref{fig:lamdamax} (lower panel) shows the comparison of $\lambda_{\rm max}$ during 60 days with (solid line) and without (dashed line) dust disruption. As shown, $\lambda_{\rm max}$ in the two cases are similar in the early times and become different when grain disruption begins. Their difference is larger at a shorter distance because grain disruption is stronger.

The dashed line shows that alignment grain size saturates earlier than disruption grain size, such that $\lambda_{\rm max}$ remains unchanged just after 10 day (for grain at 0.5 pc) and 20 day (for grain at 2 pc). The slower disruption rate at the distant cloud will makes $\lambda_{\rm max}$ to be constant longer.

\subsection{Effect of size-dependence tensile strength}
Till now, we calculated the grain disruption size, $R_{\rm V}$ and $\lambda_{\rm max}$ by assuming that the maximum tensile strength is constant for all grain sizes. A more realistic situation is that small grains would have compact structures and thus have a higher $S_{\rm max}$. To see how the size-dependent tensile strength affects our results, we now assume $S_{\rm max}=10^{\rm 7}\erg\cm^{-3}$ for grains larger than $\sim 0.05\mum$ and $S_{\rm max}=10^{9}\erg\cm^{-3}$ for grains smaller than $0.05\mum$.

\begin{figure}[t]
        \includegraphics[width=0.45\textwidth]{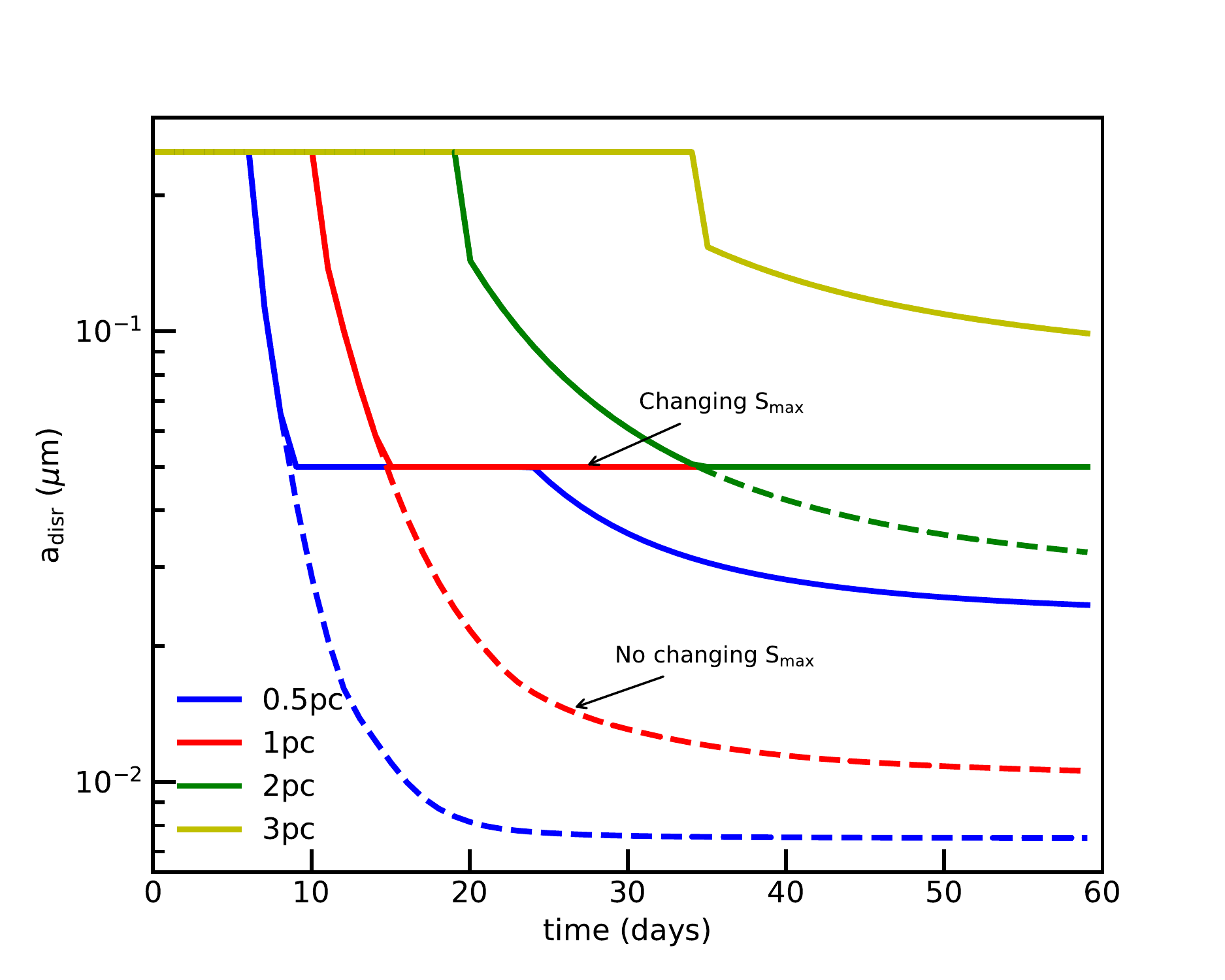}
        \includegraphics[width=0.45\textwidth]{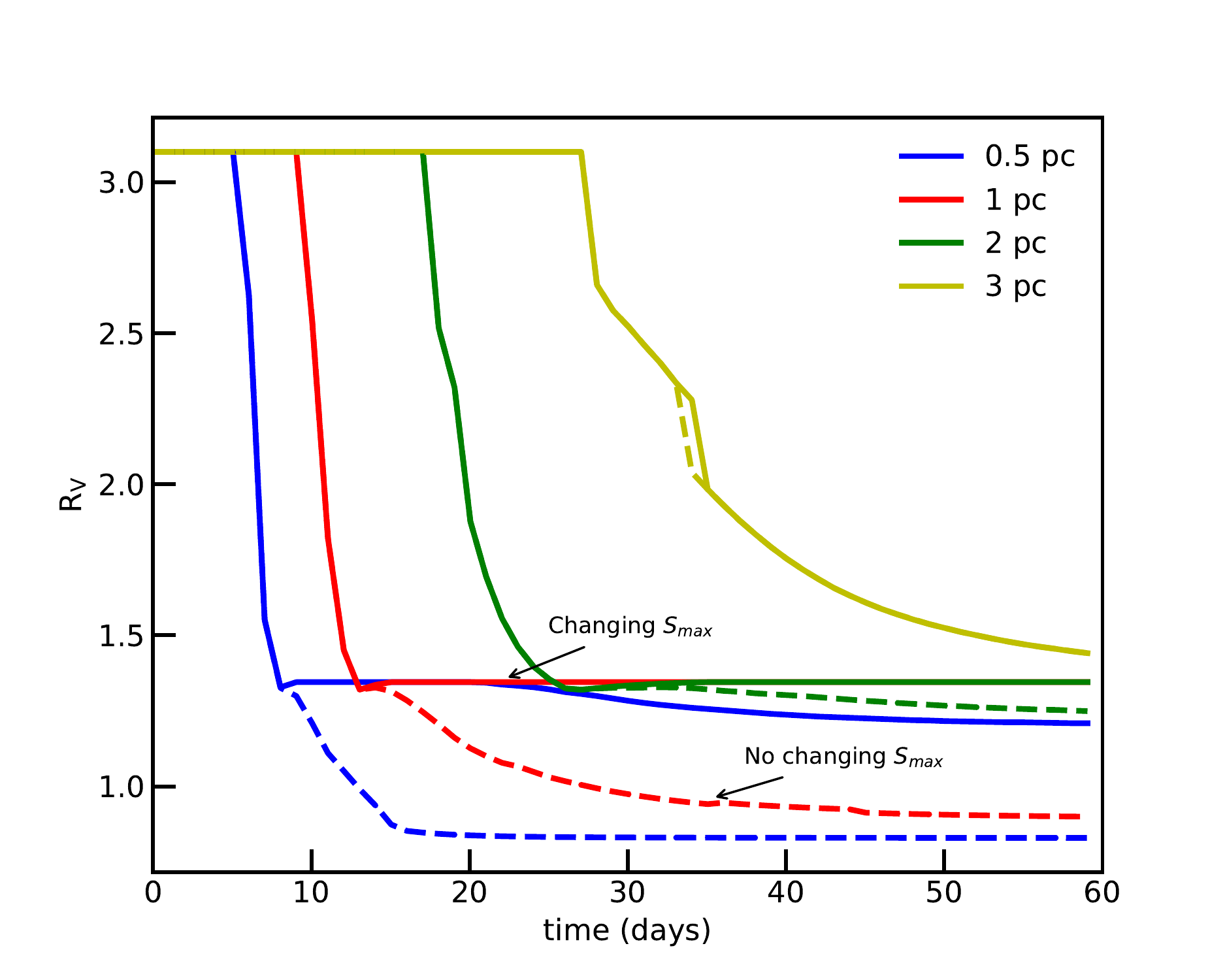}
        \includegraphics[width=0.45\textwidth]{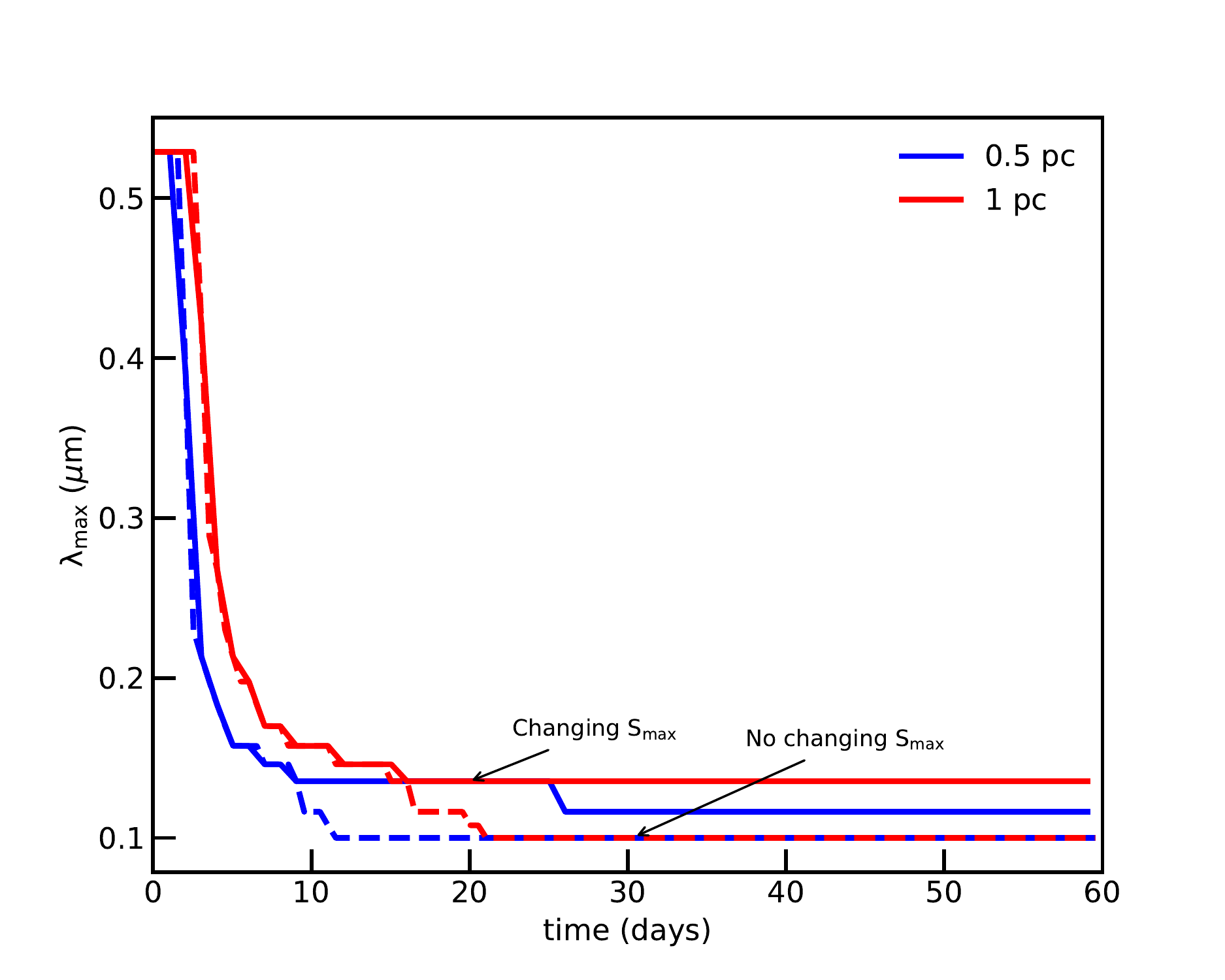}
        \caption{Comparison of $a_{\rm disr}$ (top panel), $R_{\rm V}$ (middle panel) and $\lambda_{\rm max}$ (bottom panel) vs. time with constant (dashed lines) and changing $S_{\rm max}$ (solid lines).}
           \label{fig:changeSmax}
\end{figure}

Figure \ref{fig:changeSmax} (top panel) shows the grain disruption size for constant $S_{\rm max}$ (dashed lines) and changing $S_{\rm max}$ (solid lines) for the different cloud distances. In the latter case, the grain disruption size is decreased to a final value of $0.05\mum$ if the radiation energy density is not strong enough to destroy further. 

Same as the top panel, but Figure \ref{fig:changeSmax} (middle panel) shows $R_{\rm V}$ versus time for constant (dashed lines) and changing $S_{\rm max}$ (solid lines). The $R_{\rm V}$ values in two cases decrease rapidly with time, but the latter case has higher terminal values of $R_{V}\sim 1.45$ due to higher disruption sizes (see top panel).

Figure \ref{fig:changeSmax} (bottom panel) shows the time dependence of $\lambda_{\rm max}$ during 60 days for constant (dashed lines) and changing $S_{\rm max}$ (solid lines). For the latter case, the terminal peak wavelength is slightly larger than the former case, e.g.,  $\lambda_{\rm max}=0.13\mum$ at 1 pc and $\lambda_{\rm max}=0.15\mum$ at 2 pc. From the middle and top panels, it appears that the effect of changing $S_{\rm max}$ has a more important effect on $R_{V}$ than on $\lambda_{\rm max}$.

\section{Effect of RATD on the SNe Ia light curve}\label{sec:lightcurve}
In Section \ref{sec:ext}, we show the disruption of large grains into smaller ones by RATD increases the dust extinction in the NUV band and decreases in the optical and NIR band with time. As a result, the observed radiation spectrum of SNe Ia would be different from the case where dust properties are constant, i.e., in the absence of RATD. In this section, we use the new extinction value calculated from Section \ref{sec:ext} to predict how the SNe Ia light curve changes with time in the presence of RATD.

Let us first derive a simplified intrinsic light curve of the SNe Ia. The explosion of SNe Ia releases a huge amount of energy and eject all material into space in the short time. Its radius will increase dramatically, and the temperature will drop during the free expanding phase. For our simple model, we assume the total energy $E\sim 10^{51}\erg$  is the constant during this phase, and the ejecta of mass $M_{\rm ej}=1.4M_{\odot}$ will move with the constant velocity: $v = \sqrt{2E/M_{\rm ej}} \sim 10^{4}E_{\rm 51}^{1/2}(M_{\odot}/M_{\rm ej})^{1/2} \rm km\s^{-1}$ with $E_{51}=E/10^{51}\erg$. The total bolometric luminosity of SNe Ia is given by:
\begin{align}
    L_{\rm bol} =  4 \pi R^{2} \sigma T_{\rm SN}^{4},
\end{align}
where $R=vt$ is the radius of SNe Ia after $t$ time and $T_{\rm SN}$ is its effective temperature. Using $L_{\rm bol}$ in Equation (\ref{eq:1}), the effective temperature of SNe Ia decreases with time as
\begin{align}
    T_{\rm SN}\simeq 10^{4}\left(\frac{L_{\rm bol}}{10^{9}L_{\rm \odot}}\right)^{1/4}\left(\frac{10^{4}\rm km\s^{-1}}{v}\right)^{1/2}\left(\frac{t}{1~\rm days}\right)^{1/2}\K.\label{eq:TSN}
\end{align}
 
The intensity of SNe radiation after propagating through a dusty cloud at wavelength $\lambda$ is governed by
\begin{align} \label{eq:18}
I_{\lambda}(t) = I_{\lambda}^{0}e^{-\tau(\lambda,t)},
\end{align}
where 
\begin{align}
    I_{\lambda}^{0}=\frac{2 hc^{2}}{\lambda^{5}}\frac{1}{\exp(hc/\lambda kT_{\rm SN})-1},
\end{align} 
is the intrinsic specific radiation intensity, and the optical depth $\tau(\lambda,t)=A(\lambda,t)/1.086$. Here, the infrared thermal emission from dust grains is disregarded because we are interested only in the UV-NIR spectrum of SNe Ia. With $A(\lambda,t)$ obtained in Section \ref{sec:ext}, one can calculate the observed radiation intensity via Equation (\ref{eq:18}).\footnote{Here, to illustrate the effect of time-varying dust extinction, we disregard the rising phase of the SNe light curve. With $A(\lambda,t)$ available from Section \ref{sec:ext}, one can predict the observed intensity for an accurate intrinsic intensity without difficulty.}

\begin{figure}
        \includegraphics[width=0.5\textwidth]{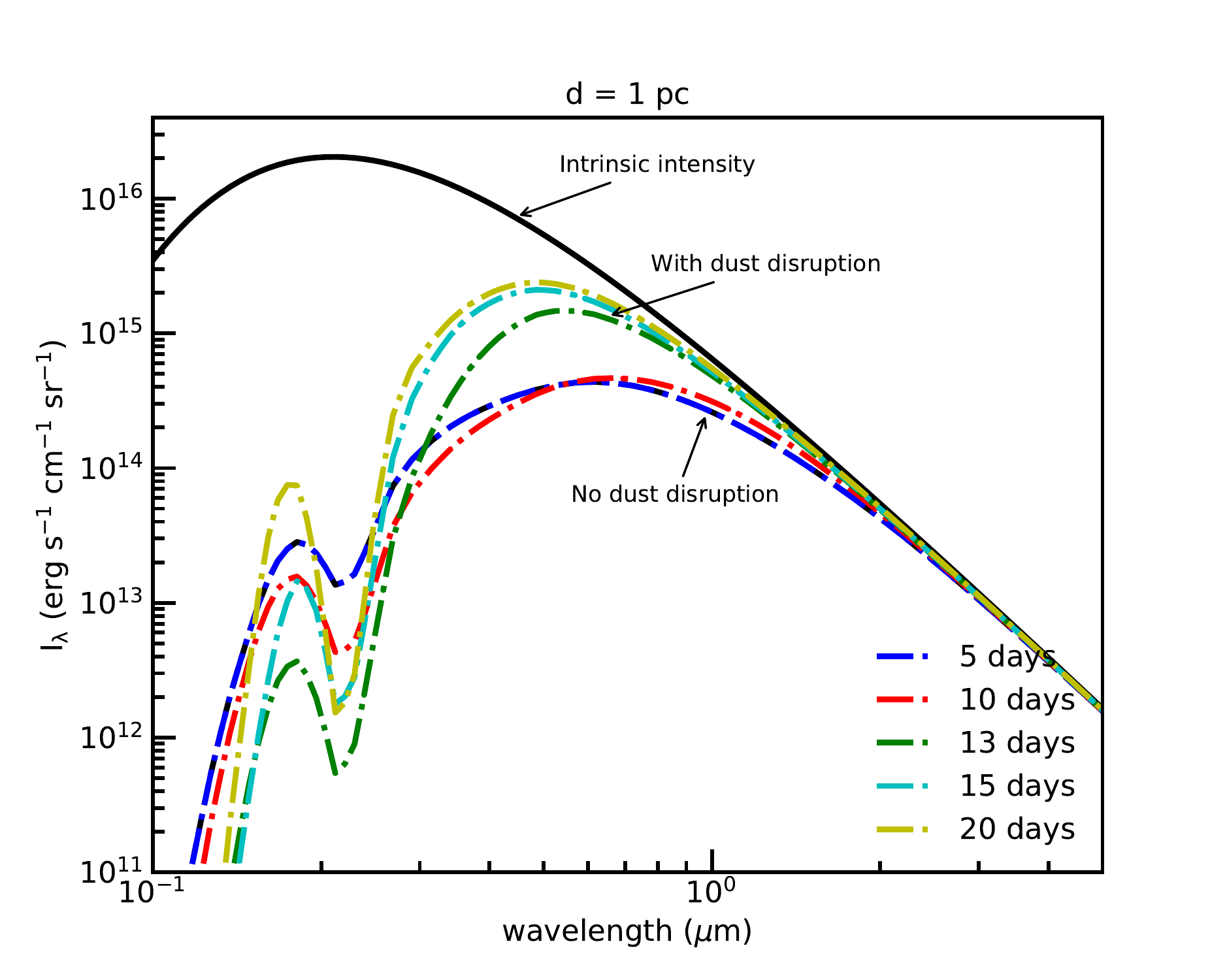}
        \caption{Comparison of the intrinsic radiation intensity from a SN Ia (black line) with the radiation intensity attenuated by dust extinction evaluated at the different times from 5 to 2 days (dashed lines), assuming the cloud distance of $d=1$ pc.}
           \label{fig:lightcurve}
\end{figure}

Figure \ref{fig:lightcurve} shows the UV-NIR spectrum of a SN Ia observed at the different times (dashed and dot-dashed lines) compared with the intrinsic spectrum (black solid line). We assume the visual extinction estimated for SN 1986G of $A_{\rm V}=2.03$ mag to derive the total gas column density $N_{\rm H} = 3.14 \times 10^{21}\cm^{-2}$. The black dashed line represents the observed SNe spectrum with constant dust extinction of $A_{\rm V}(t=0)=2.03$ (no disruption), and the color dot-dashed line shows the results when RATD is taken into account. Without RATD, the observed SNe spectrum appears constant with time but will vary in the presence of RATD. Indeed, due to RATD, large grains are disrupted in smaller grains, increasing the extinction at the NUV band, such that SNe Ia become redder. In contrast, the decrease of the extinction in the optical and NIR band due to grain disruption makes SNe Ia brighter than with time. The radiation intensity at long wavelengths ($\lambda \sim 3-5 \mum$) is not affected by dust disruption because of optically thin regime (see Section \ref{sec:ext}).

\begin{figure*}
    \centering
    \includegraphics[width=0.45\textwidth]{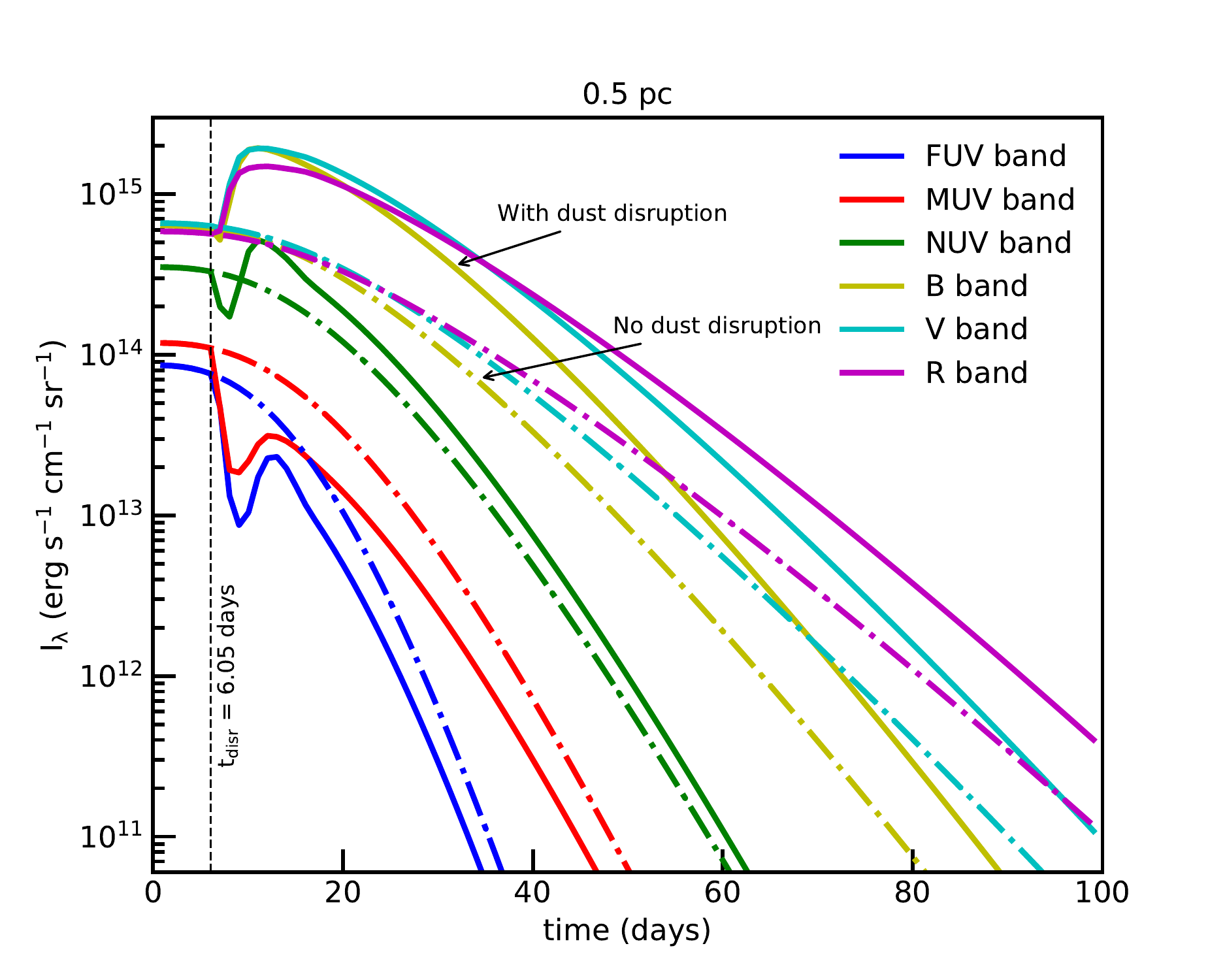}
    \includegraphics[width=0.45\textwidth]{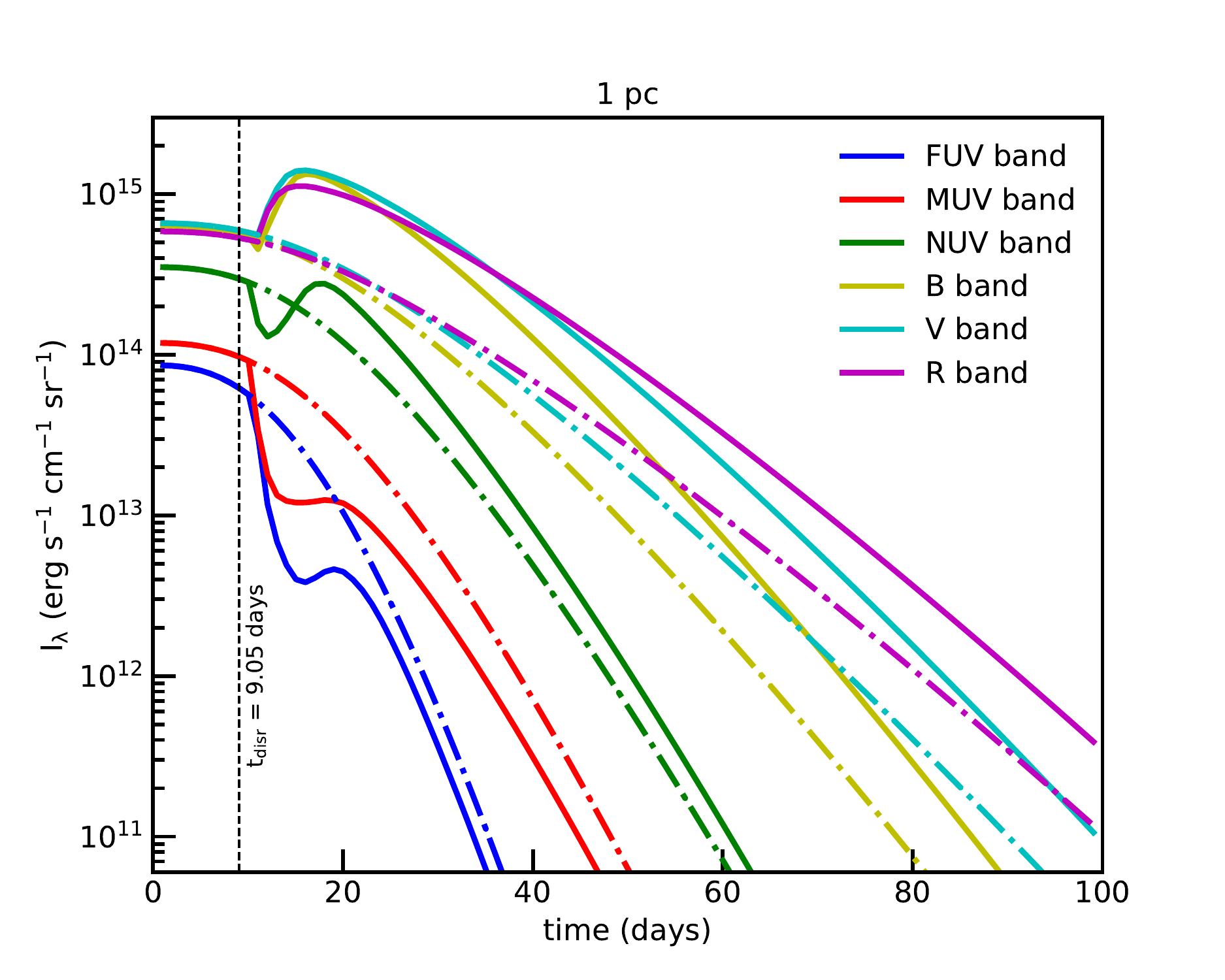}
    \includegraphics[width=0.45\textwidth]{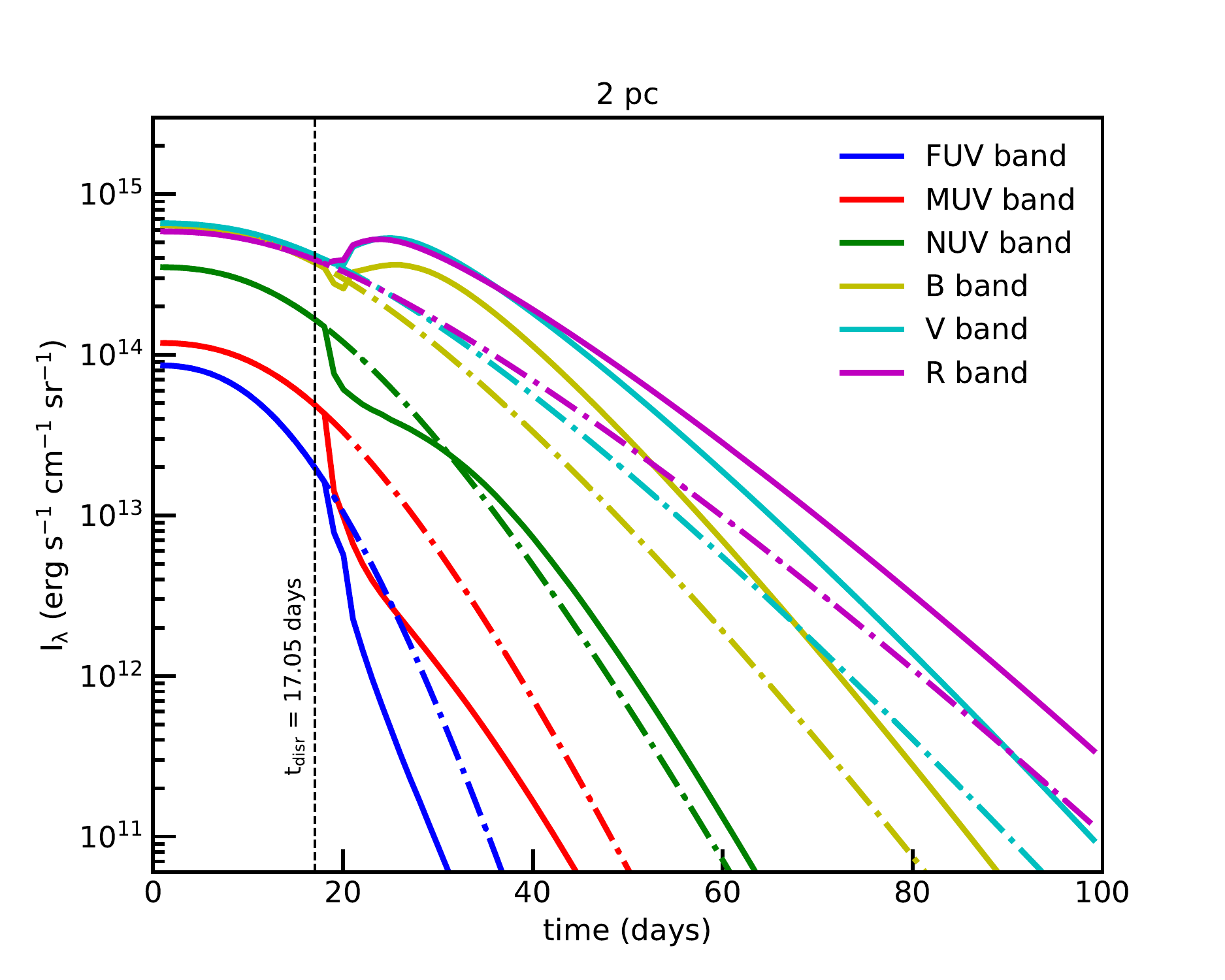}
    \includegraphics[width=0.45\textwidth]{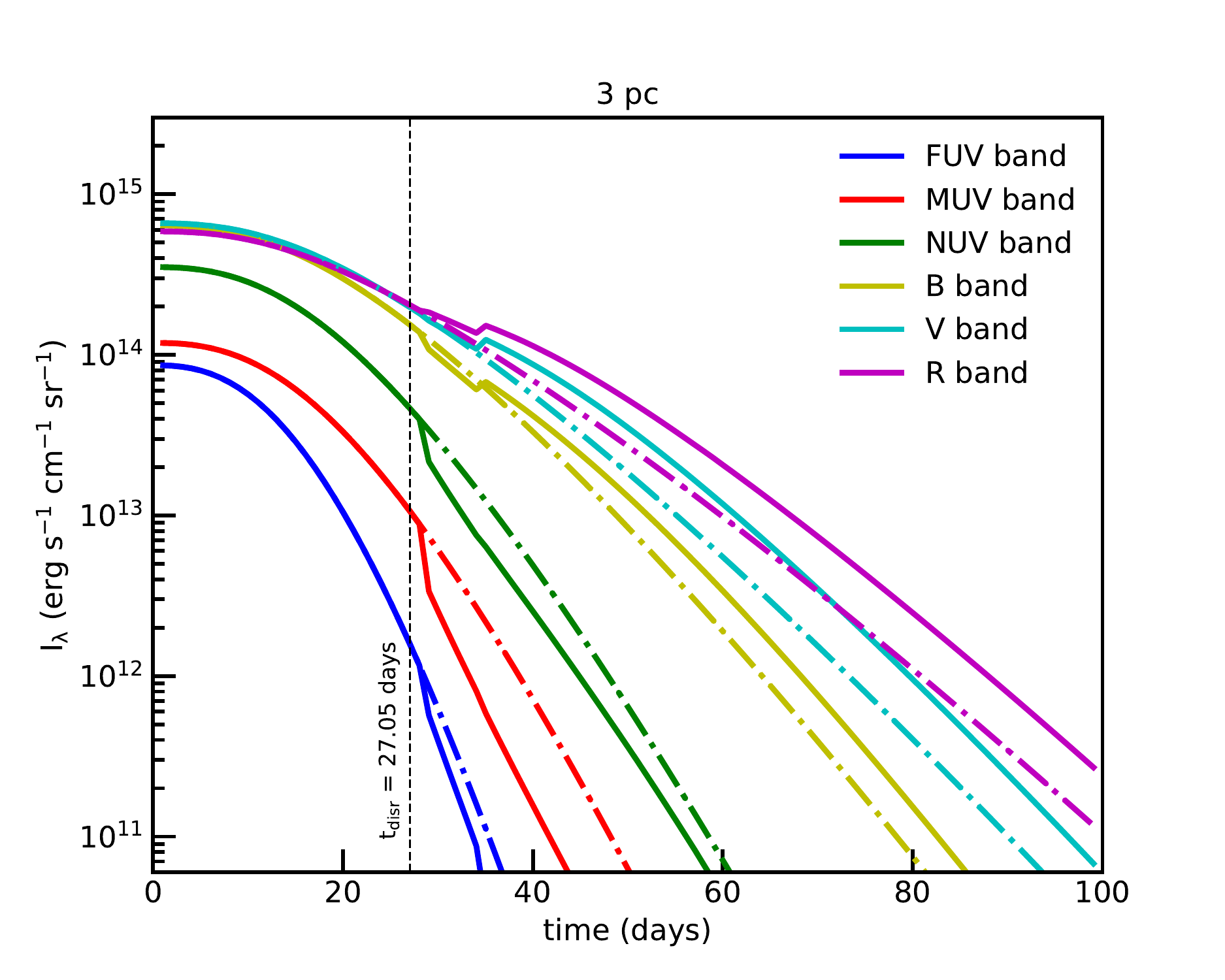}
    \caption{Specific intensity at the different bands as a function of time without (dashed lines) and with (solid lines) rotational disruption, assuming the different dust cloud distances from 0.5 pc to 3 pc. The tensile strength $S_{\rm max}=10^{7}\erg\cm^{-3}$ is considered. Vertical dotted lines mark the disruption time of graphite grains.}
       \label{fig:compareIv}
\end{figure*}

To see more detail about time variation of SNe colors, in Figure \ref{fig:compareIv}, we show the light intensity at FUV ($0.15 \mum$), MUV ($0.25 \mum$), NUV ($0.3 \mum$), U, B, V and R bands for the different dust cloud locations, assuming $S_{\rm max}=10^{7}\erg\cm^{-3}$. In the absence of RATD, the radiation intensity of SNe Ia at the different bands only decreases with time due to the decrease of $T_{\rm SN}$. However, in the presence of RATD, the radiation intensity from optical-NIR wavelengths exhibits an abrupt increase (i.e., SNe become re-brighter), whereas it has a drop at FUV-NUV bands (solid lines) due to the effect of grain disruption. The "rebrighter" time in optical-NIR is longer for more distant clouds. For instance, the re-brighter time $t_{rebrighter}\sim t_{\rm disr}\sim 10$ days for $d=0.5$ pc and $t_{re-brighter}\sim 30$ days for $d=3$ pc. 
\begin{figure*}
    \centering
    \includegraphics[width=0.45\textwidth]{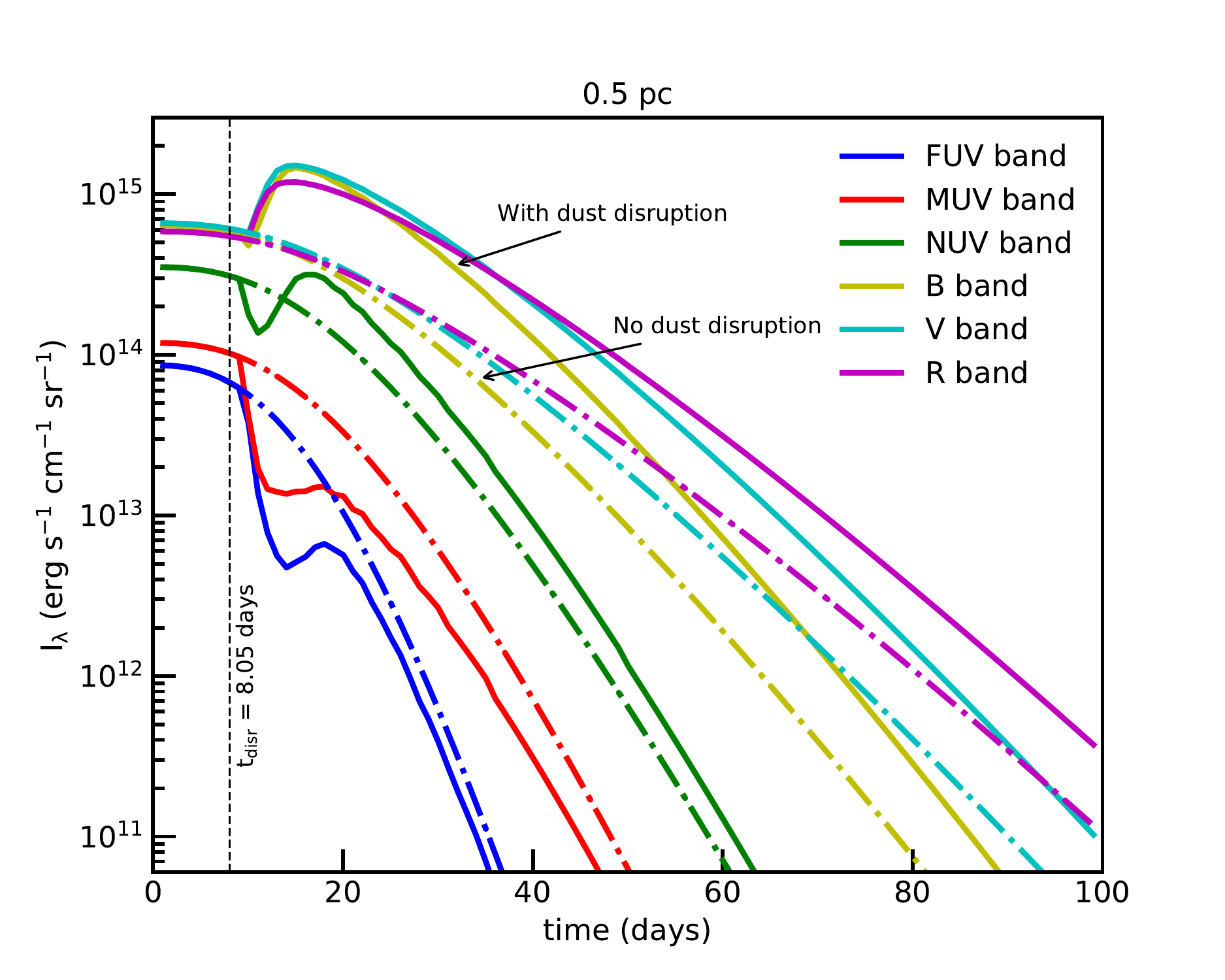}
    \includegraphics[width=0.45\textwidth]{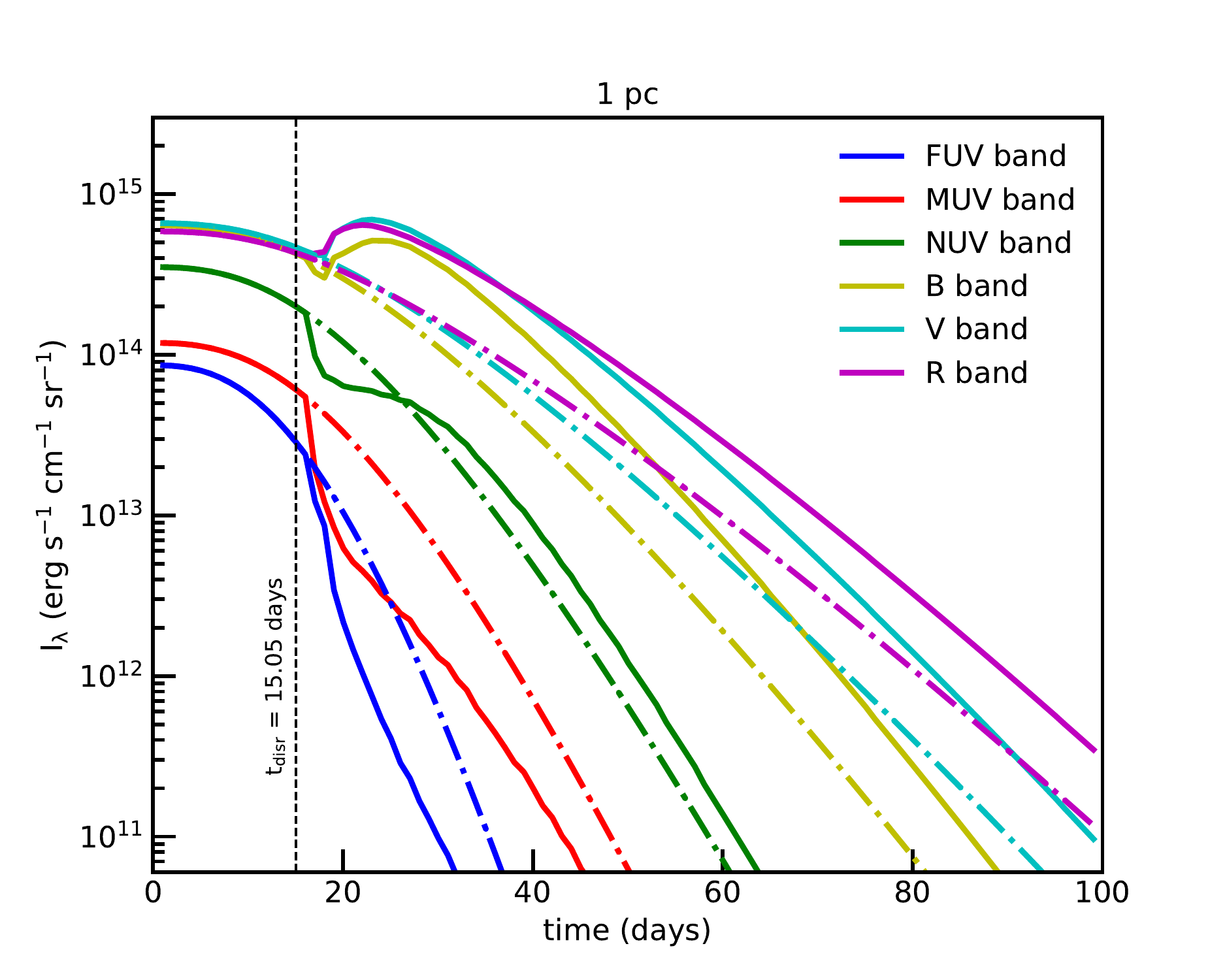}
    \includegraphics[width=0.45\textwidth]{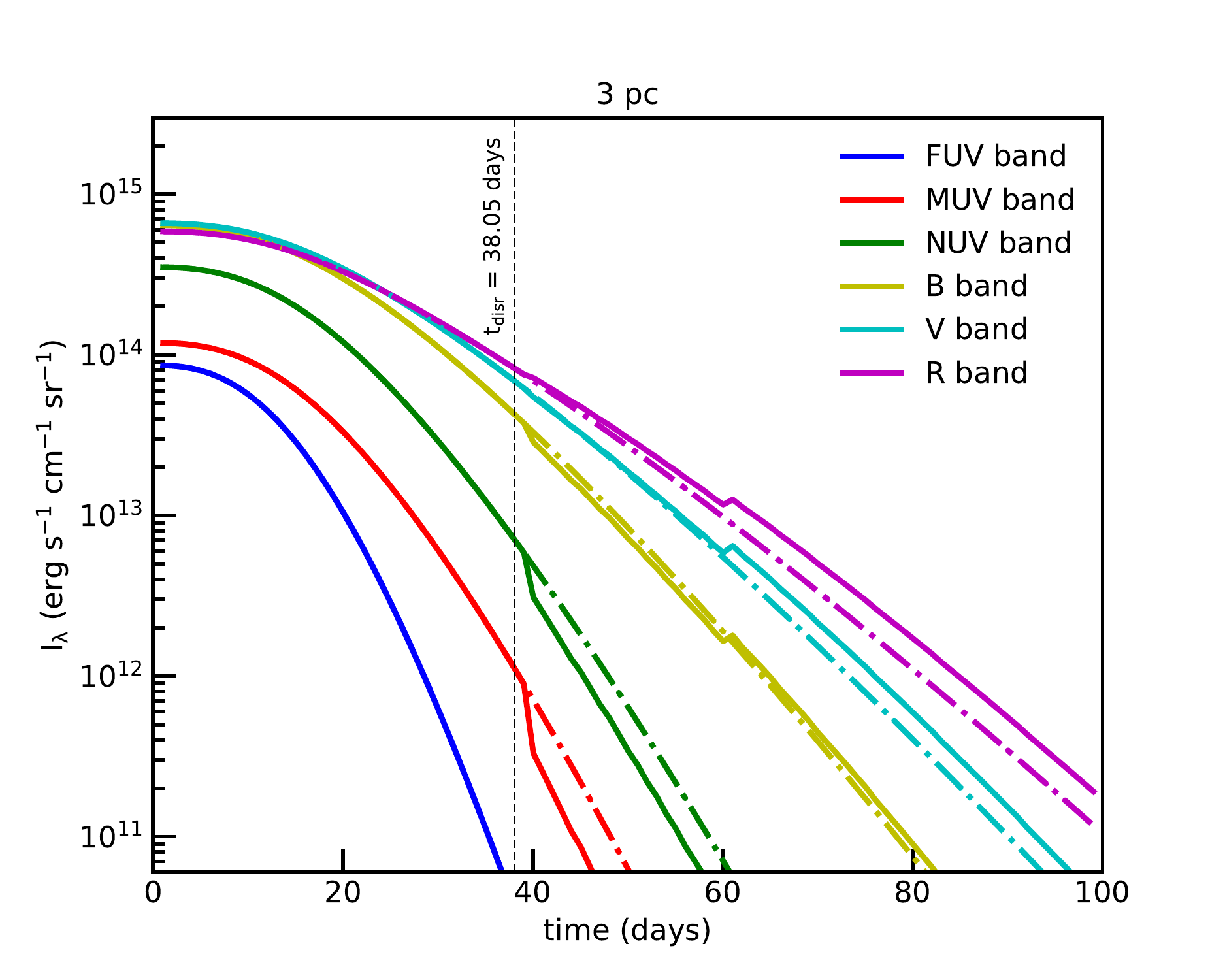}
    \includegraphics[width=0.45\textwidth]{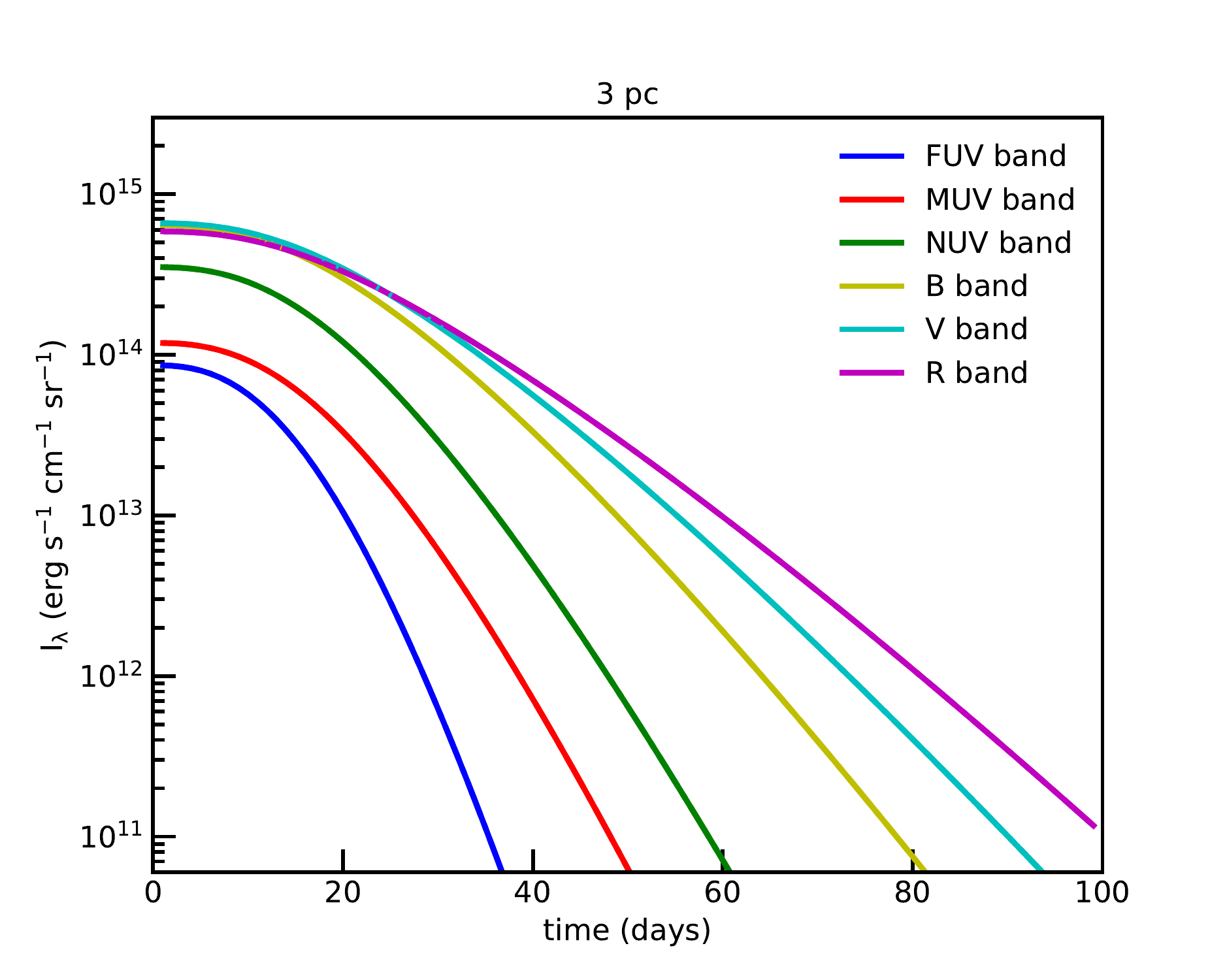}
    \caption{Same as Figure \ref{fig:compareIv} but for grain has $S_{\rm max} = 10^{8}\erg\cm^{-3}$.}
    \label{fig:compareIvs8}
\end{figure*}
 
Figure \ref{fig:compareIvs8} presents the same as Figure \ref{fig:compareIv} but with grains with $S_{\rm max}= 10^{8}\erg\cm^{-3}$. One can see that the large disruption grain size of the stronger grains gives a smaller effect on the dust extinction in UV-NIR band than the light grain. As a result, the 'rebrighter' and 'redimmer' feature due to the reduction of the extinction in the visible band and increase in the FUV-NUV band smaller and nearly no effect if the cloud is farther than 2 pc.

Note that to highlight the effect of RATD on the observed SNe Ia light curve, we have modeled the intrinsic SNe Ia light curve as a fireball with a constant expansion velocity. However, the radioactivity of synthesized Ni$^{56}$ and Co$^{56}$ (\citealt{Arnet82}) can provide more radiative energy, such that the peak luminosity of SNe Ia and the effective temperature would be different from our simplified model. Nevertheless, we expect the effect of RATD on the SNe light curve is similar.

\section{Discussion}\label{sec:discuss}
\subsection{Comparison of $R_{\rm V}$ and $\lambda_{\rm max}$ from our models with SNe Ia data}

Using RATD theory, we have computed the extinction curves $A(\lambda,t)$ by dust grains which have the grain size distribution being modified by RATD at the different times and dust cloud distances in Section \ref{sec:RV}. Our obtained results show that, due to RATD, $R_{\rm V}$ rapidly decreases with time, from an original value of $R_{\rm V}=3.1$ to small values of $R_{\rm V}\sim 1.5$ after a timescale $t_{\rm disr}\sim 50-60$ days if the dust cloud is located within $d\sim 4$ pc from the supernova (see Figure \ref{fig:Rv}). The final value $R_{V}$ achieved when RATD ceases is smaller for the dust cloud closer to the supernova, but it increases with the tensile strength of grains (see Figures \ref{fig:Rv} ). Comparing our results with the observational data of SNe Ia (see Table \ref{tab:B1}), one can see that the RATD can successfully reproduce the unusual low values of $R_{\rm V}$ observed toward many SNe Ia (see Table \ref{tab:B1}; \citealt{Burn14}; \citealt{Cikota16}).  

We also modeled the polarization of SNe Ia light by grains aligned with the magnetic fields using RAT mechanism in Section \ref{sec:pol}. Our results show that the peak wavelength $\lambda_{\rm max}$ decreases from the standard ISM value $\sim 0.55\mum$ to $\lambda_{\rm max}<0.2\mum$ due to RAT alignment after a timescale of $t_{\rm align}\sim 5-20$ days for clouds at distance within 4 pc (see Figure \ref{fig:lamdamax}, upper panel). These results can reproduce the low values of $\lambda_{\rm max}$ observed toward SNe Ia (see Table \ref{tab:B1}; \citealt{Patat15}; \citealt{Zelaya17}). We note that the characteristic timescale for achieving terminal $\lambda_{\rm max}$ is much shorter than that for terminal $R_{V}$ because $t_{\rm align}<t_{\rm disr}$.

From lower panel of Figure \ref{fig:lamdamax}, one can see that the main reason for the decrease of $\lambda_{\rm max}$ from $0.5\mum$ to $0.2 \mum$ is due to the alignment of small grains by RAT mechanism as suggested by \cite{Hoang17}. Therefore, to obtain a higher $\lambda_{\rm max}$, one can increase the critical limit for suprathermal rotation, i.e., $T_{\rm gas}$.

\begin{figure}[t]
        \includegraphics[width=0.5\textwidth]{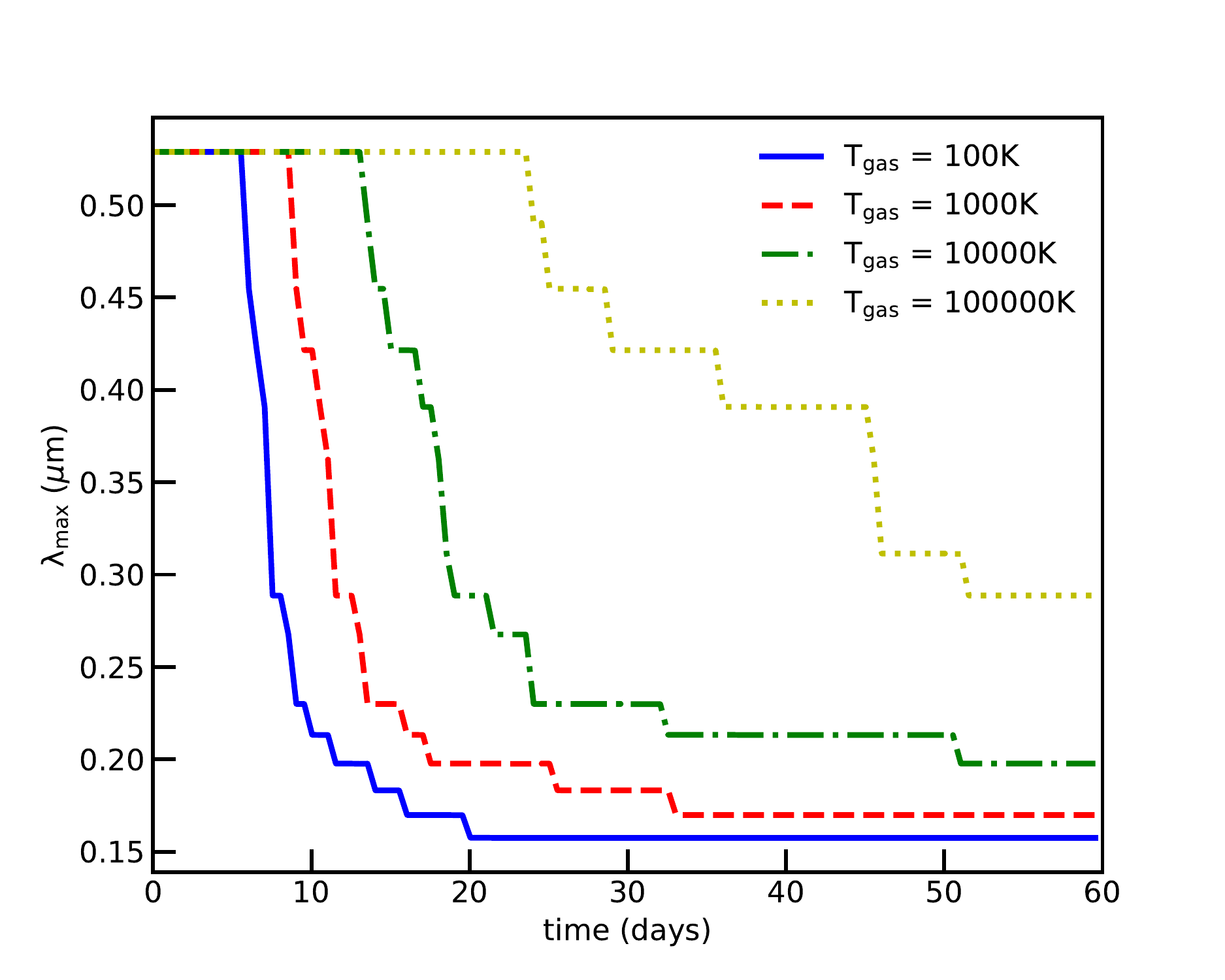}
        \caption{The peak wavelength $\lambda_{\rm max}$ during 60 days for the different gas temperature and cloud distance $d= 3$ pc, assuming $S_{\rm max}=10^{7}\erg\cm^{3}$.}
           \label{fig:max_Tgas}
\end{figure}

Figure \ref{fig:max_Tgas} shows the peak wavelength of the dust polarization for the different gas temperatures and the dust cloud at 3 pc from the source. Here the grain disruption is not considered in order to focus on the effect of grain alignment. The peak wavelength increases with increasing the gas temperatures because higher temperatures result in larger $\omega_{T}$ such that stronger radiation energy is required spin-up grains to suprathermal rotation and align grains efficiently. For example, a dust cloud at 3 pc will give $\lambda_{\rm max} \sim 0.3 \mum$ if $T_{\rm gas} = 10^{5}\K$, two times larger than the case $T_{\rm gas} \rm =10^{3}\K$. 

\subsection{Test RATD and RAT alignment mechanisms with time-variation of SNe Ia extinction and polarization}\label{sec:test}

Our modeling results of extinction by RATD reveals that the extinction and $R_{V}$ remains constant until the grain disruption occurs at $t=t_{\rm disr}$ spanning $5-30$ days depending on the distance (right panel of Figure \ref{fig:Rv}) and grain tensile strength $S_{\rm max}$ (Figure \ref{fig:Rv_Smax}). Beyond $t_{\rm disr}$, the optical-NIR extinction and $R_{V}$ decrease rapidly and achieve saturated levels after $50-60$ days for the cloud distance $d<4$ pc. Therefore, by observing SNe Ia at early times and infer time-dependence $R_{V}$, one can test the RATD mechanism.

On the other hand, our modeling of dust polarization due to grains aligned by the RAT mechanism show that the dust polarization starts to increase after a timescale corresponding to the alignment timescale, $t_{\rm align}$, which spans between $3-10$ days (see Figure \ref{fig:Plamda_t} and \ref{fig:Plamda_d}). When the RATD begins at $t\sim t_{\rm disr}> t_{\rm align}$, the polarization starts to decrease rapidly over time, and its time-variation is similar to that of the extinction. Therefore, the best strategy to test RAT alignment mechanism is to observe the SNe Ia polarization at earliest times $t<5-15$ days if the dust cloud is located within 4 pc from the source. This is important for interpreting the photometric and polarimetric data of SNe Ia with high dust reddening.

\subsection{Effect of RATD on SNe Ia colors and light curves}

Dust extinction is considered the most important systematic uncertainty for SNe Ia cosmology (\citealt{Scolnic:2019ug}). Understanding better the intrinsic light curves will provide the better study the physical properties and the evolution of SNe Ia (\citealt{Burn14}) and allows a precise measurement of cosmological constant and more precise constraint on dark energy. 

Usually, to infer the intrinsic light curve of SNe Ia, one adopts a typical model of constant dust extinction and fits the model to observational data. However, here we show that the properties of dust within 4 pc from SNe Ia are modified by supernova radiation via RATD, thus, and their extinction curves are fundamentally different from the typical dust model if there exists some dust clouds within 4 pc from the source. Therefore, to achieve an high accuracy of the intrinsic SNe Ia light curve, we suggest to conduct observations at the earliest times as possible, ideally within the first ten days after the first light when grain disruption has not occurred yet.

The situation becomes more complicated for deriving intrinsic SNe Ia colors because the intrinsic colors of SNe Ia are expected to vary with time (see \citealt{Goobar08}). Moreover, as shown in Figure \ref{fig:Rv}, the color excess $E(B-V)$ by dust extinction first increases with time and then it declines slowly. This trend may already be seen toward many SNe Ia (see \citealt{Bulla18}). 

\subsection{Constraining the distance and properties of dusty clouds toward SNe Ia with observations}
Understanding the distance and physical properties of dust clouds to SNe Ia is essential for inferring the intrinsic SN light curve as shown previously (\citealt{Goobar08}; \citealt{Bulla18}) as well as in this paper.

In Section \ref{sec:lightcurve}, we show that SNe Ia become re-brighter at optical-NIR bands accompanied by a sudden dimmer at UV bands when grain disruption begins. The rebrighter time is essentially the same as the disruption time $t_{\rm disr}$ which is a function of the cloud distance to the source (\citealt{Hoang19}). Therefore, one can constrain the cloud distance by observing the SNe Ia colors and identify the re-brighter time. On the other hand, the polarization at optical-NIR bands decreases substantially due to RATD. As a result, a simultaneous decrease in the polarization together with re-brighter would be a smoking gun for RATD and for inferring the cloud distance.

We note that the cloud distance of some SNe Ia were constrained in \cite{Bulla18} using the time-variation of color excess $E(B-V)$ caused by scattering of photons by dust. Yet, their analysis assumed a standard grain size distribution from the Milky way which cannot reproduce the unusual low values of $R_{V}$ and $\lambda_{\rm max}$. 

Physical and chemical properties of the dusty cloud near SNe Ia are also poorly known. Usually, one derives the gas column density of the dust cloud using the standard relationship $N_{\rm H}/A_{V}\approx 5.8\times 10^{21}\cm^{-2}\rm mag^{-1}/R_{V}$ (\citealt{Draine03}). Yet, the optical extinction decreases with time from the original value if the cloud is located within 4 pc from the source (see Figure \ref{fig:Alamda_t}). Therefore, the estimate of $N_{\rm H}$ using the observed $A_{V}$ would yield a lower value than the actual column density.

\subsection{On the origin of anomalous $K-\lambda_{\rm max}$ and $K-R_{V}$ relationships of SNe Ia}\label{sec:K_lambda}

The wavelength-dependence polarization of distant stars in the Milky Way is usually fitted by the Serkowski law (\citealt{Serkow75}):
\begin{align} \label{eq:22}
P_{SerK}({\lambda}) = P_{\rm max} \exp\left[-K \ln^{2}\left(\frac{\lambda_{\rm max}}{\lambda}\right)\right],
\end{align}
where $P_{\rm max}$ is the maximum degree of polarization, $\lambda_{\rm max}$ is the wavelength at the maximum polarization (hereafter peak wavelength), and $K$ is a parameter given by  (\citealt{Wilking80}; \citealt{Whit92}):
\begin{align} \label{eq:23}
K = 0.01 \pm 0.05 + (1.66 \pm 0.09)\lambda_{\rm max}.
\end{align}

Nevertheless, \cite{Patat15} and \cite{Cikota18} showed that the observational data of SNe Ia do not follow the standard relationship (Eq. \ref{eq:23}). Specifically, the $K$ value for a given small $\lambda_{\rm max}$ is much larger than predicted Equation (\ref{eq:23}). This puzzle once again reveals the abnormal properties of dust in the local environments of SNe Ia.

Our modeling results in Section \ref{sec:pol} reveal that the polarization curves tend to become narrower, i.e., the parameter $K$ tends to increase, whereas the value of $\lambda_{\rm max}$ decreases in the presence of grain disruption by RATD (see Figures \ref{fig:Plamda_t}, \ref{fig:Plamda_d} and \ref{fig:lamdamax}). To see whether the RATD mechanism can reproduce the observed anomalous relationship of $K$-$\lambda_{\rm max}$, we fit the Serkowski law (Equation \ref{eq:22}) to the polarization curve calculated by Equation (\ref{eq:16}) to obtain $K$ and $\lambda_{\rm max}$ (see Appendix \ref{apdx:fit} for more details) and compare with the observational data.

\begin{figure}[!htb]
        \includegraphics[width=0.5\textwidth]{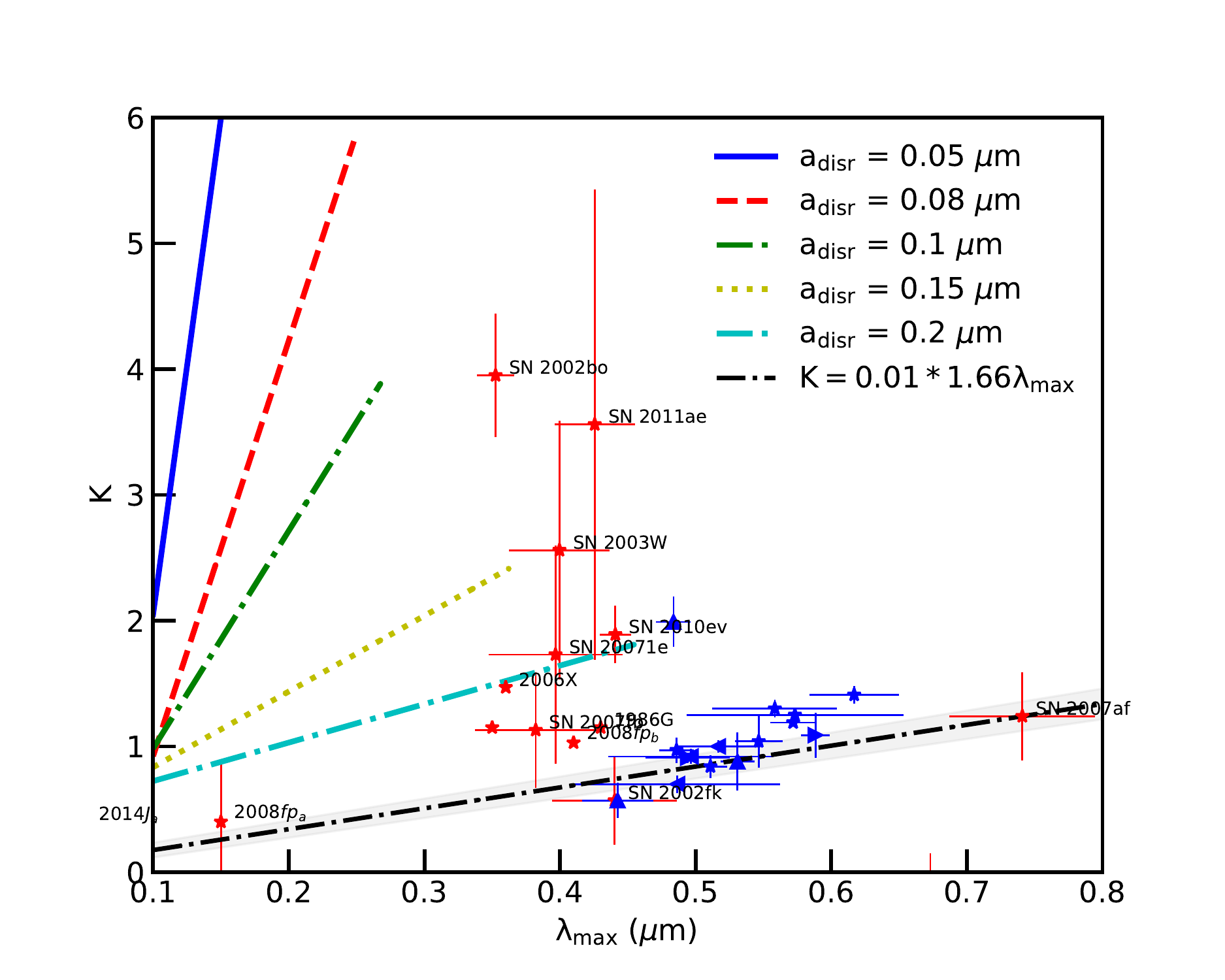}               		   
        \includegraphics[width=0.5\textwidth]{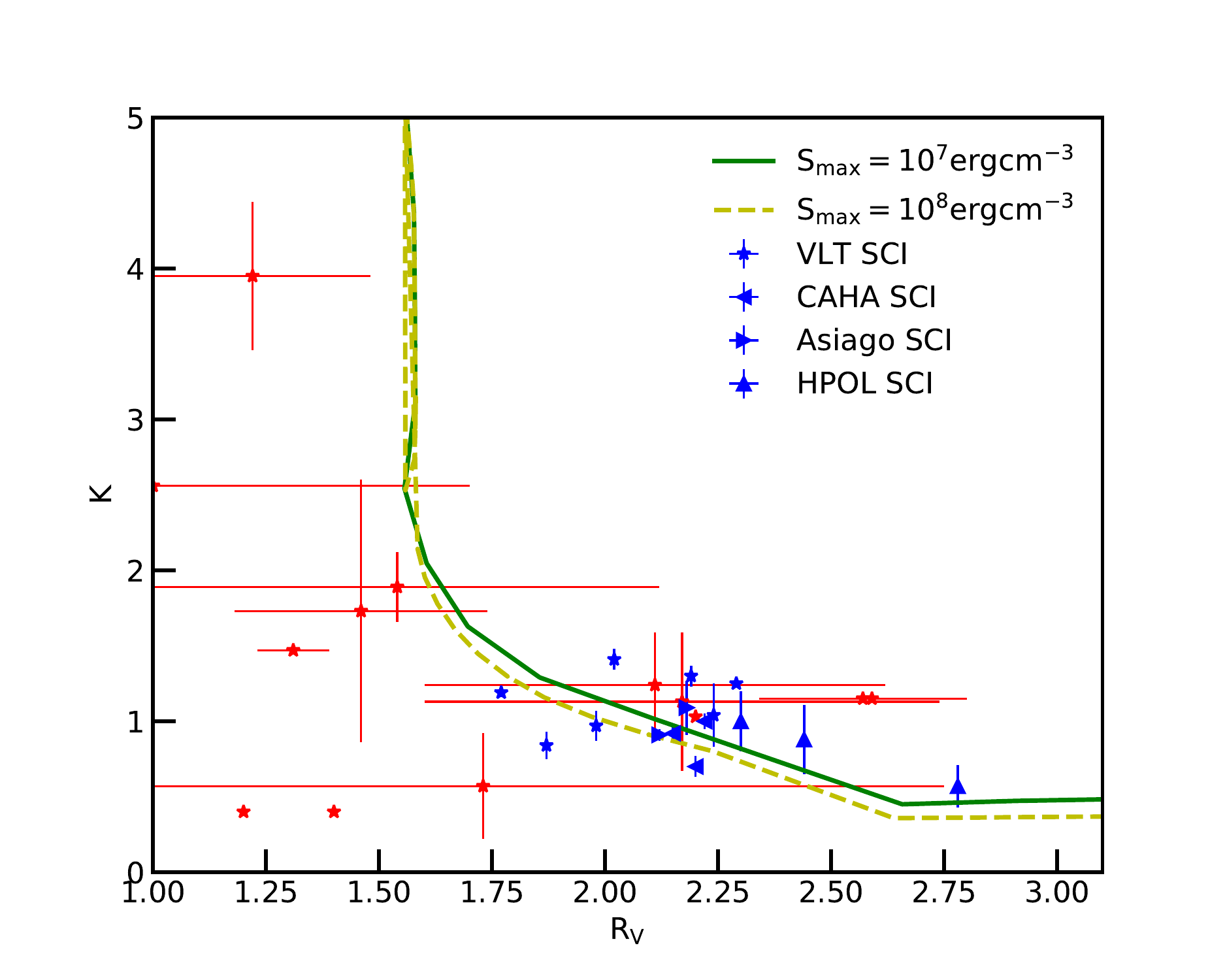}
        \caption{Upper panel: relationship between $K$ and $\lambda_{\rm max}$ predicted by our model for different disruption sizes and alignment size. The black line is the standard relationship given by Equation (\ref{eq:23}). Lower panel: $K$ vs. $R_{V}$ from our models compared with observational data for SNe Ia. Symbols show observational data for SNe Ia presented (red symbols) and normal stars in our galaxy (blue symbols) listed in Table \ref{tab:B1} and Table \ref{tab:B2}.}
           \label{fig:K_max}
\end{figure}

In Figure \ref{fig:K_max} (upper panel), we show the peak wavelength $\lambda_{\rm max}$ and the parameter $K$ for several values of the grain disruption size from $a_{\rm disr}=0.2\mum$ to $0.05\mum$. For each $a_{\rm disr}$, the alignment grain size is varied from $a_{\rm align}=0.05\mum$ to $\rm 0.002 \mum$ to account for the effect of enhanced alignment by SNe light (see Appendix \ref{apdx:fit} for more details). Observational data from SNe Ia and some stars in our galaxy  (see Table \ref{tab:B1} and \ref{tab:B2}) are shown for comparison. We see that, for a given $a_{\rm disr}$, $K$ decrease rapidly with decreasing $\lambda_{\rm max}$ due to the decrease of $a_{\rm align}$. Moreover, for a given $a_{\rm align}$, $\lambda_{\rm max}$ tends to decrease with decreasing $a_{\rm disr}$. In particular, we can see that the high $K$ values of SNe Ia can be reproduced by our models with RATD at the different $a_{\rm disr}$. For example, SN 2006X can be explained with the maximum grain size $a_{\rm disr}\sim  0.2\mum$ and alignment grain size $a_{\rm align}\sim 0.03 \mum$, or SN 2010ev can be explained with $a_{\rm disr}= 0.2\mum$ and $a_{\rm align} \approx 0.04\mum$. 

Lastly, our results in Section \ref{sec:ext} and \ref{sec:pol} show that under strong radiation field, both extinction curve and polarization curves are modified from the standard curves in the ISM. This suggests that there is some correlation between $R_{V}$ and $K$. In Figure \ref{fig:K_max} (lower panel) we show the values of $R_{V}$ and $K$ calculated during 100 days with the cloud at 1 pc for several maximum tensile strengths and compare with observation data (see Table \ref{tab:B1} and \ref{tab:B2}). The increase of $K$ with decreasing $R_{V}$ as a result of RAT alignment and RATD appears to be consistent with the observational data. 

\section{Conclusions}\label{sec:summary}
In this paper, we have studied the effects of rotational disruption and alignment of dust grains by radiative torques on the extinction, polarization, and light curves of SNe Ia. The main findings of our study are summarized as follows:

\begin{itemize}

\item[1.] Bases on the RATD mechanism, grains can be disrupted when radiative torques spin up it up to critical speed during 200 days after the supernova explosion. The grain disruption size decreases with time at the fixed distance and increases with increasing the cloud distance to the source. The higher maximum tensile strength makes grain is more difficult to disrupt and the gas density is independent to the dust disruption size in the very strong radiation field as SNe Ia.

\item[2.] Using the grain disruption size from RATD, we model the extinction of SNe light by dust absorption. We find that the conversion of large grains into smaller ones by RATD results in a rapid decrease in the optical-NIR extinction but increase in the UV extinction. This results in the rapid decrease in the the total-to-selective visual extinction ratio $R_{V}$ over a timescale of $t_{\rm disr}\sim 50$ days. This can successfully explain the unusually low values observed toward SNe Ia.

\item[3.] Using RAT theory, we study the alignment of grains by SNe radiation. We find that the intense SNe radiation can align small grains, and the alignment size decreases with time. We model the polarization of SNe light by aligned grains. We find that the polarization degree increases first due to enhanced alignment by SNe radiation and then drops rapidly when grain disruption begins. We find that the peak wavelength decreases rapidly from the original standard value to very small values of $\lambda_{\rm max}<0.4\mum$ within less than $t_{\rm align}< 10$ days.

\item[4.] We find that due to RATD and RAT alignment, the $K-\lambda_{\rm max}$ cannot be described by the standard relationship in the Galaxy, but it can qualitatively explain the $K-\lambda_{\rm max}$ reported for SNe Ia. 
 
\item[5.] We suggest a possible way to constrain the cloud distance by performing extreme early time observations. The time of the abrupt decrease in optical-NIR extinction and polarization would put a strong constraint on the cloud distance. An increase in the polarization degree and decrease in the peak wavelength is useful to test the RAT mechanism.

\item[6.] We predict an abrupt increase in the SNe brightness at optical-NIR bands but a decrease in UV extinction due to RATD. This can be tested with photometric and polarimetric observations at early times.

\item[7.] Our results suggest that to achieve a precise measurement of cosmological distances as well as precise constraint on dark energy, one needs to account for the time-variation of dust extinction if there exist dust clouds within several parsecs from the supernova.

\end{itemize}

\acknowledgments
T.H. acknowledges the support from the Basic Science Research Program through the National Research Foundation of Korea (NRF), funded by the Ministry of Education (2017R1D1A1B03035359).

\appendix
\section{Grain size distribution modified by RATD}\label{apdx:sizedist}
In Section \ref{sec:ext}, we mentioned that there are three possible ways to calculate the grain size redistribution function $dn/da$. We have shown results from model 1 where the slope is changed but the normalization constant $C$ is constant. We now study the results for model 2 where the slope $\alpha$ is kept constant but the constant $C$ changes accordingly.

With the case keeping an order is $\alpha=-3.5$ (model 2), the new constant with the new maximum size is:
\begin{align} \label{eq:20}
C_{\rm new} = C \left(\frac{a_{\rm max}^{0.5} - a_{\rm min}^{0.5}}{a_{\rm disr}^{0.5} - a_{\rm min}^{0.5}}\right).
\end{align}

\begin{figure*}[!htb]
        \includegraphics[width=0.5\textwidth]{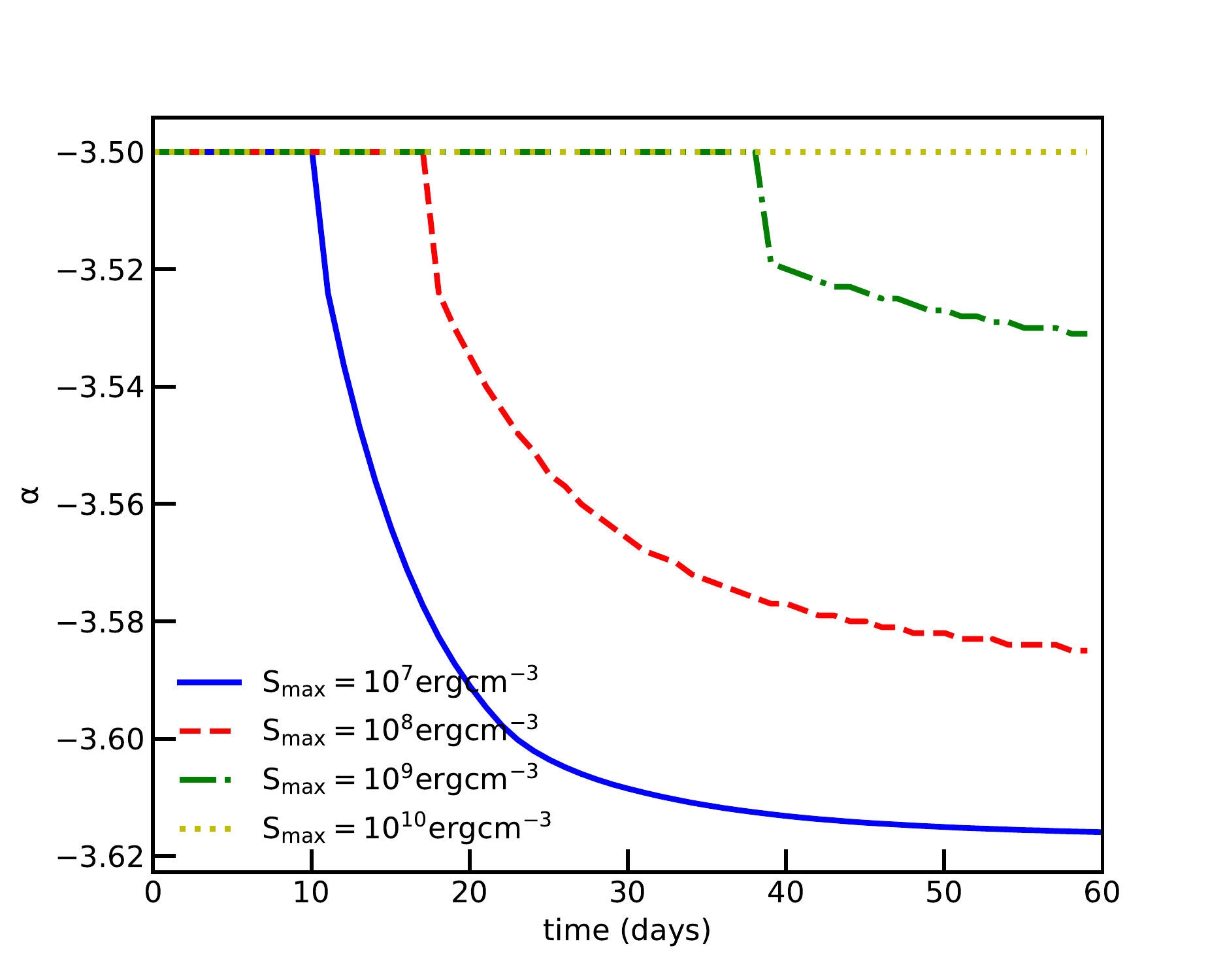}
        \includegraphics[width=0.5\textwidth]{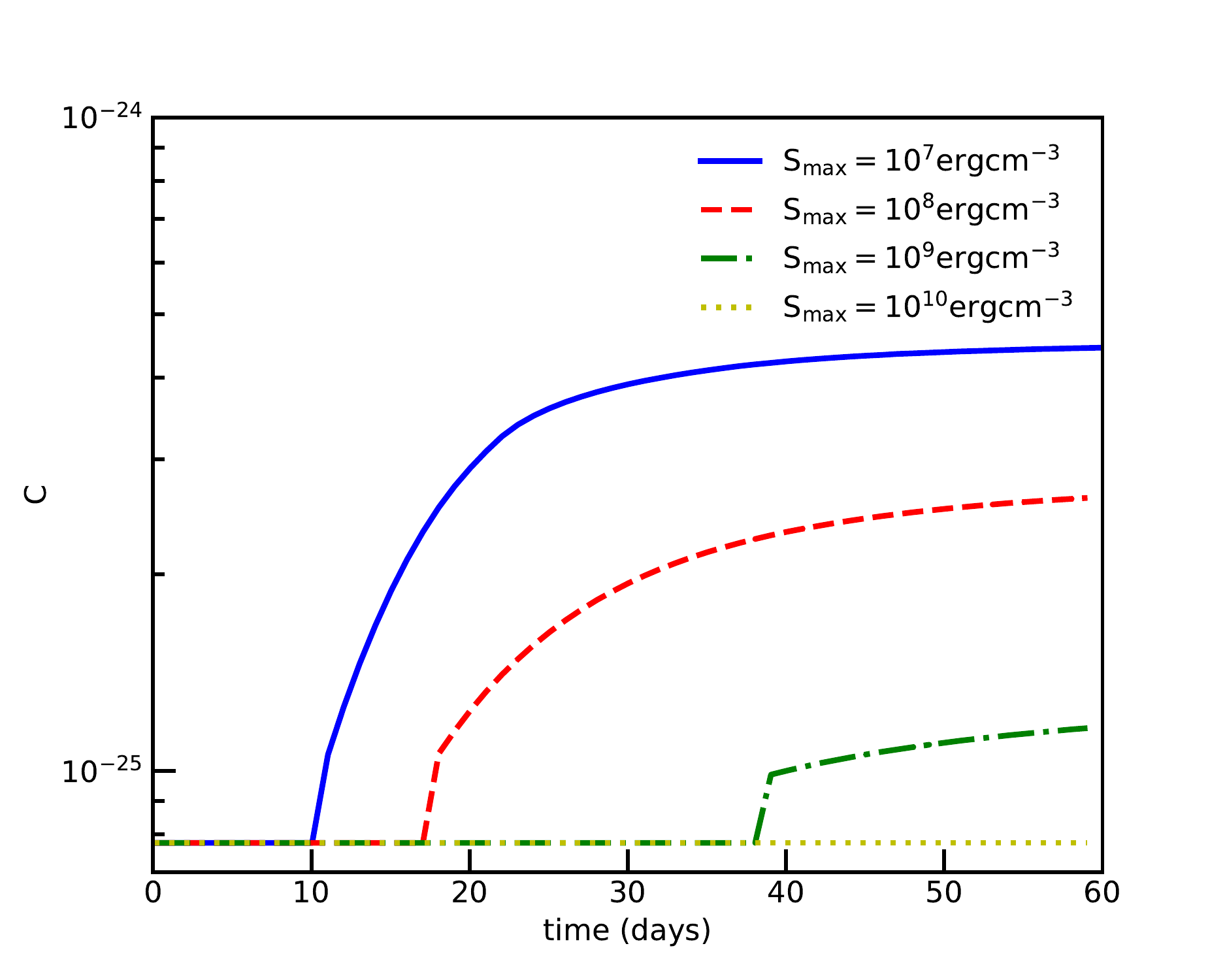}
        \caption{Variation of the slope $\alpha$ vs. time for model 1 (left panel) and $C$ vs. time for model 2 for the different values of maximum tensile strength, assuming the cloud distance of 1 pc.}
           \label{fig:A1}
\end{figure*}

Figure \ref{fig:A1} shows the value of $\alpha$ (model 1) and C (model 2) for silicate grains with time if the cloud stays at 1 pc for the different values of $S_{\rm max}$. As shown, the slope increases (i.e., becomes steeper) from the original value of $\alpha=-3.5$ to $\alpha\sim -3.62$ when RATD begins. This implies that RATD transports dust mass from large sizes to smaller sizes. The right panel shows the value of the constant normalization $C$ which increases with time due to RATD and reaches the terminal value when RATD ceases. In general, the constant varies from $10^{-25.11}\cm^{2.5}$ to $4\times 10^{-25}\cm^{2.5}$ with the cloud in 1 pc.

\begin{figure}[!htb]
        \includegraphics[width=0.5\textwidth]{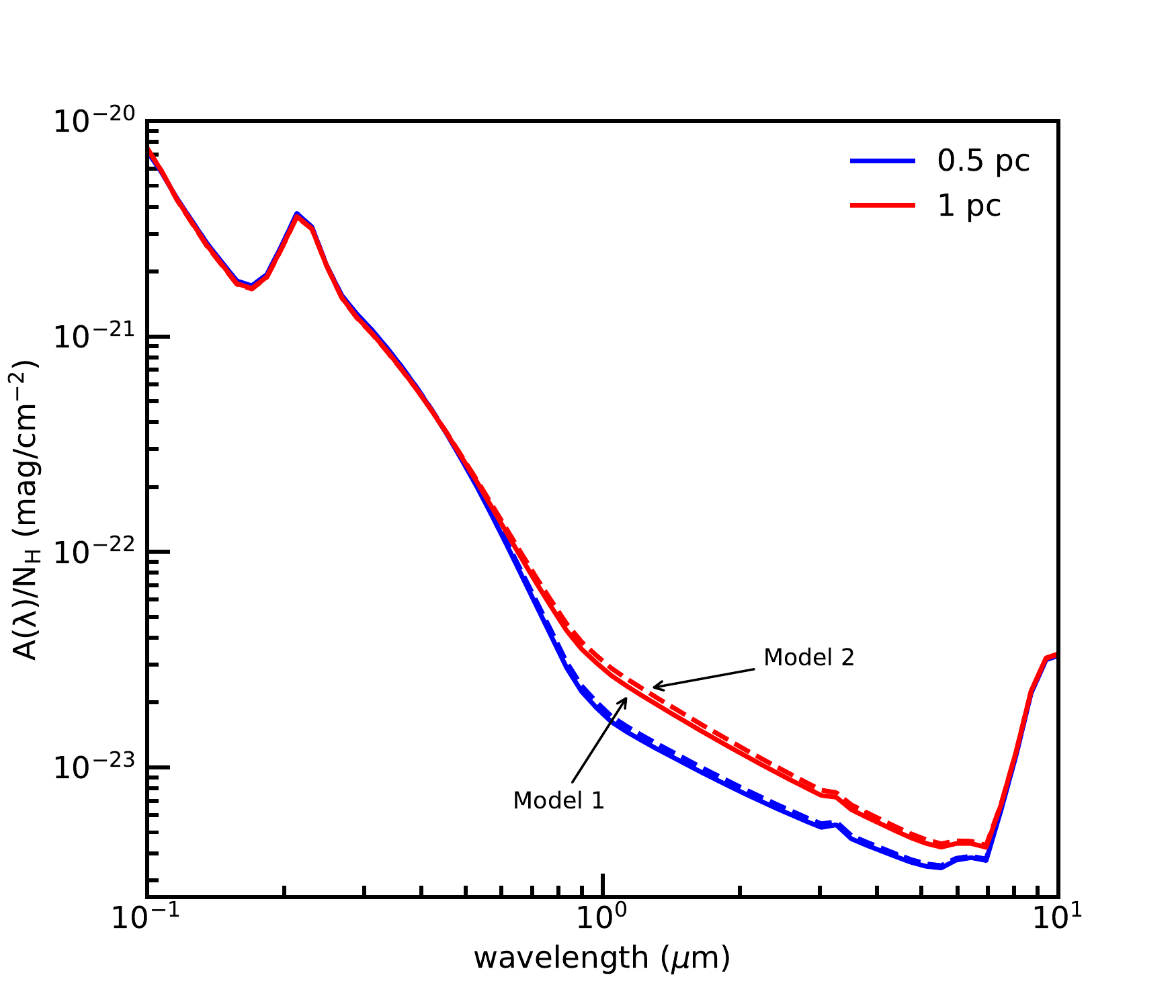}
       \includegraphics[width=0.5\textwidth]{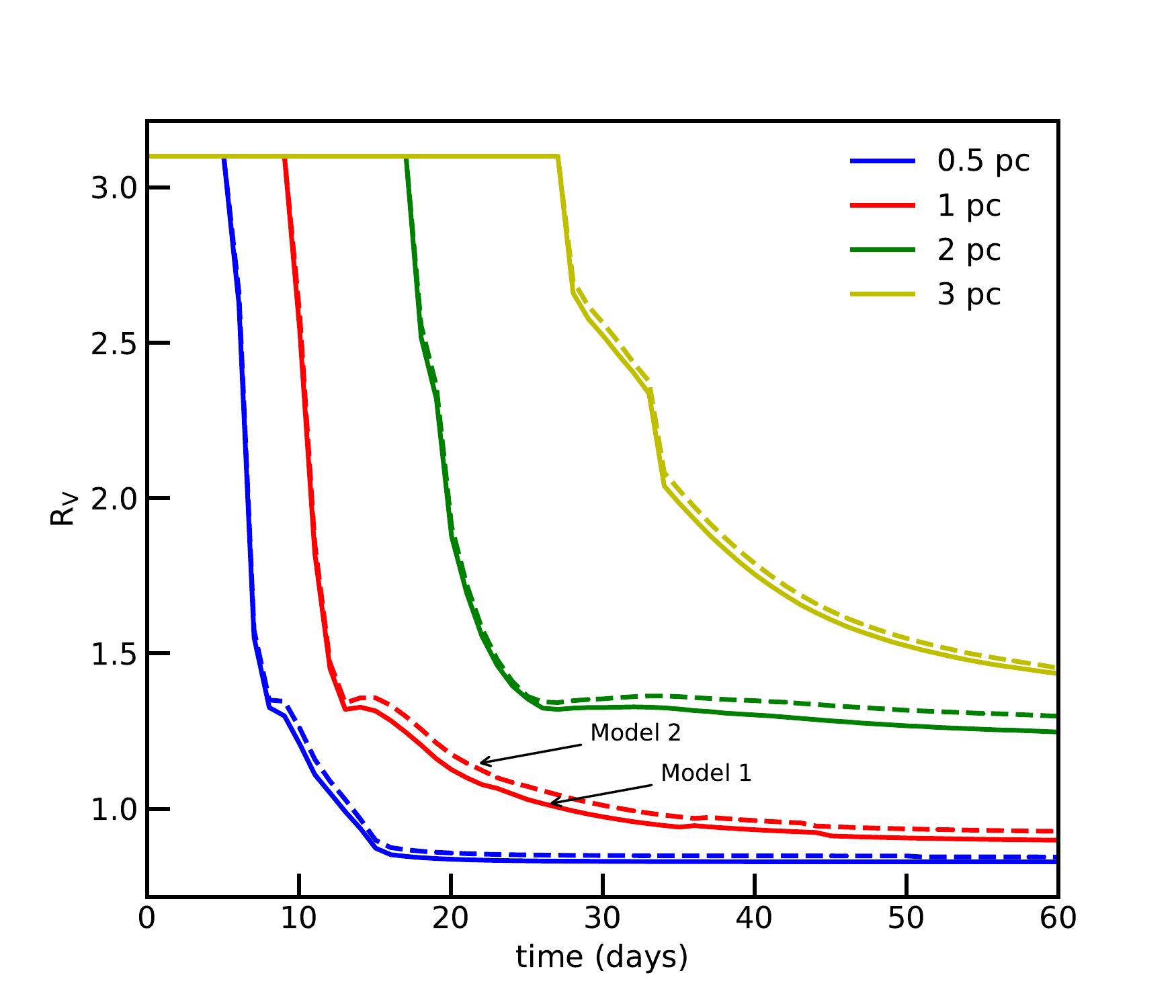}
        \caption{Left panel: extinction curve after 45 days with the cloud at 0.5 pc and 1 pc for model 1 (solid line) and model 2 (dashed line). Right panel: $R_{V}$ vs. times for the different cloud distance . The tensile strength $S_{\rm max}=10^{7}\erg\cm^{-3}$ is assumed.}
           \label{fig:A2}
\end{figure}

Figure \ref{fig:A2} (left panel) shows the extinction curve after 45 days with the cloud at 0.5 pc and 1 pc. One can see that model 1 and 2 give similar extinction, except at optical-NIR wavelengths.

Figure \ref{fig:A2} (right panel) shows the variation of $R_{\rm V}$ with time for the two models. Model 2 gives slightly lower extinction values than model 1 after grain disruption.

\section{Observational data of SNe Ia and stars}\label{apdx:obser}

Table \ref{tab:B1} shows the physical parameters for 13 SNe Ia available in the literature. 
\begin{table*}[t]
  \centering
  \begin{tabular} { cccccc }
  \hline 
  SNe & $\lambda_{\rm max}(\mum)$ & $K$ & $R_{\rm V}$ & $A_{\rm V}$(mag) & $E(B-V)$(mag) \\ 
  
  \hline
  1986G & 0.43  &  1.15  &  2.57  &  2.03  &  0.79 \\ 
  2006X & 0.36 & 1.47 & 1.31 & 1.88 & 1.44 \\
  2008fp(mod1) & 0.15 & 0.40 & 1.20 & 0.71 & 0.59 \\ 
  2008fp(mod2) & 0.41 & 1.03 & 2.20 & 0.29 & 0.13 \\   
  2014J(mod1) & 0.05 & 0.40 & 1.40 & 1.85 & 1.37 \\ 
  2014J(mod2) & 0.35 & 1.15 & 2.59 & 1.17 & 0.45 \\
  SN 2007le & 0.3967 $\pm$ 0.0494 & 1.73 $\pm$ 0.87 & 1.46 $\pm$ 0.28 & 0.54 $\pm$ 0.08 & 0.39 \\ 
  SN 2010ev & 0.4408 $\pm$ 0.0114 & 1.89 $\pm$ 0.23 & 1.54 $\pm$ 0.58 & 0.50 $\pm$ 0.18 & 0.32\\  
  SN 2007fb & 0.3821 $\pm$ 0.0447 & 1.13 $\pm$ 0.46 & 2.17 $\pm$ 0.57 & $\leq$ 0.09 & $\leq$ 0.03 \\ 
  SN 2003W & 0.3996 $\pm$ 0.0371 & 2.56 $\pm$ 1.03 & 1.00 $\pm$ 0.70 & 0.30 & 0.29 \\ 
  SN 2007af & 0.7409 $\pm$ 0.0537 & 1.24 $\pm$ 0.35 & 2.11 $\pm$ 0.51 & 0.31 & 0.15\\ 
  SN 2002fk & 0.4403 $\pm$ 0.0460 & 0.57 $\pm$ 0.35 & 1.73 $\pm$ 1.02 & 0.03 & 0.02\\ 
  SN 2002bo & 0.3525 $\pm$ 0.0137 & 3.95 $\pm$ 0.49 & 1.22 $\pm$ 0.26 & 0.62 $\pm$ 0.10 & 0.53 \\ 
  SN 2011ae & 0.4256 $\pm$ 0.0295 & 3.56 $\pm$ 1.87 & ... & ... & ...\\ 
  SN $2005hk^{a}$ & 0.6731 $\pm$ 0.2116 & -1.36 $\pm$ 1.51 & 3.1 & 0.22 $\pm$ 0.06 & 0.07\\
  \hline  
  \end{tabular}
  \caption{Physical parameters for SNe Ia (\citealt{Phillips13}; \citealt{Ama14}; \citealt{Hoang17}). SN 1986G: (\citealt{Hough87}); SN 2006X: (\citealt{Patat09}); SN 2008fp and SN 2014J: (\citealt{Cox14}; \citealt{Patat15}; \citealt{Kawa14}) and \citealt{Hoang17}). Data for the last SNe Ia are taken from \citealt{Zelaya17}.}
  \label{tab:B1}
\end{table*} 

Table \ref{tab:B2} shows the physical parameters for 15 normal stars with low $R{\rm V}$ but normal peak wavelength taken from \cite{Cikota18}.

\begin{table}[!htb]
\begin{center}
\begin{tabular}{cccccc}
\hline 
SN Name & Telescope & $\lambda_{\rm max}(\mum)$ & $K$ & $R_{\rm V}$ & $E(B-V)$\\ 
\hline 
HD 54439 & VLT & 0.4859 $\pm$ 0.0129 & 0.97 $\pm$ 0.10 & 1.98 & 0.28\\ 
HD 73420 & VLT & 0.5465 $\pm$ 0.0175 & 1.04 $\pm$ 0.21 & 2.24 & 0.37\\
HD 78785 & VLT & 0.5732 $\pm$ 0.08 & 1.25 $\pm$ 0.01 & 2.29 & 0.76\\  
HD 96042 & VLT & 0.5109 $\pm$ 0.0124 & 0.84 $\pm$ 0.09 & 1.87 & 0.48\\  
HD 141318 & VLT & 0.5719 $\pm$ 0.017 & 1.19 $\pm$ 0.03 & 1.77 & 0.30\\ 
HD 152245 & VLT & 0.6169 $\pm$ 0.033 & 1.41 $\pm$ 0.07 & 2.02 & 0.42\\  
HD 152853 & VLT & 0.5584 $\pm$ 0.046 & 1.30 $\pm$ 0.07 & 2.19 & 0.37\\ 
BD +23 3762 & CAHA & 0.4965 $\pm$ 0.061 & 0.92 $\pm$ 0.06 & 2.15 & 1.05\\ 
BD +45 3341 & CAHA & 0.5166 $\pm$ 0.031 & 1.00 $\pm$ 0.05 & 2.22 & 0.74\\
HD 28446 & CAHA & 0.4865 $\pm$ 0.076 & 0.70 $\pm$ 0.07 & 2.20 & 0.46\\
HD 194092 & Asiago & 0.5884 $\pm$ 0.0107 & 1.09 $\pm$ 0.18 & 2.18 & 0.41\\
HD 14357 & Asiago & 0.4942 $\pm$ 0.031 & 0.91 $\pm$ 0.04 & 2.12 & 0.56\\
HD 226868 & HPOL & 0.4425 $\pm$ 0.0262 & 0.57 $\pm$ 0.14 & 2.78 & 1.08\\
HD 218323 & HPOL & 0.4837 $\pm$ 0.0128 & 1.00 $\pm$ 0.20 & 2.30 & 0.90\\        
HD 217035 & HPOL & 0.5309 $\pm$ 0.0126 & 0.88 $\pm$ 0.23 & 2.44 & 0.76\\
\hline
\end{tabular}
\caption{Physical parameter for the normal stars has low value of $R_{\rm V}$, normal $\lambda_{\rm max}$ but high K observed by VLT, CAHA, Asiago and HPOL telescopes (\citealt{Cikota18}).} 
\label{tab:B2}
\end{center}
\end{table}

\section{Relationship between $K-\lambda_{\rm max}$ and $R_{V}-\lambda_{\rm max}$}\label{sec:K_RV}\label{apdx:fit}
\subsection{Fitting polarization curves with Serkowski law}

\begin{figure*}
        \includegraphics[width=0.5\textwidth]{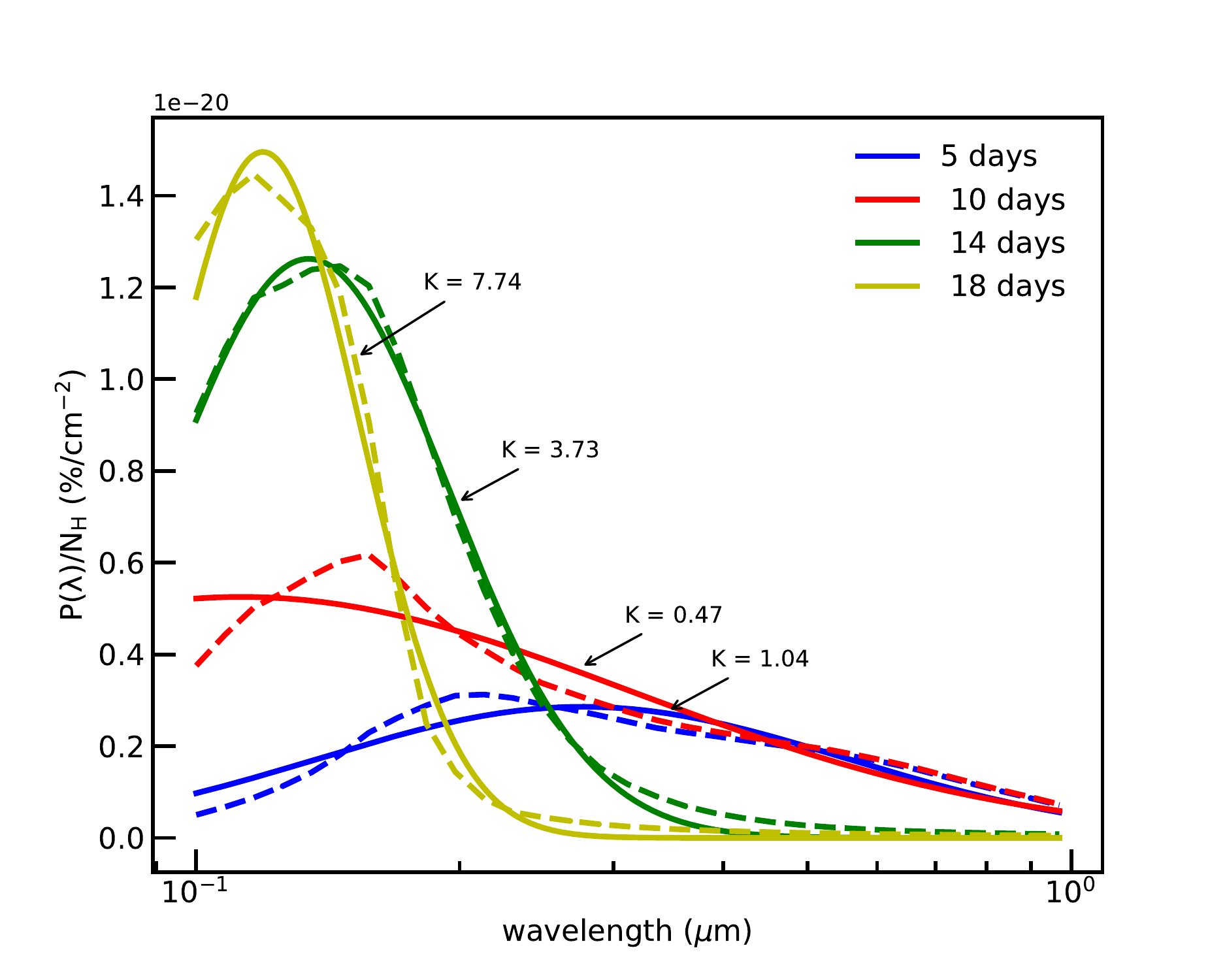}
        \includegraphics[width=0.5\textwidth]{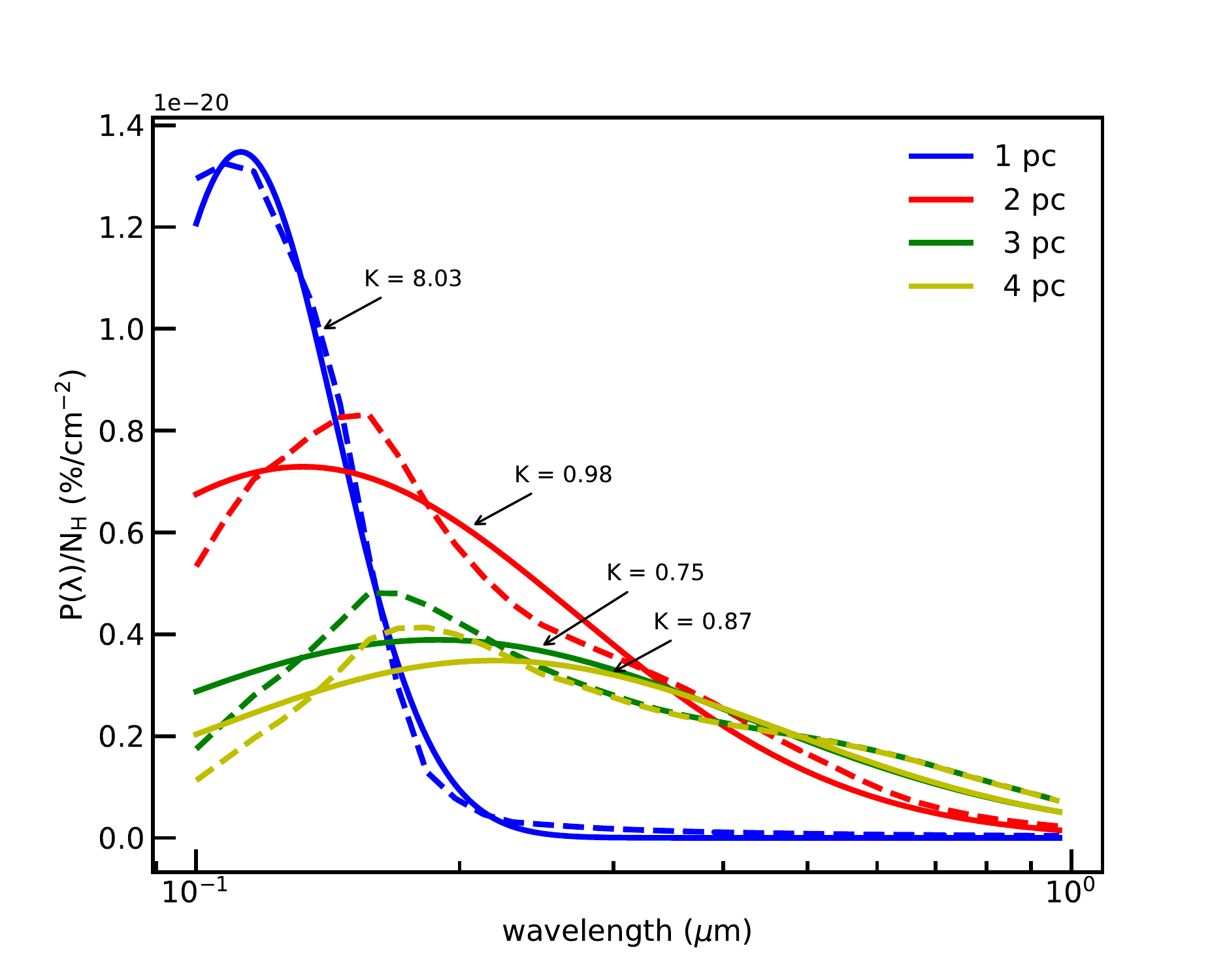}
        \caption{The best-fit Serkowski law (solid lines) for the theoretical polarization curve (dashed lines) computed at the different times for a dust cloud at 1 pc in Figure \ref{fig:Plamda_t} (left panel) and for the different cloud distances at $t= 20$ days in Figure \ref{fig:Plamda_d} (right panel).}
           \label{fig:K_lambda}
\end{figure*}

We fit the theoretical polarization curves with the Serkowski law (Eq. \ref{eq:20}) to derive the best-fit parameters $K$ and $\lambda_{\rm max}$, which are summarized in Table \ref{tab:1}. 
Figure \ref{fig:K_lambda} shows the best fit Serkowski law for the theoretical polarization curves in Figures \ref{fig:Plamda_t} and \ref{fig:Plamda_d} for the different times (left panel) and different cloud distances (right panel). One see that due to grain disruption and grain alignment by RATs begins, the $K$ values increases significantly accompanied with the decrease of $\lambda_{\rm max}$. In other word, there is a simultaneous variation in $K$ and $\lambda_{\rm max}$ with time but they follow the opposite trend.

\begin{table*}[t]
\begin{center}
\begin{tabular}{cccccccc}
\hline 
Day & $a_{\rm align} (\mum)$ & $a_{\rm disr}(\mum)$ & $P_{\rm max}/N_{\rm H} (\%\/cm^{-2})$ & $P_{\rm Serkowski}/N_{\rm H}(\%/cm^{-2})$ & $\lambda_{\rm max}(\mum)$ & $\lambda_{\rm Serkowski}(\mum)$ & $K$ \\ 
\hline 
5  & 2.065e-2 & 0.250 & 3.125e-21 & 2.858e-21 & 0.213 & 0.277 & 1.039 \\ 
10 & 7.227e-3 & 0.250 & 6.171e-21 & 5.254e-21 & 0.158 & 0.113 & 0.472 \\ 
14 & 6.093e-3 & 0.059 & 12.46e-21 & 12.62e-21 & 0.146 & 0.134 & 3.728 \\
18 & 5.754e-3 & 0.028 & 14.47e-21 & 14.95e-21 & 0.116 & 0.192 & 7.744 \\ 
\hline 
\end{tabular} 
\\
\begin{tabular}{cccccccc}
\hline 
\hline
Distance (pc) & $a_{\rm align} (\mum)$ & $a_{\rm disr}( \mum)$ & $P_{\rm max}/N_{\rm H} (\%/cm^{-2})$ & $P_{\rm Serkowski}/N_{\rm H}(\%/cm^{-2})$ & $\lambda_{\rm max}( \mum)$ & $\lambda_{\rm Serkowski}(\mum)$ & K \\ 
\hline 
1  & 5.670e-3 & 0.022 & 13.24e-21 & 13.47e-21 & 0.108 & 0.112 & 8.033 \\ 
2 & 6.870e-3 & 0.143 & 8.314e-21 & 7.291e-21 & 0.157 & 0.132 & 0.977 \\ 
3 & 1.164e-2 & 0.250 & 4.809e-21 & 3.891e-21 & 0.158 & 0.188 & 0.759 \\
4 & 1.435e-2 & 0.250 & 4.135e-21 & 3.485e-21 & 0.183 & 0.220 & 0.869 \\ 
\hline 
\end{tabular} 
\caption{Model parameters and fitting parameters of the Serkowski law to the polarization curves computed at different times for a given cloud distance (upper table) and for different cloud distance at $t=20$ days (lower table).} 
\end{center}
\label{tab:1}
\end{table*}

\subsection{$K-\lambda_{\rm max}$ and $R_{V}-\lambda_{\rm max}$}
To understand in more detail how $K$ and $\lambda_{\rm max}$ change with grain disruption and alignment by RATs, we model the dust polarization by varying $a_{\rm disr}$ from $0.2 \mum$ to $0.03 \mum$ and $a_{\rm align}$ from $0.05 \mum$ to $0.002 \mum$ and derive the best-fit $K$ and $\lambda_{\rm max}$ parameters as in the previous section.

\begin{figure*}[!htb]
             \includegraphics[width=0.5\textwidth]{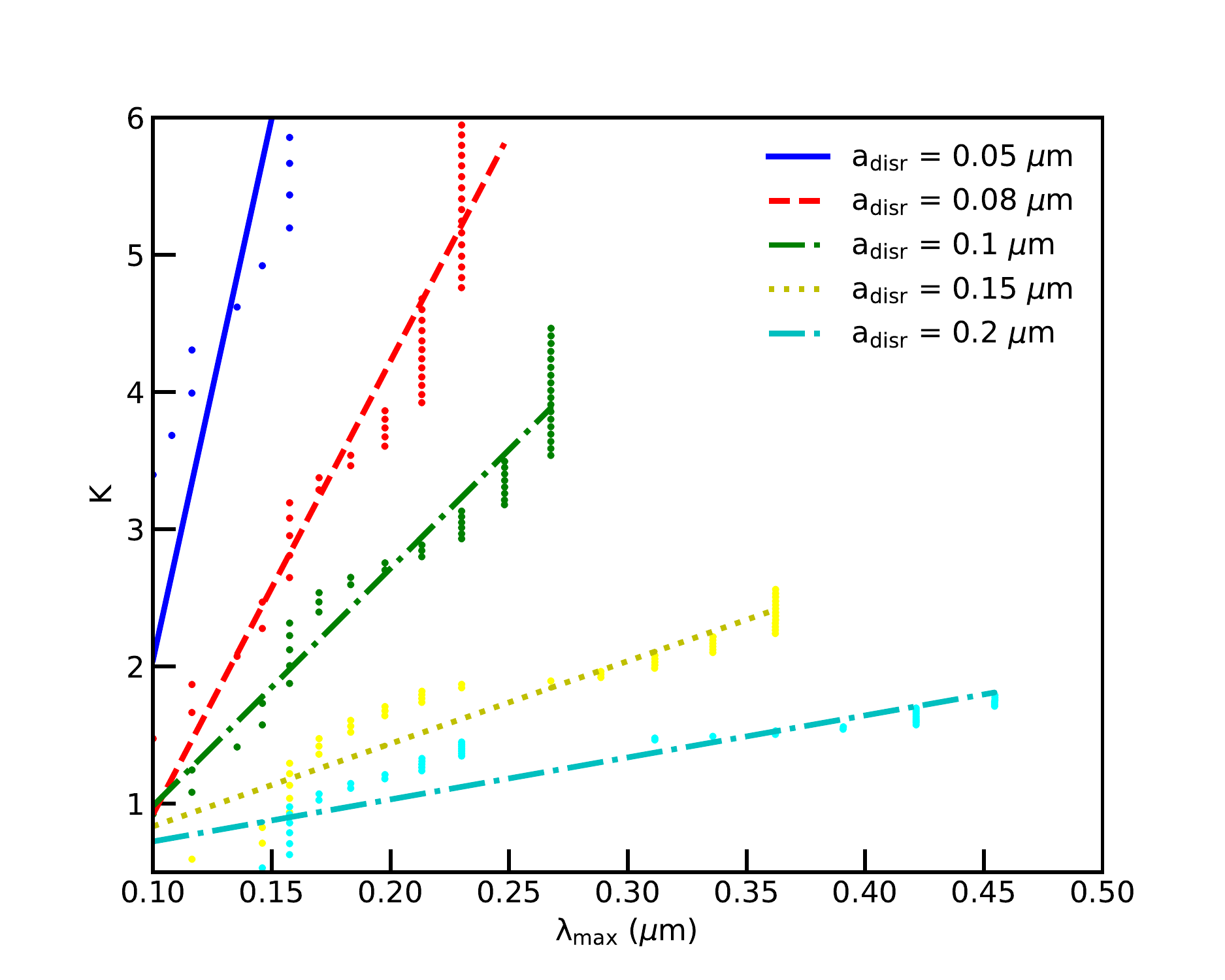}
        \includegraphics[width=0.5\textwidth]{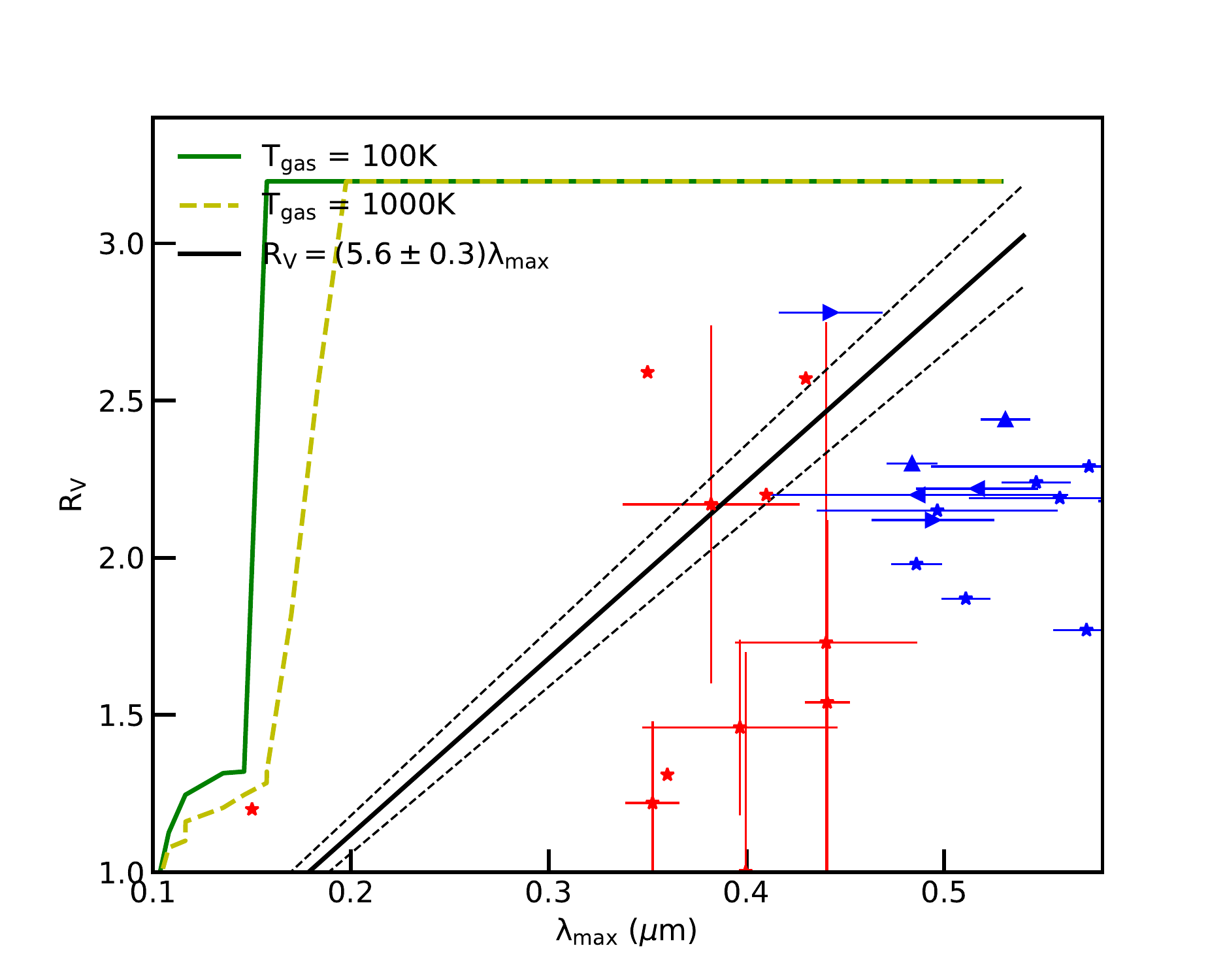}
        \caption{Left panel: $K$ vs. $\lambda_{\rm max}$ from the best-fit. Right panel: $R_{\rm V}$ vs. $\lambda_{\rm max}$ from our models over-plotted with observational data for different gas density $T_{\rm gas}$ case.}
           \label{fig:lamdamax_Rv}
\end{figure*}

In Figure \ref{fig:lamdamax_Rv} (left panel) we show the resulting $K$-$\lambda_{\rm max}$ values. Each point represents a couple ($K, \lambda_{\rm max}$) predicted by a given value of ($a_{\rm disr},a_{\rm align}$). Several values of $K$ at a given $\lambda_{\rm max}$ represents the different values of $P_{\rm max}$ when $a_{\rm align}$ decreases, and the dashed line is the best fit to these results. 

In Figure \ref{fig:lamdamax_Rv} (right panel), we plot $R_{\rm V}$ against $\lambda_{\rm max}$ evaluated at $t= 60$ days for $S_{\rm max}=10^{7}\erg\cm^{-3}$ with different gas temperatures. The black solid line shows the tentative relationship reported by \cite{1978A&A....66...57W}. As shown, both data of SNe Ia and our modeling results do not exhibit correlation between $R_{V}$ and $\lambda_{\rm max}$, in agreement with the finding in \cite{Cikota18}.

\bibliography{reference}

\begin{thebibliography}{}
\expandafter\ifx\csname natexlab\endcsname\relax\def\natexlab#1{#1}\fi

\bibitem[{{Amanullah} {et~al.}(2014){Amanullah}, {Goobar}, {Johansson},
  {Banerjee}, {Venkataraman}, {Joshi}, {Ashok}, {Cao}, {Kasliwal}, {Kulkarni},
  {Nugent}, {Petrushevska}, \& {Stanishev}}]{Ama14}
{Amanullah}, R., {Goobar}, A., {Johansson}, J., {et~al.} 2014, \apjl, 788, L21

\bibitem[{{Amanullah} {et~al.}(2015){Amanullah}, {Johansson}, {Goobar},
  {Ferretti}, {Papadogiannakis}, {Petrushevska}, {Brown}, {Cao}, {Contreras},
  {Dahle}, {Elias-Rosa}, {Fynbo}, {Gorosabel}, {Guaita}, {Hangard}, {Howell},
  {Hsiao}, {Kankare}, {Kasliwal}, {Leloudas}, {Lundqvist}, {Mattila}, {Nugent},
  {Phillips}, {Sandberg}, {Stanishev}, {Sullivan}, {Taddia}, {{\"O}stlin},
  {Asadi}, {Herrero-Illana}, {Jensen}, {Karhunen}, {Lazarevic}, {Varenius},
  {Santos}, {Sridhar}, {Wallstr{\"o}m}, \& {Wiegert}}]{Ama15}
{Amanullah}, R., {Johansson}, J., {Goobar}, A., {et~al.} 2015, \mnras, 453,
  3300

\bibitem[{{Andersson} {et~al.}(2015){Andersson}, {Lazarian}, \&
  {Vaillancourt}}]{Ander15}
{Andersson}, B.-G., {Lazarian}, A., \& {Vaillancourt}, J.~E. 2015, \araa, 53,
  501

\bibitem[{{Arnett}(1982)}]{Arnet82}
{Arnett}, W.~D. 1982, \apj, 253, 785

\bibitem[{{Brown} {et~al.}(2015){Brown}, {Smitka}, {Wang}, {Breeveld}, {de
  Pasquale}, {Hartmann}, {Krisciunas}, {Kuin}, {Milne}, {Page}, \&
  {Siegel}}]{Brown15}
{Brown}, P.~J., {Smitka}, M.~T., {Wang}, L., {et~al.} 2015, \apj, 805, 74

\bibitem[{Bulla {et~al.}(2018)Bulla, Goobar, \& Dhawan}]{Bulla18}
Bulla, M., Goobar, A., \& Dhawan, S. 2018, \mnras, 479, 3663

\bibitem[{Burns {et~al.}(2014)Burns, Stritzinger, Phillips, Hsiao, Contreras,
  Persson, Folatelli, Boldt, Campillay, Castell{\'{o}}n, Freedman, Madore,
  Morrell, Salgado, \& Suntzeff}]{Burn14}
Burns, C.~R., Stritzinger, M., Phillips, M.~M., {et~al.} 2014, \apj, 789, 32

\bibitem[{{Chiar} {et~al.}(2006){Chiar}, {Adamson}, {Whittet}, {Chrysostomou},
  {Hough}, {Kerr}, {Mason}, {Roche}, \& {Wright}}]{Chiar06}
{Chiar}, J.~E., {Adamson}, A.~J., {Whittet}, D.~C.~B., {et~al.} 2006, \apj,
  651, 268

\bibitem[{Cikota {et~al.}(2016)Cikota, Deustua, \& Marleau}]{Cikota16}
Cikota, A., Deustua, S., \& Marleau, F. 2016, \apj, 819, 152

\bibitem[{{Cikota} {et~al.}(2018){Cikota}, {Hoang}, {Taubenberger}, {Patat},
  {Mazzei}, {Cox}, {Zelaya}, {Cikota}, {Tomasella}, {Benetti}, \&
  {Rodeghiero}}]{Cikota18}
{Cikota}, A., {Hoang}, T., {Taubenberger}, S., {et~al.} 2018, \aap, 615, A42

\bibitem[{{Cox} \& {Patat}(2014)}]{Cox14}
{Cox}, N.~L.~J., \& {Patat}, F. 2014, \aap, 565, A61

\bibitem[{{Draine}(2003)}]{Draine03}
{Draine}, B.~T. 2003, \araa, 41, 241

\bibitem[{{Draine} \& {Lazarian}(1998)}]{Draine98}
{Draine}, B.~T., \& {Lazarian}, A. 1998, \apj, 508, 157

\bibitem[{{Draine} \& {Li}(2007)}]{Draine07}
{Draine}, B.~T., \& {Li}, A. 2007, \apj, 657, 810

\bibitem[{{Draine} \& {Weingartner}(1996)}]{Draine96}
{Draine}, B.~T., \& {Weingartner}, J.~C. 1996, \apj, 470, 551

\bibitem[{{Foley} {et~al.}(2014){Foley}, {Fox}, {McCully}, {Phillips}, {Sand},
  {Zheng}, {Challis}, {Filippenko}, {Folatelli}, {Hillebrandt}, {Hsiao}, {Jha},
  {Kirshner}, {Kromer}, {Marion}, {Nelson}, {Pakmor}, {Pignata}, {R{\"o}pke},
  {Seitenzahl}, {Silverman}, {Skrutskie}, \& {Stritzinger}}]{Foley14}
{Foley}, R.~J., {Fox}, O.~D., {McCully}, C., {et~al.} 2014, \mnras, 443, 2887

\bibitem[{{Goobar}(2008)}]{Goobar08}
{Goobar}, A. 2008, \apjl, 686, L103

\bibitem[{Herranen {et~al.}(2019)Herranen, Lazarian, \&
  Hoang}]{Herranen:2019kj}
Herranen, J., Lazarian, A., \& Hoang, T. 2019, \apj, 878, 0

\bibitem[{{Hillebrandt} \& {Niemeyer}(2000)}]{Hill00}
{Hillebrandt}, W., \& {Niemeyer}, J.~C. 2000, \araa, 38, 191

\bibitem[{Hoang(2017)}]{Hoang17}
Hoang, T. 2017, \apj, 836, 13

\bibitem[{{Hoang} \& {Lazarian}(2008)}]{Hoang08}
{Hoang}, T., \& {Lazarian}, A. 2008, \mnras, 388, 117

\bibitem[{{Hoang} \& {Lazarian}(2009)}]{Hoan09}
---. 2009, \apj, 695, 1457

\bibitem[{{Hoang} \& {Lazarian}(2014)}]{Hoang14}
---. 2014, \mnras, 438, 680

\bibitem[{Hoang \& Lazarian(2016)}]{Hoang16}
Hoang, T., \& Lazarian, A. 2016, \apj, 831, 159

\bibitem[{{Hoang} {et~al.}(2013){Hoang}, {Lazarian}, \& {Martin}}]{Hoang13}
{Hoang}, T., {Lazarian}, A., \& {Martin}, P.~G. 2013, \apj, 779, 152

\bibitem[{Hoang {et~al.}(2014)Hoang, Lazarian, \& Martin}]{Hoan14}
Hoang, T., Lazarian, A., \& Martin, P.~G. 2014, \apj, 790, 6

\bibitem[{{Hoang} {et~al.}(2019){Hoang}, {Tram}, {Lee}, \& {Ahn}}]{Hoang19}
{Hoang}, T., {Tram}, L.~N., {Lee}, H., \& {Ahn}, S.-H. 2019, Nature Astronomy,
  319

\bibitem[{{Hough} {et~al.}(1987){Hough}, {Bailey}, {Rouse}, \&
  {Whittet}}]{Hough87}
{Hough}, J.~H., {Bailey}, J.~A., {Rouse}, M.~F., \& {Whittet}, D.~C.~B. 1987,
  \mnras, 227, 1P

\bibitem[{{Kawabata} {et~al.}(2014){Kawabata}, {Akitaya}, {Yamanaka}, {Itoh},
  {Maeda}, {Moritani}, {Ui}, {Kawabata}, {Mori}, {Nogami}, {Nomoto}, {Suzuki},
  {Takaki}, {Tanaka}, {Ueno}, {Chiyonobu}, {Harao}, {Matsui}, {Miyamoto},
  {Nagae}, {Nakashima}, {Nakaya}, {Ohashi}, {Ohsugi}, {Komatsu}, {Sakimoto},
  {Sasada}, {Sato}, {Tanaka}, {Urano}, {Yamashita}, {Yoshida}, {Arai},
  {Ebisuda}, {Fukazawa}, {Fukui}, {Hashimoto}, {Honda}, {Izumiura}, {Kanda},
  {Kawaguchi}, {Kawai}, {Kuroda}, {Masumoto}, {Matsumoto}, {Nakaoka}, {Takata},
  {Uemura}, \& {Yanagisawa}}]{Kawa14}
{Kawabata}, K.~S., {Akitaya}, H., {Yamanaka}, M., {et~al.} 2014, \apjl, 795, L4

\bibitem[{{Lazarian} {et~al.}(2015){Lazarian}, {Andersson}, \&
  {Hoang}}]{Laza15}
{Lazarian}, A., {Andersson}, B.-G., \& {Hoang}, T. 2015, {Grain alignment: Role
  of radiative torques and paramagnetic relaxation}, ed. L.~{Kolokolova},
  J.~{Hough}, \& A.-C. {Levasseur-Regourd}, 81

\bibitem[{{Lazarian} \& {Hoang}(2007)}]{Laza07}
{Lazarian}, A., \& {Hoang}, T. 2007, \mnras, 378, 910

\bibitem[{{Lazarian} \& {Hoang}(2018)}]{Laza18}
---. 2018, arXiv e-prints, arXiv:1810.10686

\bibitem[{{Mathis} {et~al.}(1977){Mathis}, {Rumpl}, \& {Nordsieck}}]{Mathis77}
{Mathis}, J.~S., {Rumpl}, W., \& {Nordsieck}, K.~H. 1977, \apj, 217, 425

\bibitem[{{Nobili} \& {Goobar}(2008)}]{Nobi08}
{Nobili}, S., \& {Goobar}, A. 2008, \aap, 487, 19

\bibitem[{{Patat} {et~al.}(2009){Patat}, {Baade}, {H{\"o}flich}, {Maund},
  {Wang}, \& {Wheeler}}]{Patat09}
{Patat}, F., {Baade}, D., {H{\"o}flich}, P., {et~al.} 2009, \aap, 508, 229

\bibitem[{{Patat} {et~al.}(2015){Patat}, {Taubenberger}, {Cox}, {Baade},
  {Clocchiatti}, {H{\"o}flich}, {Maund}, {Reilly}, {Spyromilio}, {Wang},
  {Wheeler}, \& {Zelaya}}]{Patat15}
{Patat}, F., {Taubenberger}, S., {Cox}, N.~L.~J., {et~al.} 2015, \aap, 577, A53

\bibitem[{{Phillips} {et~al.}(2013){Phillips}, {Simon}, {Morrell}, {Burns},
  {Cox}, {Foley}, {Karakas}, {Patat}, {Sternberg}, {Williams}, {Gal-Yam},
  {Hsiao}, {Leonard}, {Persson}, {Stritzinger}, {Thompson}, {Campillay},
  {Contreras}, {Folatelli}, {Freedman}, {Hamuy}, {Roth}, {Shields}, {Suntzeff},
  {Chomiuk}, {Ivans}, {Madore}, {Penprase}, {Perley}, {Pignata}, {Preston}, \&
  {Soderberg}}]{Phillips13}
{Phillips}, M.~M., {Simon}, J.~D., {Morrell}, N., {et~al.} 2013, \apj, 779, 38

\bibitem[{Riess {et~al.}(1998)Riess, Filippenko, Challis, Clocchiatti, Diercks,
  Garnavich, Gilliland, Hogan, Jha, Kirshner, Leibundgut, Phillips, Reiss,
  Schmidt, Schommer, Smith, Spyromilio, Stubbs, Suntzeff, \&
  Tonry}]{1998AJ....116.1009R}
Riess, A.~G., Filippenko, A.~V., Challis, P., {et~al.} 1998, \apj, 116, 1009

\bibitem[{{Riess} {et~al.}(1999){Riess}, {Filippenko}, {Li}, {Treffers},
  {Schmidt}, {Qiu}, {Hu}, {Armstrong}, {Faranda}, {Thouvenot}, \&
  {Buil}}]{Riess99}
{Riess}, A.~G., {Filippenko}, A.~V., {Li}, W., {et~al.} 1999, \aj, 118, 2675

\bibitem[{Scolnic {et~al.}(2019)Scolnic, Perlmutter, Aldering, Brout, Davis,
  Filippenko, Foley, Hlozek, Hounsell, Jones, Kelly, Kessler, Kim, Rubin,
  Riess, Rodney, Roberts-Pierel, Wang, Asorey, Avelino, Bavdhankar, Brown,
  Challinor, Balland, Cooray, Dhawan, Dimitriadis, Dvorkin, Guy, Handley,
  Keeley, Kneib, LHuillier, Lattanzi, Mandel, Mertens, Rigault, Motloch,
  Mukherjee, Narayan, Nomerotski, Page, Pogosian, Puglisi, Raveri, Regnault,
  Rest, Rojas-Bravo, Sako, Shi, Sridhar, Suzuki, Tsai, Wood-Vasey, Copin, Zhao,
  \& Zhu}]{Scolnic:2019ug}
Scolnic, D., Perlmutter, S., Aldering, G., {et~al.} 2019, arXiv:1903.05128,
  1903.05128v1

\bibitem[{{Serkowski} {et~al.}(1975){Serkowski}, {Mathewson}, \&
  {Ford}}]{Serkow75}
{Serkowski}, K., {Mathewson}, D.~S., \& {Ford}, V.~L. 1975, \apj, 196, 261

\bibitem[{Wang {et~al.}(2008)Wang, Li, Filippenko, Foley, Smith, \&
  Wang}]{Wang08}
Wang, X., Li, W., Filippenko, A.~V., {et~al.} 2008, \apj, 677, 1060

\bibitem[{{Wang} {et~al.}(2012){Wang}, {Wang}, {Filippenko}, {Baron}, {Kromer},
  {Jack}, {Zhang}, {Aldering}, {Antilogus}, {Arnett}, {Baade}, {Barris},
  {Benetti}, {Bouchet}, {Burrows}, {Canal}, {Cappellaro}, {Carlberg}, {di
  Carlo}, {Challis}, {Crotts}, {Danziger}, {Della Valle}, {Fink}, {Foley},
  {Fransson}, {Gal-Yam}, {Garnavich}, {Gerardy}, {Goldhaber}, {Hamuy},
  {Hillebrandt}, {H{\"o}flich}, {Holland}, {Holz}, {Hughes}, {Jeffery}, {Jha},
  {Kasen}, {Khokhlov}, {Kirshner}, {Knop}, {Kozma}, {Krisciunas}, {Lee},
  {Leibundgut}, {Lentz}, {Leonard}, {Lewin}, {Li}, {Livio}, {Lundqvist},
  {Maoz}, {Matheson}, {Mazzali}, {Meikle}, {Miknaitis}, {Milne}, {Mochnacki},
  {Nomoto}, {Nugent}, {Oran}, {Panagia}, {Perlmutter}, {Phillips}, {Pinto},
  {Poznanski}, {Pritchet}, {Reinecke}, {Riess}, {Ruiz-Lapuente}, {Scalzo},
  {Schlegel}, {Schmidt}, {Siegrist}, {Soderberg}, {Sollerman}, {Sonneborn},
  {Spadafora}, {Spyromilio}, {Sramek}, {Starrfield}, {Strolger}, {Suntzeff},
  {Thomas}, {Tonry}, {Tornambe}, {Truran}, {Turatto}, {Turner}, {Van Dyk},
  {Weiler}, {Wheeler}, {Wood-Vasey}, {Woosley}, \& {Yamaoka}}]{Wang12}
{Wang}, X., {Wang}, L., {Filippenko}, A.~V., {et~al.} 2012, \apj, 749, 126

\bibitem[{{Waxman} \& {Draine}(2000)}]{Wax20}
{Waxman}, E., \& {Draine}, B.~T. 2000, \apj, 537, 796

\bibitem[{{Weingartner} \& {Draine}(2001)}]{Wein01}
{Weingartner}, J.~C., \& {Draine}, B.~T. 2001, \apj, 548, 296

\bibitem[{{Whittet} {et~al.}(1992){Whittet}, {Martin}, {Hough}, {Rouse},
  {Bailey}, \& {Axon}}]{Whit92}
{Whittet}, D.~C.~B., {Martin}, P.~G., {Hough}, J.~H., {et~al.} 1992, \apj, 386,
  562

\bibitem[{Whittet \& van Breda(1978)}]{1978A&A....66...57W}
Whittet, D. C.~B., \& van Breda, I.~G. 1978, A\&A, 66, 57

\bibitem[{{Wilking} {et~al.}(1980){Wilking}, {Lebofsky}, {Martin}, {Rieke}, \&
  {Kemp}}]{Wilking80}
{Wilking}, B.~A., {Lebofsky}, M.~J., {Martin}, P.~G., {Rieke}, G.~H., \&
  {Kemp}, J.~C. 1980, \apj, 235, 905

\bibitem[{Zelaya {et~al.}(2017)Zelaya, Clocchiatti, Baade, Höflich, Maund,
  Patat, Quinn, Reilly, Wang, Wheeler, Förster, \&
  Gonz{\'{a}}lez-Gait{\'{a}}n}]{Zelaya17}
Zelaya, P., Clocchiatti, A., Baade, D., {et~al.} 2017, \apj, 836, 88

\bibitem[{{Zheng} {et~al.}(2017){Zheng}, {Kelly}, \& {Filippenko}}]{Zheng17}
{Zheng}, W., {Kelly}, P.~L., \& {Filippenko}, A.~V. 2017, \apj, 848, 66

\end{thebibliography}

\end{document}